\documentclass[superscriptaddress,showpacs,twocolumn,prd]{revtex4}
\usepackage{graphicx}

\begin{document}

\onecolumngrid
\thispagestyle{empty}
\begin{flushright}
{\large 
LU TP 05-26\\
\large hep-lat/0506004\\
\large revised\\
\large July 2005}
\end{flushright}
\vskip5cm
\begin{center}
{\Large\bf
Masses and Decay Constants of Pseudoscalar Mesons to Two Loops \\[0.5cm]
       in Two-Flavor Partially Quenched Chiral Perturbation Theory}

\vskip2cm

{\large \bf Johan Bijnens and Timo A. L\"ahde}\\[1cm]
{Department of Theoretical Physics, Lund University,\\
S\"olvegatan 14A, S 223 62 Lund, Sweden}

\vskip3cm

{\large\bf Abstract}

\vskip1cm

\parbox{14cm}{\large
This paper presents a first study of the masses and decay constants of 
the charged, or flavor-off-diagonal, pseudoscalar mesons to two loops 
for two flavors of sea-quarks, in Partially Quenched Chiral Perturbation 
Theory (PQ$\chi$PT). Explicit analytical expressions up to ${\cal 
O}(p^6)$ in the momentum expansion are given. The calculations have been 
performed within the supersymmetric formulation of PQ$\chi$PT. A 
numerical analysis is done to indicate the size of the 
corrections.
}

\vskip2cm

{\large{\bf PACS}: {12.38.Gc, 12.39.Fe, 11.30.Rd} }
\end{center}
\vskip2cm
\twocolumngrid
\setcounter{page}{0}

\title{Masses and Decay Constants of Pseudoscalar Mesons to Two Loops \\
       in Two-Flavor Partially Quenched Chiral Perturbation Theory
      }

\author{Johan Bijnens}
\affiliation{Department of Theoretical Physics, Lund University,\\
S\"olvegatan 14A, S 223 62 Lund, Sweden}
\author{Timo A. L\"ahde}
\affiliation{Department of Theoretical Physics, Lund University,\\
S\"olvegatan 14A, S 223 62 Lund, Sweden}

\pacs{12.38.Gc, 12.39.Fe, 11.30.Rd}

\begin{abstract} 
This paper presents a first study of the masses and decay constants of 
the charged, or flavor-off-diagonal, pseudoscalar mesons to two loops 
for two flavors of sea-quarks, in Partially Quenched Chiral Perturbation 
Theory (PQ$\chi$PT). Explicit analytical expressions up to ${\cal 
O}(p^6)$ in the momentum expansion are given. The calculations have been 
performed within the supersymmetric formulation of PQ$\chi$PT. A 
numerical analysis is done to indicate the size of the 
corrections.
\end{abstract}

\maketitle
\section{Introduction}

At the present time, the most promising way to derive information on 
low-energy hadronic observables directly from QCD is by means of 
numerical Lattice QCD simulations. However, it is well known that
such simulations are computationally very demanding if dynamical 
sea-quark effects (unquenched Lattice QCD) are to be taken into account.
Although much progress has been made recently, simulations with
sea-quark masses that are close to the physical $u,d$ quark masses 
of a few MeV are still not available, whereas sea-quark masses down to 
several tens of MeV are becoming possible. The physical values of the 
quark masses then have to be reached by extrapolation from these 
simulations.

A generalization of QCD where the valence quark masses are different
from the sea-quark masses is possible, and is called Partially Quenched 
QCD (PQQCD). QCD may then be obtained as a continuous limit from this 
theory, in contrast to the quenched calculations where no such smooth 
connection exists. The reason for doing PQQCD Lattice simulations is 
that the valence quark masses can be varied much more efficiently than 
the sea-quark masses, which allows for more extensive studies of the 
quark mass dependence of the various observables.

The generalization of $\chi$PT to sea-quark masses different from the 
valence ones, i.e. to the partially quenched case~\cite{BG1,BG2}, 
provides a practical method for the extrapolation to physical quark 
masses from the lattice simulations of PQQCD. The quark mass dependence 
of partially quenched chiral perturbation theory (PQ$\chi$PT) 
is explicit, and thus the limit where the sea-quark masses become equal 
to the valence quark masses can be taken. As a consequence, unquenched 
$\chi$PT~\cite{GL} is included in PQ$\chi$PT, and the free 
parameters, or low-energy constants (LEC:s), of $\chi$PT can be 
directly obtained from those of PQ$\chi$PT~\cite{BG2,Sharpe1}. The need
for PQ$\chi$PT calculations to next-to-next-to-leading order (NNLO) has 
also been demonstrated by Lattice QCD simulations~\cite{Latt1,Latt2}. 

The calculation of the charged pseudoscalar meson masses and decay 
constants to one loop (NLO) in PQ$\chi$PT has been carried out in 
Refs.~\cite{BG2,Sharpe1,Sharpe2}, and first results for the mass of a 
pseudoscalar meson at two loops (NNLO) in PQ$\chi$PT have 
already appeared, for three flavors of degenerate sea-quarks~\cite{BDL}. 
The decay constants of the pseudoscalar mesons are also known
to NNLO~\cite{BL1} for three sea-quarks. A full PQ$\chi$PT calculation 
of the pseudoscalar meson masses for three flavors of sea-quarks is in 
progress~\cite{BDL2}. However, as many Lattice QCD simulations are 
performed with two rather than three sea-quark flavors, such 
expressions for the pseudoscalar meson masses and decay constants to 
NNLO are also called for.

This paper presents the full calculation of the masses and 
decay constants of the charged, or flavor off-diagonal, pseudoscalar mesons 
in NNLO PQ$\chi$PT, for two flavors of sea-quarks ($n_{\mathrm{sea}} = 
2$). 
The results are characterized by the number of nondegenerate valence and 
sea-quarks, denoted $d_{\mathrm{val}}$ and $d_{\mathrm{sea}}$, 
respectively. For 
the decay constants of the charged pseudoscalar mesons, the maximum number 
of nondegenerate valence quark masses is $d_{\mathrm{val}} = 2$. The degree 
of quark mass degeneracy in each result is sometimes also referred to with 
the notation $d_{\mathrm{val}} + d_{\mathrm{sea}}$. The decay constant of 
the charged pion in the $SU(2)$ symmetric limit for both valence and 
sea-quarks
thus corresponds to the $1 + 1$ case.

The analytical expressions for the NNLO shifts of the meson masses and decay 
constants are in general very long, but the expressions simplify 
considerably when the sea or valence quark masses become degenerate. In 
view of this, the NNLO loop results are given separately for each case of 
$d_{\mathrm{val}} + d_{\mathrm{sea}}$ considered. In the next sections, a 
short technical overview of the NNLO calculations is given, with an emphasis
on the parts not covered in the previous publications~\cite{BDL,BL1},
along with the 
full results for the NNLO contributions to the meson masses and decay
constants. Finally, a numerical analysis of the results is given, along 
with a concluding discussion.

\section{Technical Overview}

The technical aspects of the PQ$\chi$PT calculations to NLO have been 
thoroughly covered in Refs.~\cite{Sharpe1,Sharpe2}. The new parts needed 
for NNLO are treated in Refs.~\cite{BDL,BL1}. For the sake of 
completeness, some of the issues which are directly relevant for the 
present two-flavor calculations are repeated here.

Most significantly, the Lagrangians of PQ$\chi$PT at ${\cal O}(p^4)$ and
${\cal O}(p^6)$ can be directly obtained from the corresponding
Lagrangians of unquenched $n_F$ flavor $\chi$PT by replacement of 
traces with supertraces. This follows because the derivation
of these Lagrangians in Ref.~\cite{BCE1} only used relations which 
remain valid when all traces are replaced by the corresponding 
supertraces. For the definition of the supertraces in PQ$\chi$PT
see Refs~\cite{BG1,BG2,Sharpe1}. We work here in the supersymmetric 
version of PQ$\chi$PT without the singlet $\Phi_0$ as discussed in 
Ref.~\cite{Sharpe2}. The divergences of this version of
PQ$\chi$PT also follow directly from those of $n_f$ flavor $\chi$PT 
calculated in Ref.\cite{BCE2}, provided that $n_f$ is set equal to the 
number of sea-quark flavors. This is so since all the manipulations in
Ref.~\cite{BCE2} for $n_f$ flavors remain valid when traces are replaced 
with supertraces. Alternatively, this derivation of the Lagrangians of 
PQ$\chi$PT can be argued for with the Replica method of 
Ref.~\cite{replica}.

\subsection{Quark Masses and Low-Energy Constants}

The number of independent low-energy constants (LEC:s) in unquenched and 
partially quenched $\chi$PT is slightly different, but the former are 
always linear combinations of the latter. The number of LEC:s is also 
dependent on the number of flavors. Table~\ref{tab:LEC} gives an 
overview of the various sets of LEC:s and the number of parameters up to 
NNLO for the different versions of $\chi$PT and PQ$\chi$PT.

\begin{table}[h!]
\begin{center}
\caption{The relevant sets of LEC:s, where the $i+j$ notation 
indicates the number of physically relevant ($i$) and contact ($j$) 
terms in the respective Lagrangians. At NLO, the latter ones are usually 
denoted $h_i^r$ for $n_f = 2$ and $H_i^r$ for $n_f = 3$. The relations 
between the various sets of LEC:s is discussed in the text. For 
simplicity, the superscript $(npq)$ of the partially quenched 
LEC:s has been suppressed in most of this paper.}
\vspace{.3cm}
\begin{tabular}{c||c|c|c|c|c}
 & \,\,$\chi$PT\,\, & \,\,$\chi$PT\,\, & $\chi$PT 
& \,\,PQ$\chi$PT\,\, & \,\,PQ$\chi$PT\,\, \\[1mm]
$n_f$ &  2 & 3 & $n$ & 2 & 3\\ 
 & & & & & \vspace{-.2cm} \\
\hline\hline 
 & & & & & \vspace{-.2cm} \\
LO & $F,B$ & $F_0,B_0$ & $F_0^{(n)},B_0^{(n)}$ & $F,B$ & $F_0,B_0$\\
 & & & & & \vspace{-.2cm} \\ \hline
 & & & & & \vspace{-.2cm} \\
NLO & $l_i^r$ & $L_i^r$ &$ L_i^{r(n)}$
 &$ L_i^{r(2pq)}$ &$ L_i^{r(3pq)}$\\
$i+j$ & 7\,+\,3 & 10\,+\,2 & 11\,+\,2 & 11\,+\,2 & 11\,+\,2 \\
 & & & & & \vspace{-.2cm} \\ \hline
 & & & & & \vspace{-.2cm} \\
\,NNLO\, & $c_i^r$ & $C_i^r$ & $K_i^{r(n)}$ & $K_i^{r(2pq)}$ & 
$K_i^{r(3pq)}$\\
$i+j$ & \,53\,+\,4\, & \,90\,+\,4\, & \,112\,+\,3\, & \,112\,+\,3\, 
& \,112\,+\,3\,
\end{tabular}
\label{tab:LEC}
\end{center}
\end{table}

At lowest order, the LEC:s of PQ$\chi$PT with $n$ flavors of 
sea-quarks are the same as those of $n$ flavor unquenched $\chi$PT. At 
NLO with $n$ flavors of sea-quarks, they are $L^{r(npq)}_0$ through 
$L^{r(npq)}_{10}$. The two possible extra terms $L_{11}^{r(npq)}$ 
and $L_{12}^{r(npq)}$ can be removed using field 
redefinitions~\cite{BCE1,BCE2}. For the case of three sea-quarks, the 
$L_i^r$ are simple linear combinations of the $L_i^{r(3pq)}$, and these 
relations can be found in Refs.~\cite{BL1,BCE1}. Similarly at NNLO, the 
$C_i^r$ are linear combinations of the $K_i^{r(3pq)}$, and the 
corresponding relations can be found in Ref.~\cite{BCE1}. For three 
flavors of sea-quarks, one simply takes 
$L_{11}^{r(3pq)} = L_{12}^{r(3pq)} = 0$.

The unquenched two-flavor Lagrangian as defined in the second
paper of Ref.~\cite{GL}
differs by an $L_{11}^r$ type term. At NLO, this makes no difference
since that term does not contribute, but in order to get the correct
values at NNLO for the $l_i^r,c_i^r$ and the $L_i^{r(2pq)},K_i^{r(2pq)}$
one should take $L^{r(2pq)}_{11} = -l^r_4 / 4$ and $L^{r(2pq)}_{12} = 0$.
The relations between the NLO LEC:s are given in Table~\ref{reltab}, and 
similar linear relations between the NNLO LEC:s $K^{r(2pq)}_i$ of 
PQ$\chi$PT and the $c^r_i$ of unquenched two-flavor $\chi$PT may be 
found in Ref.~\cite{BCE1}. The equations given in the remaining sections 
of this paper only deal with $n_f = 2$ PQ$\chi$PT, and therefore the 
superscript $(2pq)$ of the $L_i^{r(2pq)}$ and the $K_i^{r(2pq)}$ have 
been suppressed in most of the formulas in the following sections.

\begin{table}[h!]
\begin{center}
\caption{The correspondence between the NLO low-energy constants $L^r_i$ of 
two-flavor PQ$\chi$PT and the $l^r_i$ of standard two-flavor $\chi$PT. 
The relations relevant for the NNLO low-energy 
constants $K^r_i$ are given in Ref.~\cite{BCE1}.}
\vspace{.3cm}
\begin{tabular}{c||c}
$n_f = 2$ $\chi$PT & $n_f = 2$ PQ$\chi$PT \\ & \vspace{-.2cm} \\
\hline\hline 
& \vspace{-.2cm} \\
$l^r_1$ & $ -2\,L^{r(2pq)}_0 + 4\,L^{r(2pq)}_1 + 2\,L^{r(2pq)}_3 $ \\
$l^r_2$ & $ 4\,L^{r(2pq)}_0 + 4\,L^{r(2pq)}_2 $ \\
$l^r_3$ & $ -8\,L^{r(2pq)}_4 - 4\,L^{r(2pq)}_5 + 16\,L^{r(2pq)}_6 + 8\,L^{r(2pq)}_8 $ \\
$l^r_4$ & $ 8\,L^{r(2pq)}_4 + 4\,L^{r(2pq)}_5 $ \\
$l^r_5$ & $ L^{r(2pq)}_{10} $ \\
$l^r_6$ & $ -2\,L^{r(2pq)}_9 $ \\
$l^r_7$ & $ -16\,L^{r(2pq)}_7 - 8\,L^{r(2pq)}_8 $
\end{tabular}
\label{reltab}
\end{center}
\end{table}

The input quark masses $m_i$ enter into the calculation in terms of the 
lowest order squared meson masses $\chi_i$, which are defined as usual 
in $\chi$PT, by $\chi_i = 2B\,m_i$. In the present calculations, there 
are two valence inputs $\chi_1,\chi_2$, two sea inputs $\chi_3,\chi_4$, 
and two ghost inputs $\chi_5,\chi_6$. In order for the disconnected 
valence quark loops to be canceled, the masses of the ghost quarks are 
always equal to those of the corresponding valence quarks, such that 
$\chi_5 = \chi_1$ and $\chi_6 = \chi_2$. Explicitly, for 
$d_{\mathrm{val}}=1$, we have $\chi_1 = \chi_2$ and for 
$d_{\mathrm{val}}=2$, we have $\chi_1 \ne \chi_2$. Similarly, for the 
sea-quark sector $d_{\mathrm{sea}}=1$ implies $\chi_3 = \chi_4$, while 
for $d_{\mathrm{sea}}=2$ we have $\chi_3 \ne \chi_4$. The most general 
results (in the quark masses) for $n_{\mathrm{sea}} = 2$ are therefore 
those with $d_{\mathrm{val}}=2$ and $d_{\mathrm{sea}}=2$.

\subsection{Chiral Logarithms}

The finite parts of the one-loop integrals which are encountered in the 
NNLO calculation of the masses and decay constants of the pseudoscalar 
mesons may be expressed in terms of the chiral logarithms $A,B$ and $C$. 
Of these, the logarithms $A$ are defined as
\begin{eqnarray}
\bar A(\chi_i) &=& -\:\pi_{16}\:\chi_i\log(\chi_i/\mu^2),
\label{Alog} \\
\bar A(\chi_i,\varepsilon) &=& \bar A(\chi_i)^2 / (2\pi_{16}\,\chi_i) 
\nonumber \\
&+& \pi_{16}\,\chi_i\,(\pi^2/12 + 1/2),
\label{Alogeps}
\end{eqnarray}
where $\pi_{16} = 1/(16 \pi^2)$. Note that the chiral logarithms which 
depend on $\varepsilon$ are naturally generated by the process of 
dimensional regularization and can be reexpressed in terms of the 
standard logarithms $A$ and $B$. However, they have been retained in the 
final results for the sake of notational efficiency. The finite parts of 
the logarithms $B$ are
\begin{eqnarray}
\bar B(\chi_i,\chi_i,0) &=& -\:\pi_{16}\left(1 + \log(\chi_i/\mu^2) 
\right),
\label{Blog} \\
\bar B(\chi_i,\chi_j,0) &=& -\:\pi_{16}\, 
\frac{\chi_i\log(\chi_i/\mu^2)
- \chi_j\log(\chi_j/\mu^2)}{\chi_i - \chi_j}, \nonumber \\
\bar B(\chi_i,\chi_i,0,\varepsilon) &=&
\bar A(\chi_i)\bar B(\chi_i,\chi_i,0) / (\pi_{16}\,\chi_i)
\label{Blogeps} \\
&& - \:\bar A(\chi_i)^2 / (2\pi_{16}\,\chi^2_i)
\nonumber \\
&& + \:\pi_{16}\,(\pi^2/12 + 1/2), \nonumber \\
\bar B(\chi_i,\chi_j,0,k) &=& \chi_i \bar B(\chi_i,\chi_j,0) + \bar 
A(\chi_j),
\label{Blogk}
\end{eqnarray}
where the $\bar B(\chi_i,\chi_j,0,k)$ integrals, which are symmetric 
under the interchange of $\chi_i$ and $\chi_j$, have been introduced in 
order to make the symmetries in the end results more explicit. Finally the 
remaining one-loop integral $C$ is given by
\begin{eqnarray}
\bar C(\chi_i,\chi_i,\chi_i,0) &=& -\:\pi_{16}/(2 \chi_i).
\label{Clog}
\end{eqnarray}
In all of these cases, the dependence on the subtraction scale $\mu$ 
has been moved into the loop integrals.

\subsection{Two-loop Integrals}

The finite two-loop integrals $H^F,H_1^F,H_{21}^F$ that appear in 
the NNLO calculation of the meson masses and decay constants may be 
evaluated using the methods of Ref.~\cite{ABT1}. Note that the 
corresponding primed integrals $H^{F'},H_1^{F'},H_{21}^{F'}$ indicate 
differentiation with respect to $p^2$ and appear only for the decay 
constants. The notation used for the $H$ integrals in this paper is 
otherwise similar to that of Ref.~\cite{ABT1}, except that an extra 
integer argument now indicates the propagator structure, such that e.g.
\begin{equation}
H^F(\chi_i,\chi_j,\chi_k,p^2) \rightarrow 
H^F(n,\chi_i,\chi_j,\chi_k,p^2).
\end{equation}
The case of $n = 1$ indicates that the integral consists of
single propagators only, as in Ref.~\cite{ABT1}, whereas \mbox{$n =
2$} indicates that the first propagator appears squared and \mbox{$n =
3$} that the second propagator appears squared. The cases with two
double propagators that can appear in the calculations are \mbox{$n =
5$}, for which the first and second propagators appear squared, and
\mbox{$n = 7$} for which the second and third propagators are squared.
Explicit expressions for \mbox{$n = 1$} can be found in
Ref.~\cite{ABT1}, and the other cases may be obtained by
differentiation with respect to the masses of those expressions.

\subsection{Notation for Meson Masses and Decay Constants}

As explained in great detail in Refs.~\cite{Sharpe1,Sharpe2,BL1}, the neutral 
meson propagators in PQ$\chi$PT have a complicated structure. In order 
to obtain a result which can be expressed in terms of the chiral 
logarithms and two-loop integrals outlined in the previous subsection, 
that propagator has to be treated by partial fractioning, which produces 
expressions that consist of sums of single and double poles. The 
residues, which are rational functions of the sea and valence quark 
masses, pile up and produce very long and awkward end results. With a 
major effort, these can be simplified and in some cases the expressions 
have been compressed by more than an order of magnitude.

The residues $R$ which are naturally generated in the 
calculations are, for $d_{\mathrm{sea}}=2$, the single-pole 
residues $R_{jk}^{i}$ and the double-pole residue $R_i^d$. The residues 
simplify when pairs of sea-quark masses become degenerate, and hence the 
number of indices in the single-pole residue depends on the degree of 
degeneracy in the sea-quark masses. The residues relevant for the 
present two-flavor PQ$\chi$PT calculations may be written in terms of 
the quantities $R^z_{a\ldots b}$, which are defined as
\begin{eqnarray}
R^z_{ab} &=& \chi_a - \chi_b, \nonumber \\
R^z_{abc} &=& \frac{\chi_a - \chi_b}{\chi_a - \chi_c}, \nonumber \\
R^z_{abcd} &=& \frac{(\chi_a - \chi_b)(\chi_a - \chi_c)}
{\chi_a - \chi_d},
\label{RSfunc}
\end{eqnarray}
and so on. Thus $R^z_{a\ldots b}$ has the same dimension as $\chi_i$ for
an even number of indices and is dimensionless for an odd number of indices.
It should be noted that the $R^z$ notation may also appear
independently in the final results. In such cases, the $R^z$ have been 
generated by simplification procedures. For $d_{\mathrm{sea}} = 2$, the 
residues which originate in the partial fractioning of the neutral meson 
propagator are
\begin{eqnarray}
R_{jk}^{i} &=& R^z_{i34jk}, \nonumber \\
R_{i}^{d} &=& R^z_{i34\pi},\nonumber \\
R_{i}^{c} &=& R^i_{3\pi} + R^i_{4\pi} - R^i_{\pi\pi},
\end{eqnarray}
where the neutral pion mass in the sea-quark sector is simply given by 
$\chi_\pi = \chi_{34} = \left(\chi_3+\chi_4\right)/2$. 
Because of this trivial relation, the above residue 
notation is highly redundant, a fact which has been exploited in the 
simplification of the end results. The quantity $R^v$, which is 
defined as
\begin{eqnarray}
R^v_{ijk} &=& R^i_{jj} + R^i_{kk} - 2 R^i_{jk}
\end{eqnarray}
appears throughout the final results and has therefore been given a separate
name. For the case of $d_{\mathrm{sea}} = 1$, all residues associated with 
the sea-quark sector reduce to numbers. Some nontrivial residues 
may still appear if the valence quarks are nondegenerate, according to
\begin{eqnarray}
R_{j}^{i} &=& R^z_{i3j}, \nonumber \\
R_{i}^{d} &=& R^z_{i3}\,.
\end{eqnarray}
For convenience, $R_{i}^{d}$ is also used for $d_{\mathrm{val}} = 1$
in order to maintain a notation similar to the more general cases.

\subsection{Summation Conventions}

The expressions for the meson masses and decay constants are symmetric, 
within the sea and valence quark sectors, under the interchange of quark 
masses, and may thus be compactified by the introduction of 
summation conventions which exploit these symmetries. The summations in the 
valence quark sector are similar to those in the three-flavor work of 
Ref.~\cite{BL1}, whereas those of the sea-quark sector are considerably 
simpler in the two-flavor case.

The sea-quark sector of the results involves the summation index $s$, 
which may appear for the quark masses $\chi_i$, and among the indices of 
the residue functions $R$. A simple example is $\chi_s = \chi_3 + 
\chi_4$. On the other hand, if the term in question consists of products 
of integrals, residues and quark masses, then the summation sign always 
applies to the term as a whole, such that e.g.
\begin{eqnarray}
\bar A(\chi_s)\,R^1_{\eta s}\,\chi_s &=& \sum_{i\,=\,3,4}
\bar A(\chi_i)\,R^1_{\eta i}\,\chi_i.
\end{eqnarray}
Any contribution which is written in terms of the summation index 
$s$ explicitly fulfills the required symmetry properties in the
sea-quark sector. 

The summation indices $p$ and $q$, which refer to the valence quark sector, 
are only needed for the case of $d_{\mathrm{val}} = 2$ and then always
occur for the squared valence quark masses $\chi_1$ and $\chi_2$. If the 
index $q$ is present, there will always be an index $p$ and the resulting 
sum is over the pairs $(p,q) = (1,2)$ and
$(p,q) = (2,1)$. If only $p$ is present, the sum is over the indices
$1$ and~$2$. An example is
\begin{eqnarray}
\bar A(\chi_p)\,R^p_{q\pi}\,\chi_p &=&
\bar A(\chi_1)\,R^1_{2\pi}\,\chi_1 \:+\: [1\leftrightarrow 2].
\end{eqnarray}
Any contribution written in terms of the $(p,q)$ notation is thus symmetric
under the interchange of the valence quark masses $\chi_1$ and $\chi_2$.

\section{Two-Flavor Pseudoscalar Meson Decay Constants}

\begin{figure}[h!]
\begin{center}
\includegraphics[width=\columnwidth]{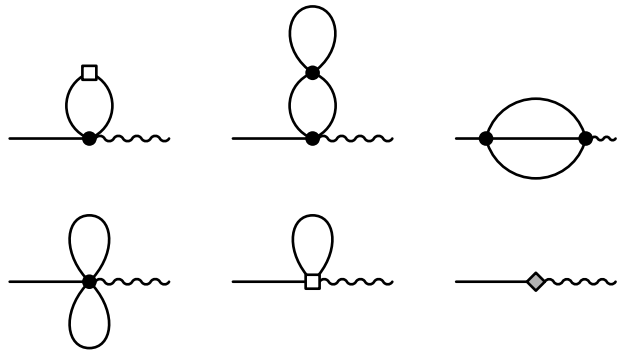}
\caption{Feynman diagrams at ${\mathcal O}(p^6)$ or two-loop for the
matrix element of the axial current operator. Filled circles denote
vertices of the ${\mathcal L}_2$ Lagrangian, whereas open squares and
shaded diamonds denote vertices of the ${\mathcal L}_4$ and
${\mathcal L}_6$ Lagrangians, respectively.}
\label{decfig}
\end{center}
\end{figure}

The decay constants $F_a$ of the pseudoscalar mesons are obtained
from the definition
\begin{equation}
\langle 0| A_a^\mu(0) |\phi_a(p)\rangle = i\sqrt{2}\,p^\mu\,F_a,
\label{decdef}
\end{equation}
in terms of the axial current operator $A_a^\mu$. The diagrams which 
contribute to that operator at ${\mathcal O}(p^6)$, or NNLO, are shown 
in Fig.~\ref{decfig}. Diagrams at ${\mathcal O}(p^2)$ and ${\mathcal 
O}(p^4)$ also contribute to Eq.~(\ref{decdef}) via the renormalization 
of the pseudoscalar meson wave function $\phi_a(p)$. The decay constants 
are functions of the valence inputs $\chi_1,\chi_2$ and the sea inputs 
$\chi_3,\chi_4$, which are defined in terms of the quark masses through 
$\chi_i = 2 B\,m_i$, and of the quantities $\chi_{ij} = 
(\chi_i+\chi_j)/2$, which correspond to the lowest order charged meson 
masses. Other parameters include the decay constant in the chiral limit 
($F$), the quark condensate in the chiral limit, via $\langle{\bar 
q}q\rangle = - B F^2$, and the LEC:s of ${\cal O}(p^4)$ and ${\cal 
O}(p^6)$, i.e. the $L_i^{r(2pq)}$ and the $K_i^{r(2pq)}$, which for 
simplicity are denoted by $L_i^r$ and $K_i^r$.

The decay constants of the pseudoscalar mesons are given in the form
\begin{equation}
F_{\mathrm{phys}} = F \left[ 1 + \frac{\delta^{(4)\mathrm{vs}}}{F^2}
+ \frac{\delta^{(6)\mathrm{vs}}_{\mathrm{ct}}
+ \delta^{(6)\mathrm{vs}}_{\mathrm{loops}}}{F^4}
+ \mathcal{O}(p^8) \right],
\label{delteq}
\end{equation}
where the ${\cal O}(p^4)$ and ${\cal O}(p^6)$ contributions have been
separated. The contribution from the ${\cal O}(p^6)$ counterterms 
$\delta^{(6)}_{\mathrm{ct}}$ has also been separated from the remaining 
part $\delta^{(6)}_{\mathrm{loops}}$, which consists mostly of the 
contributions from the chiral loops of ${\cal O}(p^6)$. It should be 
noted that $\delta^{(6)}_{\mathrm{loops}}$ also receives a contribution 
from the chiral loops and counterterms of ${\cal O}(p^4)$, 
which originate in the expansion of Eq.~(\ref{decdef}) to ${\cal 
O}(p^6)$. The superscripts (v) and (s) indicate the values of 
$d_{\mathrm{val}}$ and $d_{\mathrm{sea}}$, respectively. As expected, 
the infinities in all expressions for the decay constant have canceled. 
The appearance of 'unphysical' $\bar B$ logarithms is, in part, due to 
the partial quenching, and to the fact that the results have been 
expressed in terms of the lowest order masses rather than the 
full physical ones. Thus not all of them correspond to the quenched 
chiral logarithms which are ill-behaved in the chiral limit.

\subsection{Results for $d_{\mathrm{val}} = 1$}

The NLO result for $d_{\mathrm{val}} = 1$ is short, and will
only be given for $d_{\mathrm{sea}} = 2$. The result for
$d_{\mathrm{sea}} = 1$ can readily be derived from that expression. The
combined NLO result (loops and counterterms) for $d_{\mathrm{sea}} = 2$, is
\begin{eqnarray} 
\delta^{(4)12} & = & 8\,L^r_{4}\,\chi_{34}
+ 4\,L^r_{5}\,\chi_1
\:+\,1/2\,\bar{A}(\chi_{1s}),
\label{F0_NLO_nf2_12} 
\end{eqnarray} 
and the contribution from the ${\cal O}(p^6)$ counterterms to the NNLO 
expression for the decay constant is
\begin{eqnarray} 
\delta^{(6)12}_{\mathrm{ct}} & = & 8\,K^r_{19}\,\chi_1^2
\:+\,16\,K^r_{20}\,\chi_1 \chi_{34}
\:+\,8\,K^r_{21}\,\chi_s^2 \nonumber \\
&& + \,\:32\,K^r_{22}\,\chi_{34}^2
\:+\,8\,K^r_{23}\,\chi_1^2.
\label{F0tree_NNLO_nf2_12} 
\end{eqnarray} 
At NNLO, the chiral loops form, by far, the largest contribution to the 
decay constant. As a straightforward derivation of the results for 
$d_{\mathrm{sea}} = 1$ from the $d_{\mathrm{sea}} = 2$ case is tedious 
and complicated, the expressions for these two cases are given 
separately below. 

\begin{widetext}

\begin{eqnarray} 
\delta^{(6)11}_{\mathrm{loops}} & = & \pi_{16}\,L^r_{0}\,\left[ - 4\,\chi_1 \chi_3 
+ 2\,\chi_1^2 - \chi_3^2 \right]
\:-\,2\,\pi_{16}\,L^r_{1}\,\chi_1^2
\:-\,\pi_{16}\,L^r_{2}\,\left[ \chi_1^2 + 3 \chi_3^2 \right] \nonumber \\
&& + \,\:\pi_{16}\,L^r_{3}\,\left[ - 3\,\chi_1 \chi_3 + 5/2\,\chi_1^2 - 1/2\,\chi_3^2 \right] 
\:+\,\pi_{16}^2\,\left[ - 23/96\,\chi_1 \chi_3 + 1/64\,\chi_1^2 
- 47/192\,\chi_3^2 \right]
\:-\,32\,L^r_{4}L^r_{5}\,\chi_1 \chi_3 \nonumber \\
&& - \,\:32\,L^{r2}_{4}\,\chi_3^2
\:-\,8\,L^{r2}_{5}\,\chi_1^2
\:+\,8\,L^{r2}_{11}\,\chi_1^2
\:+\,6\,\bar{A}(\chi_1)\,L^r_{0}\,\left[ \chi_1 + R^d_1 \right]
\:-\,4\,\bar{A}(\chi_1)\,L^r_{1}\,\chi_1
\:-\,10\,\bar{A}(\chi_1)\,L^r_{2}\,\chi_1 \nonumber \\
&& + \,\:6\,\bar{A}(\chi_1)\,L^r_{3}\,\left[ \chi_1 + R^d_1 \right]
\:-\,2\,\bar{A}(\chi_1)\,L^r_{5}\,\chi_1
\:-\,1/2\,\bar{A}(\chi_1) \bar{B}(\chi_{13},\chi_{13},0)\,\chi_{13}
\:+\,1/8\,\bar{A}(\chi_1,\varepsilon)\,\pi_{16}\,\chi_3 \nonumber \\
&& - \,\:\bar{A}(\chi_{13})\,\pi_{16}\,\left[ 1/6\,\chi_1 + 2/3\,\chi_3 \right]
\:-\,8\,\bar{A}(\chi_{13})\,L^r_{0}\,\chi_{13}
\:-\,20\,\bar{A}(\chi_{13})\,L^r_{3}\,\chi_{13}
\:-\,8\,\bar{A}(\chi_{13})\,L^r_{4}\,\chi_3
\:+\,4\,\bar{A}(\chi_{13})\,L^r_{5}\,\chi_1 \nonumber \\
&& +\,\:\bar{A}(\chi_{13},\varepsilon)\,\pi_{16}\,\chi_3
\:-\,24\,\bar{A}(\chi_3)\,L^r_{1}\,\chi_3 
\:-\,6\,\bar{A}(\chi_3)\,L^r_{2}\,\chi_3 
\:+\,12\,\bar{A}(\chi_3)\,L^r_{4}\,\chi_3
\:+\,3/8\,\bar{A}(\chi_3,\varepsilon)\,\pi_{16}\,\chi_3 \nonumber \\
&& + \,\:6\,\bar{B}(\chi_1,\chi_1,0)\,L^r_{0}\,R^d_1\,\chi_1
\:+\,6\,\bar{B}(\chi_1,\chi_1,0)\,L^r_{3}\,R^d_1\,\chi_1
\:-\,2\,\bar{B}(\chi_1,\chi_1,0)\,L^r_{5}\,R^d_1\,\chi_1 \nonumber \\
&& - \,\:1/8\,\bar{B}(\chi_1,\chi_1,0,\varepsilon)\,\pi_{16}\,R^d_1\,\chi_1
\:-\,16\,\bar{B}(\chi_{13},\chi_{13},0)\,L^r_{4}\,\chi_{13} \chi_3
\:-\,8\,\bar{B}(\chi_{13},\chi_{13},0)\,L^r_{5}\,\chi_{13}^2 \nonumber \\
&& + \,\:32\,\bar{B}(\chi_{13},\chi_{13},0)\,L^r_{6}\,\chi_{13} \chi_3
\:+\,16\,\bar{B}(\chi_{13},\chi_{13},0)\,L^r_{8}\,\chi_{13}^2
\:-\,1/8\,H^{F}(1,\chi_1,\chi_{13},\chi_{13},\chi_1)\,\chi_3 \nonumber \\
&& - \,\:3/8\,H^{F}(1,\chi_{13},\chi_{13},\chi_3,\chi_1)\,\chi_3
\:+\,1/8\,H^{F}(2,\chi_1,\chi_{13},\chi_{13},\chi_1)\,R^d_1\,\chi_1
\:+\,5/12\,H^{F'}(1,\chi_1,\chi_1,\chi_1,\chi_1)\,\chi_1^2 \nonumber \\
&& + \,\:H^{F'}(1,\chi_1,\chi_{13},\chi_{13},\chi_1)\,\left[ 1/8\,\chi_1 \chi_3 
- 1/2\,\chi_1^2 \right]
\:+\,3/8\,H^{F'}(1,\chi_{13},\chi_{13},\chi_3,\chi_1)\,\chi_1 \chi_3 \nonumber \\
&& + \,\:1/2\,H^{F'}(2,\chi_1,\chi_1,\chi_1,\chi_1)\,R^d_1\,\chi_1^2
\:+\,3/8\,H^{F'}(2,\chi_1,\chi_{13},\chi_{13},\chi_1)\,R^d_1\,\chi_1^2 \nonumber \\
&& + \,\:1/4\,H^{F'}(5,\chi_1,\chi_1,\chi_1,\chi_1)\,(R^d_1)^2 \chi_1^2
\:-\,2\,H^{F'}_1(3,\chi_{13},\chi_1,\chi_{13},\chi_1)\,R^d_1\,\chi_1^2 \nonumber \\
&& + \,\:3/8\,H^{F'}_{21}(1,\chi_1,\chi_{13},\chi_{13},\chi_1)\,\chi_1^2
\:+\,9/8\,H^{F'}_{21}(1,\chi_3,\chi_{13},\chi_{13},\chi_1)\,\chi_1^2 \nonumber \\
&& - \,\:3/8\,H^{F'}_{21}(2,\chi_1,\chi_{13},\chi_{13},\chi_1)\,R^d_1\,\chi_1^2,
\label{F0loop_NNLO_nf2_11} 
\end{eqnarray} 

\begin{eqnarray} 
\delta^{(6)12}_{\mathrm{loops}} & = & \pi_{16}\,L^r_{0}\,\left[ - 4\,\chi_\pi \chi_1 
- 3\,\chi_\pi^2 + 2\,\chi_1^2 + 2\,\chi_3 \chi_4 \right]
\:-\,2\,\pi_{16}\,L^r_{1}\,\chi_1^2
\:-\,\pi_{16}\,L^r_{2}\,\left[ 3\,\chi_\pi^2 + \chi_1^2 \right] \nonumber \\
&& + \,\:\pi_{16}\,L^r_{3}\,\left[ - 3\,\chi_\pi \chi_1 - 2\,\chi_\pi^2 + 5/2\,\chi_1^2 
+ 3/2\,\chi_3 \chi_4 \right]
\:+\,\pi_{16}^2\,\left[ - 23/96\,\chi_\pi \chi_1 - 1/3\,\chi_\pi^2 + 1/64\,\chi_1^2 
+ 17/192\,\chi_3 \chi_4 \right] \nonumber \\
&& - \,\:32\,L^r_{4}L^r_{5}\,\chi_1 \chi_{34}
\:-\,32\,L^{r2}_{4}\,\chi_{34}^2
\:-\,8\,L^{r2}_{5}\,\chi_1^2
\:+\,8\,L^{r2}_{11}\,\chi_1^2
\:+\,6\,\bar{A}(\chi_\pi)\,L^r_{0}\,R^\pi_{11}\,\chi_\pi
\:-\,8\,\bar{A}(\chi_\pi)\,L^r_{1}\,\chi_\pi \nonumber \\
&& - \,\:2\,\bar{A}(\chi_\pi)\,L^r_{2}\,\chi_\pi
\:+\,6\,\bar{A}(\chi_\pi)\,L^r_{3}\,R^\pi_{11}\,\chi_\pi
\:+\,4\,\bar{A}(\chi_\pi)\,L^r_{4}\,\chi_\pi
\:-\,2\,\bar{A}(\chi_\pi)\,L^r_{5}\,R^\pi_{11}\,\chi_1 \nonumber \\
&& - \,\:1/4\,\bar{A}(\chi_\pi) \bar{B}(\chi_{1s},\chi_{1s},0)\,R^\pi_{1s}\,\chi_{1s} 
\:+\,1/8\,\bar{A}(\chi_\pi,\varepsilon)\,\pi_{16}\,R^c_1\,\chi_\pi
\:+\,6\,\bar{A}(\chi_1)\,L^r_{0}\,\left[ R^c_1\,\chi_1 + R^d_1 \right]
\:-\,4\,\bar{A}(\chi_1)\,L^r_{1}\,\chi_1 \nonumber \\
&& - \,\:10\,\bar{A}(\chi_1)\,L^r_{2}\,\chi_1
\:+\,6\,\bar{A}(\chi_1)\,L^r_{3}\,\left[ R^c_1\,\chi_1 + R^d_1 \right]
\:-\,2\,\bar{A}(\chi_1)\,L^r_{5}\,R^c_1\,\chi_1
\:-\,1/4\,\bar{A}(\chi_1) \bar{B}(\chi_{1s},\chi_{1s},0)\,R^1_{s\pi}\,\chi_{1s} 
\nonumber \\ 
&& + \,\:1/8\,\bar{A}(\chi_1,\varepsilon)\,\pi_{16}\,R^1_{\pi\pi}\,\chi_{34}
\:-\,16\,\bar{A}(\chi_{34})\,L^r_{1}\,\chi_{34}
\:-\,4\,\bar{A}(\chi_{34})\,L^r_{2}\,\chi_{34}
\:+\,8\,\bar{A}(\chi_{34})\,L^r_{4}\,\chi_{34} \nonumber \\
&& + \,\:1/4\,\bar{A}(\chi_{34},\varepsilon)\,\pi_{16}\,\chi_{34}
\:-\,\bar{A}(\chi_{1s})\,\pi_{16}\,\left[ 1/4\,\chi_{34} + 1/6\,\chi_{1s} \right]
\:-\,4\,\bar{A}(\chi_{1s})\,L^r_{0}\,\chi_{1s}
\:-\,10\,\bar{A}(\chi_{1s})\,L^r_{3}\,\chi_{1s} \nonumber \\
&& - \,\:4\,\bar{A}(\chi_{1s})\,L^r_{4}\,\chi_{34}
\:+\:2\,\bar{A}(\chi_{1s})\,L^r_{5}\,\chi_1
\:+\,1/4\,\bar{A}(\chi_{1s},\varepsilon)\,\pi_{16}\,\left[ \chi_{34} + \chi_s \right] 
\:+\,6\,\bar{B}(\chi_1,\chi_1,0)\,L^r_{0}\,R^d_1\,\chi_1 \nonumber \\
&& + \,\:6\,\bar{B}(\chi_1,\chi_1,0)\,L^r_{3}\,R^d_1\,\chi_1
\:-\:2\,\bar{B}(\chi_1,\chi_1,0)\,L^r_{5}\,R^d_1\,\chi_1
\:-\,1/8\,\bar{B}(\chi_1,\chi_1,0,\varepsilon)\,\pi_{16}\,R^d_1\,\chi_1 \nonumber \\
&& - \,\:8\,\bar{B}(\chi_{1s},\chi_{1s},0)\,L^r_{4}\,\chi_{34} \chi_{1s}
\:-\:4\,\bar{B}(\chi_{1s},\chi_{1s},0)\,L^r_{5}\,\chi_{1s}^2
\:+\,16\,\bar{B}(\chi_{1s},\chi_{1s},0)\,L^r_{6}\,\chi_{34} \chi_{1s} \nonumber \\
&& + \,\:8\,\bar{B}(\chi_{1s},\chi_{1s},0)\,L^r_{8}\,\chi_{1s}^2
\:+\,1/16\,H^{F}(1,\chi_\pi,\chi_{1s},\chi_{1s},\chi_1)\,R^v_{\pi 1s}\,\chi_\pi 
\:-\,H^{F}(1,\chi_1,\chi_{1s},\chi_{1s},\chi_1)\,\left[ 1/8\,R^1_{s\pi}\,\chi_{1s} 
\right. \nonumber \\
&& - \,\:1/16 \left. R^c_1\,\chi_1 \right]
\:-\,1/4\,H^{F}(1,\chi_{13},\chi_{14},\chi_{34},\chi_1)\,\chi_{34}
\:+\,1/16\,H^{F}(2,\chi_1,\chi_{1s},\chi_{1s},\chi_1)\,R^d_1\,\chi_1 \nonumber \\
&& + \,\:1/4\,H^{F'}(1,\chi_\pi,\chi_\pi,\chi_1,\chi_1)\,(R^\pi_{11})^2 \chi_1^2 
\:+\,1/2\,H^{F'}(1,\chi_\pi,\chi_1,\chi_1,\chi_1)\,R^\pi_{11} R^c_1\,\chi_1^2 
\nonumber \\
&& - \,\:H^{F'}(1,\chi_\pi,\chi_{1s},\chi_{1s},\chi_1)\,\left[ 1/4\,R^\pi_{11}\,\chi_1^2 
+ 1/16\,R^v_{\pi 1s}\,\chi_\pi \chi_1 \right]
\:+\,H^{F'}(1,\chi_1,\chi_1,\chi_1,\chi_1)\,\left[ 1/6\,\chi_1^2 \right. \nonumber \\
&& + \,\:1/4 \left. (R^c_1)^2 \chi_1^2 \right] 
\:+\,H^{F'}(1,\chi_1,\chi_{1s},\chi_{1s},\chi_1)\,\left[ 1/8\,R^1_{s\pi}\,\chi_1 \chi_{1s} 
- 1/2\,R^1_{s\pi}\,\chi_1^2 + 3/16\,R^c_1\,\chi_1^2 \right] \nonumber \\
&& + \,\:1/4\,H^{F'}(1,\chi_{13},\chi_{14},\chi_{34},\chi_1)\,\chi_1 \chi_{34}
\:+\,1/2\,H^{F'}(2,\chi_1,\chi_\pi,\chi_1,\chi_1)\,R^\pi_{11} R^d_1\,\chi_1^2 \nonumber \\ 
&& + \,\:1/2\,H^{F'}(2,\chi_1,\chi_1,\chi_1,\chi_1)\,R^c_1 R^d_1\,\chi_1^2
\:+\,3/16\,H^{F'}(2,\chi_1,\chi_{1s},\chi_{1s},\chi_1)\,R^d_1\,\chi_1^2 \nonumber \\
&& + \,\:1/4\,H^{F'}(5,\chi_1,\chi_1,\chi_1,\chi_1)\,(R^d_1)^2 \chi_1^2
\:-\,1/2\,H^{F'}_1(1,\chi_\pi,\chi_{1s},\chi_{1s},\chi_1)\,R^\pi_{1s} R^z_{1s\pi}\,\chi_1^2
\nonumber \\
&& - \,\:H^{F'}_1(1,\chi_{1s},\chi_{1s},\chi_1,\chi_1)\,R^\pi_{1s} R^z_{1s\pi}\,\chi_1^2 
\:-\,H^{F'}_1(3,\chi_{1s},\chi_1,\chi_{1s},\chi_1)\,R^d_1\,\chi_1^2 \nonumber \\
&& - \,\:3/16\,H^{F'}_{21}(1,\chi_\pi,\chi_{1s},\chi_{1s},\chi_1)\,R^v_{\pi 
1s}\,\chi_1^2
\:+\,H^{F'}_{21}(1,\chi_1,\chi_{1s},\chi_{1s},\chi_1)\,\left[ 
3/8\,R^1_{s\pi}\,\chi_1^2 - 3/16\,R^c_1\,\chi_1^2 \right] \nonumber \\
&& + \,\:3/4\,H^{F'}_{21}(1,\chi_{34},\chi_{13},\chi_{14},\chi_1)\,\chi_1^2
\:-\,3/16\,H^{F'}_{21}(2,\chi_1,\chi_{1s},\chi_{1s},\chi_1)\,R^d_1\,\chi_1^2.
\label{F0loop_NNLO_nf2_12} 
\end{eqnarray} 

\end{widetext}

\subsection{Results for $d_{\mathrm{val}} = 2$}

In general, the expressions for $d_{\mathrm{val}} = 2$ are longer than those 
for $d_{\mathrm{val}} = 1$, as the number of independent mass combinations 
that can appear in the integrals is significantly larger. The most 
general two-flavor expression for the pseudoscalar meson decay constant is 
the result with $d_{\mathrm{val}} = 2$ and $d_{\mathrm{sea}} = 2$. 
However, in contrast with the results for $d_{\mathrm{val}} = 1$, the 
limit where the sea-quark masses become degenerate is nontrivial even at 
NLO. In the notation of Eq.~(\ref{delteq}), the NLO contribution for 
$d_{\mathrm{val}} = 2$ and $d_{\mathrm{sea}} = 2$ is
\begin{eqnarray}
\delta^{(4)22} & = & 8\,L^r_{4}\,\chi_{34}
\:+\,4\,L^r_{5}\,\chi_{12}
\:+\,\bar{A}(\chi_p)\,\left[ 1/4\,R^p_{q\pi} \right. \nonumber \\
&& - \,\:1/8 \left. R^c_p \right]
\:+\,1/4\,\bar{A}(\chi_{ps})
\:-\,1/8\,\bar{A}(\chi_\pi)\,R^v_{\pi 12} \nonumber \\
&& - \,\:1/8\,\bar{B}(\chi_p,\chi_p,0)\,R^d_p.
\label{F0_NLO_nf2_22} 
\end{eqnarray} 
The NLO contribution for $d_{\mathrm{val}} = 2$ and $d_{\mathrm{sea}} = 
1$ may be obtained from Eq.~(\ref{F0_NLO_nf2_22}) by carefully taking 
the limit $\chi_4 \rightarrow \chi_3$. Effectively this means dropping 
the $\bar{A}(\chi_\pi)$ term and replacing $\bar{A}(\chi_{ps}) 
\rightarrow 2\bar{A}(\chi_{p3})$. Furthermore the remaining residues 
reduce such that $R^c_p \rightarrow 1$ and $R^p_{q\pi} 
\rightarrow R^p_q$. The contribution 
from the ${\cal O}(p^6)$ counterterms to the NNLO 
expression is
\begin{eqnarray} 
\delta^{(6)22}_{\mathrm{ct}} & = & 4\,K^r_{19}\,\chi_p^2
\:+\,16\,K^r_{20}\,\chi_{12} \chi_{34}
\:+\,8\,K^r_{21}\,\chi_s^2 \nonumber \\
&& + \,\:32\,K^r_{22}\,\chi_{34}^2
\:+\,8\,K^r_{23}\,\chi_1 \chi_2,
\label{F0tree_NNLO_nf2_22} 
\end{eqnarray} 
from which the corresponding expression for $d_{\mathrm{val}} = 2$ and 
$d_{\mathrm{sea}} = 1$ may readily be inferred. The remaining NNLO 
contributions with $d_{\mathrm{val}} = 2$ are, for $d_{\mathrm{sea}} = 
1$ and $d_{\mathrm{sea}} = 2$, 

\begin{widetext}
\begin{eqnarray} 
\delta^{(6)21}_{\mathrm{loops}} & = & \pi_{16}\,L^r_{0}\,\left[ - 2\,\chi_1 \chi_2 
- 4\,\chi_{12} \chi_3 + 4\,\chi_{12}^2 - \chi_3^2 \right]
\:-\,2\,\pi_{16}\,L^r_{1}\,\chi_{12}^2
\:-\,\pi_{16}\,L^r_{2}\,\left[ \chi_{12}^2 + 3 \chi_3^2 \right] \nonumber \\
&& + \,\:\pi_{16}\,L^r_{3}\,\left[ - 3/2\,\chi_1 \chi_2 - 3\,\chi_{12} \chi_3 + 
4\,\chi_{12}^2 
- 1/2\,\chi_3^2 \right] \:+\,\pi_{16}^2\,\left[ - 17/192\,\chi_1 \chi_2 - 23/96\,\chi_{12} 
\chi_3 + 5/48\,\chi_{12}^2 \right. \nonumber \\
&& - \,\:47/192 \left. \chi_3^2 \right]
\:-\,32\,L^r_{4}L^r_{5}\,\chi_{12} \chi_3
\:-\,32\,L^{r2}_{4}\,\chi_3^2 \:-\,8\,L^{r2}_{5}\,\chi_{12}^2
\:+\,8\,L^{r2}_{11}\,\chi_{12}^2
\:+\,\bar{A}(\chi_p)\,\pi_{16}\,\left[ 1/48\,\chi_q + 1/48\,\chi_{12} \right. \nonumber \\
&& + \,\:1/16 \left. \chi_3 - 1/12\,R^p_q\,\chi_{12} - 1/8\,R^p_q\,\chi_3 \right]
\:+\,\bar{A}(\chi_p)\,L^r_{0}\,\left[ \chi_p + 4\,R^p_q\,\chi_p + R^d_p \right]
\:+\,\bar{A}(\chi_p)\,L^r_{3}\,\left[ 5/2\,\chi_p + R^p_q\,\chi_p \right. \nonumber \\
&& + \,\:5/2 \left. R^d_p \right] 
\:+\,\bar{A}(\chi_p)\,L^r_{4}\,\left[ \chi_3 - 2\,R^p_q\,\chi_3 \right]
\:-\,\bar{A}(\chi_p)\,L^r_{5}\,\left[ 1/2\,\chi_{12} + R^p_q\,\chi_{12} \right]
\:+\,\bar{A}(\chi_p)^2\,\left[ 9/128 - 1/32\,R^p_q R^q_p \right] \nonumber \\
&& - \,\:\bar{A}(\chi_p) \bar{A}(\chi_{p3})\,\left[ 1/16 + 1/12\,R^p_q \right]
\:-\,\bar{A}(\chi_p) \bar{A}(\chi_{q3})\,\left[ 3/16 + 1/12\,R^p_q \right]
\:+\,1/8\,\bar{A}(\chi_p) \bar{A}(\chi_{12}) \nonumber \\ 
&& + \,\:\bar{A}(\chi_p) \bar{B}(\chi_p,\chi_p,0)\,\left[ 3/8\,\chi_p - 
1/8\,R^p_q\,\chi_p - 1/32\,R^p_q R^d_p + 1/64\,R^d_p \right]
\:-\,\bar{A}(\chi_p) \bar{B}(\chi_q,\chi_q,0)\,\left[ 1/32\,R^p_q R^d_q \right. \nonumber \\
&& - \,\:1/64 \left. R^d_q \right] 
\:-\,1/4\,\bar{A}(\chi_p) \bar{B}(\chi_{p3},\chi_{p3},0)\,\chi_{p3}
\:-\,1/8\,\bar{A}(\chi_p) \bar{B}(\chi_1,\chi_2,0)\,R^q_p\,\chi_p \nonumber \\ 
&& + \,\:1/8\,\bar{A}(\chi_p) \bar{C}(\chi_p,\chi_p,\chi_p,0)\,R^d_p\,\chi_p
\:-\,1/8\,\bar{A}(\chi_p,\varepsilon)\,\pi_{16}\,\left[ \chi_p - R^p_q\,\chi_3 \right] 
\:+\,\bar{A}(\chi_{p3})\,\pi_{16}\,\left[ 1/12\,\chi_{p3} - 1/4\,\chi_{q3} \right. \nonumber \\
&& - \,\:1/4 \left. \chi_3 \right] 
\:-\,4\,\bar{A}(\chi_{p3})\,L^r_{0}\,\chi_{p3}
\:-\,10\,\bar{A}(\chi_{p3})\,L^r_{3}\,\chi_{p3}
\:-\,4\,\bar{A}(\chi_{p3})\,L^r_{4}\,\chi_3
\:+\,2\,\bar{A}(\chi_{p3})\,L^r_{5}\,\chi_{12}
\:-\,1/8\,\bar{A}(\chi_{p3})^2 \nonumber \\ 
&& + \,\:\bar{A}(\chi_{p3}) \bar{B}(\chi_p,\chi_p,0)\,\left[ 1/8\,\chi_p 
- 5/8\,\chi_{p3} \right] 
\:-\,1/16\,\bar{A}(\chi_{p3}) \bar{B}(\chi_q,\chi_q,0)\,R^d_q
\:+\,1/6\,\bar{A}(\chi_{p3}) \bar{B}(\chi_1,\chi_2,0)\,\chi_3 \nonumber \\
&& + \,\:1/3\,\bar{A}(\chi_{p3}) \bar{B}(\chi_1,\chi_2,0,k)
\:+\,1/2\,\bar{A}(\chi_{p3},\varepsilon)\,\pi_{16}\,\chi_3
\:-\,\bar{A}(\chi_1) \bar{A}(\chi_2)\,\left[ 1/64 - 1/16\,R^1_2 R^2_1 \right]
\:-\,4\,\bar{A}(\chi_{12})\,L^r_{1}\,\chi_{12} \nonumber \\
&& - \,\:10\,\bar{A}(\chi_{12})\,L^r_{2}\,\chi_{12}
\:+\,1/8\,\bar{A}(\chi_{12})^2 
\:-\,1/2\,\bar{A}(\chi_{12}) \bar{B}(\chi_1,\chi_2,0,k)
\:+\,1/4\,\bar{A}(\chi_{12},\varepsilon)\,\pi_{16}\,\chi_{12} \nonumber \\
&& + \,\:1/4\,\bar{A}(\chi_{13}) \bar{A}(\chi_{23})
\:-\,24\,\bar{A}(\chi_3)\,L^r_{1}\,\chi_3
\:-\,6\,\bar{A}(\chi_3)\,L^r_{2}\,\chi_3
\:+\,12\,\bar{A}(\chi_3)\,L^r_{4}\,\chi_3
\:+\,3/16\,\bar{A}(\chi_3) \bar{B}(\chi_p,\chi_p,0)\,\chi_3 \nonumber \\
&& - \,\:3/8\,\bar{A}(\chi_3) \bar{B}(\chi_1,\chi_2,0)\,\chi_3
\:+\,3/8\,\bar{A}(\chi_3,\varepsilon)\,\pi_{16}\,\chi_3
\:+\,\bar{B}(\chi_p,\chi_p,0)\,\pi_{16}\,\left[ 1/96\,R^d_p\,\chi_p 
+ 1/32\,R^d_p\,\chi_q \right. \nonumber \\
&& + \,\:1/16 \left. R^d_p\,\chi_3 \right] 
\:+\,\:\bar{B}(\chi_p,\chi_p,0)\,L^r_{0}\,R^d_p\,\chi_p
\:+\,5/2\,\bar{B}(\chi_p,\chi_p,0)\,L^r_{3}\,R^d_p \chi_p
\:+\,\bar{B}(\chi_p,\chi_p,0)\,L^r_{4}\,\left[ 2\,\chi_p \chi_3 \right. \nonumber \\
&& - \,\:4 \left. R^p_q\,\chi_p \chi_3 + 3\,R^d_p\,\chi_3 \right]
\:+\,\bar{B}(\chi_p,\chi_p,0)\,L^r_{5}\,\left[ \chi_p^2 - 2\,R^p_q\,\chi_1 \chi_2 
- 1/2\,R^d_p\,\chi_{12} \right]
\:+\,\bar{B}(\chi_p,\chi_p,0)\,L^r_{6}\,\left[ 4\,\chi_3^2 \right. \nonumber \\
&& - \,\:8 \left. R^q_p\,\chi_p \chi_3 \right] 
\:+\,4\,\bar{B}(\chi_p,\chi_p,0)\,L^r_{7}\,(R^d_p)^2
\:+\,\bar{B}(\chi_p,\chi_p,0)\,L^r_{8}\,\left[ 2\,\chi_3^2 - 4\,R^q_p\,\chi_p^2 \right]
\:+\,\bar{B}(\chi_p,\chi_p,0)^2\,\left[ 1/8\,R^q_p R^d_p\,\chi_p \right. \nonumber \\
&& + \,\:1/128 \left. (R^d_p)^2 \right] 
\:-\,1/8\,\bar{B}(\chi_p,\chi_p,0) \bar{B}(\chi_1,\chi_2,0)\,R^q_p R^d_p\,\chi_p 
\:+\,1/8\,\bar{B}(\chi_p,\chi_p,0) \bar{C}(\chi_p,\chi_p,\chi_p,0)\,(R^d_p)^2 \chi_p 
\nonumber \\ 
&& - \,\:1/8\,\bar{B}(\chi_p,\chi_p,0,\varepsilon)\,\pi_{16}\,R^d_p\,\chi_{p3} 
\:-\,8\,\bar{B}(\chi_{p3},\chi_{p3},0)\,L^r_{4}\,\chi_{p3} \chi_3
\:-\,4\,\bar{B}(\chi_{p3},\chi_{p3},0)\,L^r_{5}\,\chi_{p3}^2 \nonumber \\
&& + \,\:16\,\bar{B}(\chi_{p3},\chi_{p3},0)\,L^r_{6}\,\chi_{p3} \chi_3 
\:+\,8\,\bar{B}(\chi_{p3},\chi_{p3},0)\,L^r_{8}\,\chi_{p3}^2  
\:+\,1/64\,\bar{B}(\chi_1,\chi_1,0) \bar{B}(\chi_2,\chi_2,0)\,R^d_1 R^d_2 \nonumber \\
&& - \,\:8\,\bar{B}(\chi_1,\chi_2,0)\,L^r_{7}\,R^d_1 R^d_2
\:-\,4\,\bar{B}(\chi_1,\chi_2,0)\,L^r_{8}\,R^d_1 R^d_2 
\:+\,4\,\bar{C}(\chi_p,\chi_p,\chi_p,0)\,L^r_{4}\,R^d_p\,\chi_p \chi_3 \nonumber \\
&& + \,\:2\,\bar{C}(\chi_p,\chi_p,\chi_p,0)\,L^r_{5}\,R^d_p\,\chi_p^2
\:-\,8\,\bar{C}(\chi_p,\chi_p,\chi_p,0)\,L^r_{6}\,R^d_p\,\chi_p \chi_3 
\:-\,4\,\bar{C}(\chi_p,\chi_p,\chi_p,0)\,L^r_{8}\,R^d_p\,\chi_p^2 \nonumber \\
&& + \,\:H^{F}(1,\chi_p,\chi_p,\chi_{12},\chi_{12})\,\left[ 1/8\,\chi_p - 
5/64\,\chi_{12} + 1/16\,R^p_q R^q_p\,\chi_{12} \right]
\:+\,H^{F}(1,\chi_p,\chi_{13},\chi_{23},\chi_{12})\,\left[ - 1/16\,\chi_p 
\right. \nonumber \\
&& + \,\:1/16\,\chi_q - 1/8 \left. R^p_q\,\chi_3 \right]
\:+\,H^{F}(1,\chi_1,\chi_{12},\chi_2,\chi_{12})\,\left[ 1/32\,\chi_{12} 
- 1/8\,R^1_2 R^2_1\,\chi_{12} \right] \nonumber \\
&& - \,\:1/8\,H^{F}(1,\chi_{12},\chi_{12},\chi_{12},\chi_{12})\,\chi_{12} 
\:-\,3/8\,H^{F}(1,\chi_{13},\chi_{23},\chi_3,\chi_{12})\,\chi_3 \nonumber \\
&& + \,\:H^{F}(2,\chi_p,\chi_p,\chi_{12},\chi_{12})\,\left[ 1/16\,R^p_q 
R^d_p\,\chi_{12} - 1/32\,R^d_p\,\chi_{12} \right]
\:+\,H^{F}(2,\chi_p,\chi_{12},\chi_q,\chi_{12})\,\left[ 1/16\,R^q_p R^d_p\,\chi_{12} 
\right. \nonumber \\
&& - \,\:1/32 \left. R^d_p\,\chi_{12} \right]
\:+\,1/8\,H^{F}(2,\chi_p,\chi_{13},\chi_{23},\chi_{12})\,R^d_p\,\chi_{p3} 
\:-\,1/64\,H^{F}(5,\chi_p,\chi_p,\chi_{12},\chi_{12})\,(R^d_p)^2 \chi_{12} \nonumber \\
&& - \,\:1/32\,H^{F}(5,\chi_1,\chi_2,\chi_{12},\chi_{12})\,R^d_1 R^d_2\,\chi_{12}
\:-\,H^{F'}(1,\chi_p,\chi_p,\chi_{12},\chi_{12})\,\left[ 1/8\,\chi_p \chi_{12} 
+ 15/64\,\chi_{12}^2 - 3/16\,R^p_q\,\chi_{12}^2 \right. \nonumber \\
&& + \,\:1/16 \left. (R^p_q)^2 \chi_{12}^2 \right]
\:+\,H^{F'}(1,\chi_p,\chi_{13},\chi_{23},\chi_{12})\,\left[ 1/8\,\chi_{12}^2 
+ 3/8\,R^q_p\,\chi_p \chi_{12} + 1/8\,R^q_p\,\chi_q \chi_{12} 
- 1/8\,R^q_p\,\chi_{12} \chi_3 \right] \nonumber \\
&& - \,\:H^{F'}(1,\chi_1,\chi_{12},\chi_2,\chi_{12})\,\left[ 1/32\,\chi_{12}^2 
- 3/8\,R^1_2 R^2_1\,\chi_{12}^2 \right]
\:+\,1/8\,H^{F'}(1,\chi_{12},\chi_{12},\chi_{12},\chi_{12})\,\chi_{12}^2 \nonumber \\ 
&& + \,\:3/8\,H^{F'}(1,\chi_{13},\chi_{23},\chi_3,\chi_{12})\,\chi_{12} \chi_3 
\:-\,H^{F'}(2,\chi_p,\chi_p,\chi_{12},\chi_{12})\,\left[ 1/16\,R^p_q R^d_p\,\chi_{12}^2 
- 5/32\,R^d_p\,\chi_{12}^2 \right] \nonumber \\
&& - \,\:H^{F'}(2,\chi_p,\chi_{12},\chi_q,\chi_{12})\,\left[ 1/16\,R^q_p 
R^d_p\,\chi_{12}^2 - 1/32\,R^d_p\,\chi_{12}^2 \right]
\:-\,1/8\,H^{F'}(2,\chi_p,\chi_{13},\chi_{23},\chi_{12})\,R^d_p\,\chi_{p3} \chi_{12} 
\nonumber \\
&& + \,\:5/64\,H^{F'}(5,\chi_p,\chi_p,\chi_{12},\chi_{12})\,(R^d_p)^2 \chi_{12}^2 
\:+\,1/32\,H^{F'}(5,\chi_1,\chi_2,\chi_{12},\chi_{12})\,R^d_1 R^d_2\,\chi_{12}^2 
\nonumber \\
&& + \,\:H^{F'}_1(1,\chi_p,\chi_p,\chi_{12},\chi_{12})\,\left[ 5/4\,\chi_{12}^2 
- 1/2\,R^p_q R^q_p\,\chi_{12}^2 \right]
\:-\,H^{F'}_1(1,\chi_p,\chi_{13},\chi_{23},\chi_{12})\,R^q_p\,\chi_{12}^2 \nonumber \\
&& - \,\:H^{F'}_1(1,\chi_{p3},\chi_{q3},\chi_p,\chi_{12})\,\chi_{12}^2
\:+\,H^{F'}_1(1,\chi_{12},\chi_1,\chi_2,\chi_{12})\,\left[ 1/4\,\chi_{12}^2 
- 1/2\,R^1_2 R^2_1\,\chi_{12}^2 \right] \nonumber \\
&& - \,\:1/4\,H^{F'}_1(3,\chi_{12},\chi_p,\chi_p,\chi_{12})\,R^q_p R^d_p\,\chi_{12}^2
\:+\,1/4\,H^{F'}_1(3,\chi_{12},\chi_p,\chi_q,\chi_{12})\,R^q_p R^d_p\,\chi_{12}^2 
\nonumber \\
&& - \,\:1/8\,H^{F'}_1(7,\chi_{12},\chi_p,\chi_p,\chi_{12})\,(R^d_p)^2 \chi_{12}^2 
\:-\,3/8\,H^{F'}_{21}(1,\chi_p,\chi_p,\chi_{12},\chi_{12})\,\chi_{12}^2 \nonumber \\
&& + \,\:3/8\,H^{F'}_{21}(1,\chi_p,\chi_{13},\chi_{23},\chi_{12})\,R^q_p\,\chi_{12}^2 
\:+\,3/8\,H^{F'}_{21}(1,\chi_{p3},\chi_{q3},\chi_p,\chi_{12})\,R^p_q\,\chi_{12}^2 
\nonumber \\
&& - \,\:3/8\,H^{F'}_{21}(1,\chi_{p3},\chi_{q3},\chi_q,\chi_{12})\,R^p_q\,\chi_{12}^2 
\:+\,H^{F'}_{21}(1,\chi_{12},\chi_p,\chi_p,\chi_{12})\,\left[ 15/64\,\chi_{12}^2 
- 3/16\,R^p_q R^q_p\,\chi_{12}^2 \right] \nonumber \\
&& - \,\:H^{F'}_{21}(1,\chi_{12},\chi_1,\chi_2,\chi_{12})\,\left[ 3/32\,\chi_{12}^2 
- 3/8\,R^1_2 R^2_1\,\chi_{12}^2 \right]
\:+\,3/8\,H^{F'}_{21}(1,\chi_{12},\chi_{12},\chi_{12},\chi_{12})\,\chi_{12}^2
\nonumber \\
&& + \,\:9/8\,H^{F'}_{21}(1,\chi_3,\chi_{13},\chi_{23},\chi_{12})\,\chi_{12}^2
\:-\,3/8\,H^{F'}_{21}(3,\chi_{p3},\chi_p,\chi_{q3},\chi_{12})\,R^d_p\,\chi_{12}^2 
\nonumber \\
&& - \,\:H^{F'}_{21}(3,\chi_{12},\chi_p,\chi_p,\chi_{12})\,\left[ 3/16\,R^p_q 
R^d_p\,\chi_{12}^2 - 3/32\,R^d_p\,\chi_{12}^2 \right] 
\:-\,H^{F'}_{21}(3,\chi_{12},\chi_p,\chi_q,\chi_{12})\,\left[ 3/16\,R^q_p R^d_p\,\chi_{12}^2 
\right. \nonumber \\
&& - \,\:3/32 \left. R^d_p\,\chi_{12}^2 \right]
\:+\,3/64\,H^{F'}_{21}(7,\chi_{12},\chi_p,\chi_p,\chi_{12})\,(R^d_p)^2 \chi_{12}^2 
\:+\,3/32\,H^{F'}_{21}(7,\chi_{12},\chi_1,\chi_2,\chi_{12})\,R^d_1 R^d_2\,\chi_{12}^2,
\label{F0loop_NNLO_nf2_21} 
\end{eqnarray} 

\begin{eqnarray} 
\delta^{(6)22}_{\mathrm{loops}} & = & \pi_{16}\,L^r_{0}
\,\left[ - 4\,\chi_\pi \chi_{12} - 3\,\chi_\pi^2 - 2\,\chi_1 \chi_2 + 4\,\chi_{12}^2 
+ 2\,\chi_3 \chi_4 \right]
\:-\,2\,\pi_{16}\,L^r_{1}\,\chi_{12}^2
\:-\,\pi_{16}\,L^r_{2}\,\left[ 3\,\chi_\pi^2 + \chi_{12}^2 \right] 
\nonumber \\
&& +\,\:\pi_{16}\,L^r_{3}\,\left[ - 3\,\chi_\pi \chi_{12} - 2\,\chi_\pi^2 
- 3/2\,\chi_1 \chi_2 + 4\,\chi_{12}^2 + 3/2\,\chi_3 \chi_4 \right]
\:+\,\pi_{16}^2\,\left[ - 23/96\,\chi_\pi \chi_{12} - 1/3\,\chi_\pi^2 
- 17/192\,\chi_1 \chi_2 \right. \nonumber \\
&& +\,\:5/48 \left. \chi_{12}^2 + 17/192\,\chi_3 \chi_4 \right]
\:-\,32\,L^r_{4}L^r_{5}\,\chi_{12} \chi_{34} \:-\,32\,L^{r2}_{4}\,\chi_{34}^2 
\:-\,8\,L^{r2}_{5}\,\chi_{12}^2 
\:+\,8\,L^{r2}_{11}\,\chi_{12}^2 \nonumber \\
&& + \,\:\bar{A}(\chi_p)\,\pi_{16}\left[ - 1/12\,R^p_{q\pi}\,\chi_{12} - 
1/8\,R^p_{q\pi}\,\chi_{34} + 1/48\,R^c_p\,\chi_q 
+ 1/48\,R^c_p\,\chi_{12} + 1/16\,R^c_p\,\chi_{34} \right] \nonumber \\
&& + \,\:\bar{A}(\chi_p)\,L^r_{0}\,\left[ 4\,R^p_{q\pi}\,\chi_p 
+ R^c_p\,\chi_p + R^d_p \right] 
\:+\,\bar{A}(\chi_p)\,L^r_{3}\,\left[ R^p_{q\pi}\,\chi_p + 5/2\,R^c_p\,\chi_p 
+ 5/2\,R^d_p \right] \nonumber \\ 
&& - \,\:\bar{A}(\chi_p)\,L^r_{4}\,\left[ 2\,R^p_{q\pi}\,\chi_{34} - R^c_p\,\chi_{34} 
\right] 
\:-\,\bar{A}(\chi_p)\,L^r_{5}\,\left[ R^p_{q\pi}\,\chi_{12} 
+ 1/2\,R^c_p\,\chi_{12} \right] 
\:+\,\bar{A}(\chi_p)^2\,\left[ 1/16 + 1/32\,(R^p_{q\pi})^2 \right. \nonumber \\
&& - \,\:1/32\,R^p_{q\pi} R^c_p + 1/128 \left. (R^c_p)^2 \right]
\:-\,\bar{A}(\chi_p) \bar{A}(\chi_{ps})\,\left[ 1/24\,R^p_{q\pi} + 5/48\,R^p_{s\pi} 
- 7/96\,R^c_p \right] \nonumber \\
&& - \,\:\bar{A}(\chi_p) \bar{A}(\chi_{qs})\,\left[ 1/24\,R^p_{q\pi} 
+ 1/16\,R^p_{s\pi} + 1/32\,R^c_p \right] 
\:-\,\bar{A}(\chi_p) \bar{A}(\chi_\pi)\,\left[ 1/32\,R^p_{q\pi} R^v_{\pi 12} 
- 1/64\,R^c_p R^v_{\pi 12} \right] \nonumber \\
&& + \,\:1/8\,\bar{A}(\chi_p) \bar{A}(\chi_{12})
\:+\,1/12\,\bar{A}(\chi_p) \bar{A}(\chi_{34})\,R^\pi_{pp}
\:+\,\bar{A}(\chi_p) \bar{B}(\chi_p,\chi_p,0)\,\left[ 1/4\,\chi_p - 1/8\,R^p_{q\pi} R^c_p 
\,\chi_p - 1/32\,R^p_{q\pi} R^d_p \right. \nonumber \\
&& + \,\:1/8\,(R^c_p)^2 \chi_p + 1/64 \left. R^c_p R^d_p \right]
\:+\,1/8\,\bar{A}(\chi_p) \bar{B}(\chi_p,\chi_\pi,0)\,R^z_{q\pi p} R^c_p R^v_{\pi 12}\, 
\chi_p \:-\,\bar{A}(\chi_p) \bar{B}(\chi_q,\chi_q,0)\,\left[ 1/32\,R^p_{q\pi} R^d_q 
\right. \nonumber \\
&& - \,\:1/64 \left. R^c_p R^d_q \right]
\:-\,1/8\,\bar{A}(\chi_p) \bar{B}(\chi_{ps},\chi_{ps},0)\,R^p_{s\pi}\,\chi_{ps}
\:-\,1/8\,\bar{A}(\chi_p) \bar{B}(\chi_1,\chi_2,0)\,R^q_{p\pi} R^c_p\,\chi_p \nonumber \\
&& + \,\:1/8\,\bar{A}(\chi_p) \bar{C}(\chi_p,\chi_p,\chi_p,0)\,R^c_p R^d_p\,\chi_p 
\:+\,\bar{A}(\chi_p,\varepsilon)\,\pi_{16}\,\left[ - 1/8\,\chi_p + 
1/8\,R^p_{q\pi}\,\chi_{34} + 1/8\,R^\pi_{pp}\,\chi_{34} \right] \nonumber \\
&& + \,\:\bar{A}(\chi_{ps})\,\pi_{16}\,\left[ 1/24\,\chi_{ps} - 1/8\,\chi_{qs} 
- 1/8\,\chi_{34} \right]
\:-\,2\,\bar{A}(\chi_{ps})\,L^r_{0}\,\chi_{ps}
\:-\,5\,\bar{A}(\chi_{ps})\,L^r_{3}\,\chi_{ps}
\:-\,2\,\bar{A}(\chi_{ps})\,L^r_{4}\,\chi_{34} \nonumber \\
&& + \,\:\bar{A}(\chi_{ps})\,L^r_{5}\,\chi_{12}
\:-\,1/32\,\bar{A}(\chi_{ps})^2
\:+\,\bar{A}(\chi_{ps}) \bar{A}(\chi_\pi)\,\left[ 7/96\,R^\pi_{pp} - 5/48\,R^\pi_{ps} 
- 1/32\,R^\pi_{qq} + 5/48\,R^\pi_{qs} - 1/24\,R^\pi_{12} \right] \nonumber \\
&& + \,\:\bar{A}(\chi_{ps}) \bar{B}(\chi_p,\chi_p,0)\,\left[ 1/16\,R^p_{s\pi}\,\chi_p 
- 5/16\,R^p_{s\pi}\,\chi_{ps} \right]
\:-\,\bar{A}(\chi_{ps}) \bar{B}(\chi_p,\chi_\pi,0)\,\left[ 1/12\,R^\pi_{ps} R^z_{qp\pi} 
\,\chi_p \right. \nonumber \\
&& + \,\:1/6 \left. R^\pi_{ps} R^z_{qp\pi}\,\chi_{ps} \right]
\:-\,1/32\,\bar{A}(\chi_{ps}) \bar{B}(\chi_q,\chi_q,0)\,R^d_q
\:+\,1/12\,\bar{A}(\chi_{ps}) \bar{B}(\chi_1,\chi_2,0)\,R^q_{s\pi}\,\chi_s \nonumber \\
&& + \,\:1/6\,\bar{A}(\chi_{ps}) \bar{B}(\chi_1,\chi_2,0,k)\,R^q_{s\pi} 
\:+\,1/8\,\bar{A}(\chi_{ps},\varepsilon)\,\pi_{16}\,\left[\chi_{34} + \chi_s \right] 
\:-\,1/16\,\bar{A}(\chi_{p3}) \bar{A}(\chi_{p4}) \nonumber \\
&& + \,\:1/16\,\bar{A}(\chi_{p3}) \bar{A}(\chi_{q4})
\:+\,\bar{A}(\chi_\pi)\,\pi_{16}\,\left[ 1/24\,R^v_{\pi 12}\,\chi_\pi 
+ 1/16\,R^v_{\pi 12}\,\chi_{12} \right]
\:+\,\bar{A}(\chi_\pi)\,L^r_{0}\,\left[ 6\,R^\pi_{12}\,\chi_\pi 
+ R^v_{\pi 12}\,\chi_\pi \right] \nonumber \\
&& - \,\:8\,\bar{A}(\chi_\pi)\,L^r_{1}\,\chi_\pi
\:-\,2\,\bar{A}(\chi_\pi)\,L^r_{2}\,\chi_\pi
\:+\,\bar{A}(\chi_\pi)\,L^r_{3}\,\left[ 6\,R^\pi_{12}\,\chi_\pi 
+ 5/2\,R^v_{\pi 12}\,\chi_\pi \right]
\:+\,\bar{A}(\chi_\pi)\,L^r_{4}\,\left[ 4\,\chi_\pi 
+ R^v_{\pi 12}\,\chi_\pi \right] \nonumber \\
&& - \,\:\bar{A}(\chi_\pi)\,L^r_{5}\,\left[ 2\,R^\pi_{12}\,\chi_{12} + 1/2\,R^v_{\pi 12} 
\,\chi_{12} \right] \:+\,1/128\,\bar{A}(\chi_\pi)^2\,(R^v_{\pi 12})^2 
\:+\,1/12\,\bar{A}(\chi_\pi) \bar{A}(\chi_{34})\,R^v_{\pi 12} \nonumber \\
&& + \,\:\bar{A}(\chi_\pi) \bar{B}(\chi_p,\chi_p,0)\,\left[ - 1/8\,R^p_{q\pi} 
R^\pi_{pp}\,\chi_p + 3/16\,R^c_p\,\chi_p - 1/8\,(R^c_p)^2 \chi_p 
- 1/16\,R^d_p + 1/64\,R^d_p R^v_{\pi 12} \right] \nonumber \\ 
&& - \,\:1/8\,\bar{A}(\chi_\pi) \bar{B}(\chi_p,\chi_\pi,0)\,R^p_{\pi\pi} R^\pi_{12} 
R^z_{pq\pi}\,\chi_p \:-\,1/8\,\bar{A}(\chi_\pi) \bar{B}(\chi_{ps},\chi_{ps},0)\, 
R^\pi_{ps}\,\chi_{ps} \nonumber \\
&& - \,\:1/16\,\bar{A}(\chi_\pi) \bar{B}(\chi_\pi,\chi_\pi,0)\,R^v_{\pi 12}\,\chi_{34}
\:-\,1/8\,\bar{A}(\chi_\pi) \bar{B}(\chi_1,\chi_2,0)\,R^1_{\pi\pi} R^2_{\pi\pi} 
\,\chi_{34} \nonumber \\
&& + \,\:1/8\,\bar{A}(\chi_\pi) \bar{C}(\chi_p,\chi_p,\chi_p,0)\,R^\pi_{pp} R^d_p\,\chi_p 
\:+\,1/8\,\bar{A}(\chi_\pi,\varepsilon)\,\pi_{16}\,\left[ \chi_\pi - R^\pi_{12}\,\chi_\pi 
- R^v_{\pi 12}\,\chi_\pi \right] \nonumber \\
&& + \,\:\bar{A}(\chi_1) \bar{A}(\chi_2)\left[ - 1/32\,R^p_{q\pi} R^c_q + 1/16\,R^1_{2\pi} 
R^2_{1\pi} + 1/64\,R^c_1 R^c_2 \right]
\:-\,4\,\bar{A}(\chi_{12})\,L^r_{1}\,\chi_{12}
\:-\,10\,\bar{A}(\chi_{12})\,L^r_{2}\,\chi_{12} \nonumber \\
&& + \,\:1/8\,\bar{A}(\chi_{12})^2
\:-\,1/2\,\bar{A}(\chi_{12}) \bar{B}(\chi_1,\chi_2,0,k)
\:+\,1/4\,\bar{A}(\chi_{12},\varepsilon)\,\pi_{16}\,\chi_{12}
\:-\,16\,\bar{A}(\chi_{34})\,L^r_{1}\,\chi_{34} \nonumber \\
&& - \,\:4\,\bar{A}(\chi_{34})\,L^r_{2}\,\chi_{34}
\:+\,8\,\bar{A}(\chi_{34})\,L^r_{4}\,\chi_{34}
\:+\,1/8\,\bar{A}(\chi_{34}) \bar{B}(\chi_p,\chi_p,0)\,R^p_{\pi\pi}\,\chi_{34} \nonumber \\ 
&& + \,\:\bar{A}(\chi_{34}) \bar{B}(\chi_p,\chi_\pi,0)\,\left[ 1/12\,R^z_{q\pi p} 
R^v_{\pi 12} 
\,\chi_p - 1/6\,R^z_{q\pi p} R^v_{\pi 12}\,\chi_{34} \right]
\:-\,1/6\,\bar{A}(\chi_{34}) \bar{B}(\chi_p,\chi_\pi,0,k)\,R^z_{q\pi p} R^v_{\pi 12}
\nonumber \\
&& + \,\:1/8\,\bar{A}(\chi_{34}) \bar{B}(\chi_\pi,\chi_\pi,0)\,R^v_{\pi 12}\,\chi_{34}
\:+\,\bar{A}(\chi_{34}) \bar{B}(\chi_1,\chi_2,0)\left[ - 1/4\,\chi_{34} + 1/6\,R^\pi_{12} 
\,\chi_{12} - 1/4\,R^\pi_{12}\,\chi_{34} \right] \nonumber \\
&& - \,\:1/6\,\bar{A}(\chi_{34}) \bar{B}(\chi_1,\chi_2,0,k)\,R^\pi_{12}
\:+\,1/4\,\bar{A}(\chi_{34},\varepsilon)\,\pi_{16}\,\chi_{34}
\:+\,1/16\,\bar{A}(\chi_{1s}) \bar{A}(\chi_{2s}) \nonumber \\
&& + \,\:\bar{B}(\chi_p,\chi_p,0)\,\pi_{16}\left[ 1/96\,R^d_p\,\chi_p + 
1/32\,R^d_p\,\chi_q + 1/16\,R^d_p\,\chi_{34} \right]
\:+\,\bar{B}(\chi_p,\chi_p,0)\,L^r_{0}\,R^d_p\,\chi_p \nonumber \\
&& + \,\:5/2\,\bar{B}(\chi_p,\chi_p,0)\,L^r_{3}\,R^d_p\,\chi_p
\:+\,\bar{B}(\chi_p,\chi_p,0)\,L^r_{4}\,\left[ - 4\,R^p_{q\pi}\,\chi_p \chi_{34} 
+ 2\,R^c_p\,\chi_p \chi_{34} + 3\,R^d_p\,\chi_{34} \right] \nonumber \\
&& + \,\:\bar{B}(\chi_p,\chi_p,0)\,L^r_{5}\,\left[ - 2\,R^p_{q\pi}\,\chi_1 \chi_2 
+ R^c_p\,\chi_p^2 - 1/2\,R^d_p\,\chi_{12} \right]
\:+\,\bar{B}(\chi_p,\chi_p,0)\,L^r_{6}\,\left[ 8\,R^p_{q\pi}\,\chi_p \chi_{34} 
- 4\,R^c_p\,\chi_p \chi_{34} \right. \nonumber \\ 
&& - \,\:4\,R^d_p \left. \chi_{34} \right]
\:+\,4\,\bar{B}(\chi_p,\chi_p,0)\,L^r_{7}\,(R^d_p)^2
\:+\,\bar{B}(\chi_p,\chi_p,0)\,L^r_{8}\,\left[ 4\,R^p_{q\pi}\,\chi_1 \chi_2 
- 2\,R^c_p\,\chi_p^2 + 2\,(R^d_p)^2 \right] \nonumber \\
&& - \,\:\bar{B}(\chi_p,\chi_p,0)^2\,\left[ 1/8\,R^p_{q\pi} R^d_p\,\chi_p 
- 1/8\,R^c_p R^d_p\,\chi_p - 1/128\,(R^d_p)^2 \right] \nonumber \\
&& + \,\:1/8\,\bar{B}(\chi_p,\chi_p,0) \bar{B}(\chi_p,\chi_\pi,0)\,R^z_{q\pi p} R^d_p 
R^v_{\pi 12}\,\chi_p
\:-\,1/8\,\bar{B}(\chi_p,\chi_p,0) \bar{B}(\chi_1,\chi_2,0)\,R^q_{p\pi} R^d_p\,\chi_p
\nonumber \\
&& + \,\:1/8\,\bar{B}(\chi_p,\chi_p,0) \bar{C}(\chi_p,\chi_p,\chi_p,0)\,(R^d_p)^2 \chi_p 
\:-\,1/16\,\bar{B}(\chi_p,\chi_p,0,\varepsilon)\,\pi_{16}\,\left[ R^d_p\,\chi_p 
+ R^d_p\,\chi_{34} \right] \nonumber \\
&& + \,\:8\,\bar{B}(\chi_p,\chi_\pi,0)\,L^r_{7}\,R^z_{qp\pi} R^d_p R^z_{\pi 34p}
\:+\,4\,\bar{B}(\chi_p,\chi_\pi,0)\,L^r_{8}\,R^z_{qp\pi} R^d_p R^z_{\pi 34p}
\:-\,\:4\,\bar{B}(\chi_{ps},\chi_{ps},0)\,L^r_{4}\,\chi_{ps} \chi_{34} \nonumber \\
&& - \,\:2\,\bar{B}(\chi_{ps},\chi_{ps},0)\,L^r_{5}\,\chi_{ps}^2
\:+\,8\,\bar{B}(\chi_{ps},\chi_{ps},0)\,L^r_{6}\,\chi_{ps} \chi_{34}
\:+\,4\,\bar{B}(\chi_{ps},\chi_{ps},0)\,L^r_{8}\,\chi_{ps}^2 \nonumber \\
&& + \,\:2\,\bar{B}(\chi_\pi,\chi_\pi,0)\,L^r_{4}\,R^v_{\pi 12}\,\chi_\pi^2
\:+\,\bar{B}(\chi_\pi,\chi_\pi,0)\,L^r_{5}\,R^v_{\pi 12}\,\chi_\pi^2
\:-\,4\,\bar{B}(\chi_\pi,\chi_\pi,0)\,L^r_{6}\,R^v_{\pi 12}\,\chi_\pi^2 \nonumber \\
&& - \,\:4\,\bar{B}(\chi_\pi,\chi_\pi,0)\,L^r_{7}\,(R^z_{\pi 4})^2 R^v_{\pi 12} 
\:-\,\bar{B}(\chi_\pi,\chi_\pi,0)\,L^r_{8}\,\left[ R^v_{\pi 12}\,\chi_3^2 
+ R^v_{\pi 12}\,\chi_4^2 \right] \nonumber \\
&& + \,\:1/64\,\bar{B}(\chi_1,\chi_1,0) \bar{B}(\chi_2,\chi_2,0)\,R^d_1 R^d_2
\:-\,8\,\bar{B}(\chi_1,\chi_2,0)\,L^r_{7}\,R^d_1 R^d_2
\:-\,4\,\bar{B}(\chi_1,\chi_2,0)\,L^r_{8}\,R^d_1 R^d_2 \nonumber \\
&& + \,\:4\,\bar{C}(\chi_p,\chi_p,\chi_p,0)\,L^r_{4}\,R^d_p\,\chi_p \chi_{34}
\:+\,2\,\bar{C}(\chi_p,\chi_p,\chi_p,0)\,L^r_{5}\,R^d_p\,\chi_p^2
\:-\,8\,\bar{C}(\chi_p,\chi_p,\chi_p,0)\,L^r_{6}\,R^d_p\,\chi_p \chi_{34} \nonumber \\
&& - \,\:4\,\bar{C}(\chi_p,\chi_p,\chi_p,0)\,L^r_{8}\,R^d_p\,\chi_p^2
\:+\,H^{F}(1,\chi_p,\chi_p,\chi_{12},\chi_{12})\,\left[ 1/8\,\chi_p - 1/16\,\chi_{12} 
- 1/16\,(R^p_{q\pi})^2 \chi_{12} \right. \nonumber \\
&& + \,\:1/16 \left. R^p_{q\pi} R^c_p\,\chi_{12} - 1/64\,(R^c_p)^2 \chi_{12} \right]
\:+\,H^{F}(1,\chi_p,\chi_{1s},\chi_{2s},\chi_{12})\,\left[ - 1/8\,R^p_{q\pi}\,\chi_{qs} 
+ 1/16\,R^p_{q\pi}\,\chi_{12} \right. \nonumber \\
&& - \,\:3/32 \left. R^p_{s\pi}\,\chi_p + 1/32\,R^p_{s\pi}\,\chi_q 
+ 1/16\,R^c_p\,\chi_{ps} \right]
\:-\,1/8\,H^{F}(1,\chi_{p3},\chi_{q4},\chi_{34},\chi_{12})\,\chi_{34} \nonumber \\
&& + \,\:H^{F}(1,\chi_\pi,\chi_p,\chi_{12},\chi_{12})\left[ 1/16\,R^p_{q\pi} R^v_{\pi 12} 
\,\chi_{12} - 1/32\,R^c_p R^v_{\pi 12}\,\chi_{12} \right]
\:-\,1/64\,H^{F}(1,\chi_\pi,\chi_\pi,\chi_{12},\chi_{12})\,(R^v_{\pi 12})^2 \chi_{12}
\nonumber \\  
&& + \,\:1/32\,H^{F}(1,\chi_\pi,\chi_{1s},\chi_{2s},\chi_{12})\,\left[ - R^\pi_{ps}\,\chi_p 
+ R^\pi_{ps}\,\chi_q + R^v_{\pi ps}\,\chi_\pi + R^v_{\pi 12}\,\chi_s \right]
\nonumber \\ 
&& + \,\:H^{F}(1,\chi_1,\chi_{12},\chi_2,\chi_{12})\,\left[ 1/16\,R^p_{q\pi} 
R^c_q\,\chi_{12} - 1/8\,R^1_{2\pi} R^2_{1\pi}\,\chi_{12} 
- 1/32\,R^c_1 R^c_2\,\chi_{12} \right] \nonumber \\
&& - \,\:1/8\,H^{F}(1,\chi_{12},\chi_{12},\chi_{12},\chi_{12})\,\chi_{12}
\:+\,H^{F}(2,\chi_p,\chi_p,\chi_{12},\chi_{12})\,\left[ 1/16\,R^p_{q\pi} R^d_p\,\chi_{12} 
- 1/32\,R^c_p R^d_p\,\chi_{12} \right] \nonumber \\
&& - \,\:1/32\,H^{F}(2,\chi_p,\chi_\pi,\chi_{12},\chi_{12})\,R^d_p R^v_{\pi 12}\,\chi_{12}
\:+\,H^{F}(2,\chi_p,\chi_{12},\chi_q,\chi_{12})\,\left[ 1/16\,R^q_{p\pi} R^d_p\,\chi_{12} 
- 1/32\,R^c_q R^d_p\,\chi_{12} \right] \nonumber \\
&& + \,\:1/16\,H^{F}(2,\chi_p,\chi_{1s},\chi_{2s},\chi_{12})\,R^d_p\,\chi_{ps}
\:-\,1/64\,H^{F}(5,\chi_p,\chi_p,\chi_{12},\chi_{12})\,(R^d_p)^2 \chi_{12} \nonumber \\
&& - \,\:1/32\,H^{F}(5,\chi_1,\chi_2,\chi_{12},\chi_{12})\,R^d_1 R^d_2\,\chi_{12}
\:+\,H^{F'}(1,\chi_p,\chi_p,\chi_{12},\chi_{12})\,\left[ - 1/8\,\chi_p \chi_{12} 
- 3/16\,\chi_{12}^2 \right. \nonumber \\
&& - \,\:1/16\,(R^p_{q\pi})^2 \chi_{12}^2 
+ 3/16\,R^p_{q\pi} R^c_p\,\chi_{12}^2 - 3/64 \left. (R^c_p)^2 \chi_{12}^2 \right]
\:+\,H^{F'}(1,\chi_p,\chi_{1s},\chi_{2s},\chi_{12})\,\left[ - 5/16\,R^p_{q\pi}\,\chi_q 
\chi_{12} \right. \nonumber \\
&& + \,\:1/8\,R^p_{q\pi}\,\chi_{ps} \chi_{12}
+ 3/32\,R^p_{s\pi}\,\chi_q \chi_{12} + 7/32\,R^p_{s\pi}\,\chi_{12} \chi_s 
- 1/16\,R^c_p \left. \chi_{ps} \chi_{12} \right] \nonumber \\
&& + \,\:1/8\,H^{F'}(1,\chi_{p3},\chi_{q4},\chi_{34},\chi_{12})\,\chi_{12} \chi_{34}
\:+\,H^{F'}(1,\chi_\pi,\chi_p,\chi_{12},\chi_{12})\,\left[ 1/4\,R^p_{q\pi} R^\pi_{12} 
\,\chi_{12}^2 - 1/16\,R^p_{q\pi} R^v_{\pi 12}\,\chi_{12}^2 \right. \nonumber \\
&& + \,\:1/8\,R^\pi_{pp} R^c_p\,\chi_{12}^2 
+ 1/32\,R^c_p R^v_{\pi 12} \left. \chi_{12}^2 \right]
\:+\,H^{F'}(1,\chi_\pi,\chi_\pi,\chi_{12},\chi_{12})\,\left[ 1/4\,(R^\pi_{12})^2 
\chi_{12}^2 + 1/4\,R^\pi_{12} R^v_{\pi 12}\,\chi_{12}^2 \right. \nonumber \\
&& + \,\:5/64 \left. (R^v_{\pi 12})^2 \chi_{12}^2 \right]
\:+\,H^{F'}(1,\chi_\pi,\chi_{1s},\chi_{2s},\chi_{12})\,\left[ - 3/32\,R^\pi_{ps}\,\chi_p 
\chi_{12} - 5/32\,R^\pi_{ps}\,\chi_q \chi_{12} 
+ 1/4\,R^\pi_{12}\,\chi_{12}^2 \right. \nonumber \\
&& - \,\:1/32\,R^v_{\pi ps}\,\chi_\pi \chi_{12} 
- 1/32\,R^v_{\pi 12} \left. \chi_{12} \chi_s \right]
\:+\,H^{F'}(1,\chi_1,\chi_{12},\chi_2,\chi_{12})\,\left[ - 1/16\,R^p_{q\pi} R^c_q 
\,\chi_{12}^2 + 3/8\,R^1_{2\pi} R^2_{1\pi}\,\chi_{12}^2 \right. \nonumber \\
&& + \,\:1/32\,R^c_1 R^c_2 \left. \chi_{12}^2 \right]
\:+\,1/8\,H^{F'}(1,\chi_{12},\chi_{12},\chi_{12},\chi_{12})\,\chi_{12}^2
\:+\,H^{F'}(2,\chi_p,\chi_p,\chi_{12},\chi_{12})\,\left[ - 1/16\,R^p_{q\pi} R^d_p 
\,\chi_{12}^2 \right. \nonumber \\
&& + \,\:5/32\,R^c_p R^d_p \left. \chi_{12}^2 \right]
\:+\,H^{F'}(2,\chi_p,\chi_\pi,\chi_{12},\chi_{12})\,\left[ 1/8\,R^\pi_{pp} R^d_p 
\,\chi_{12}^2 + 1/32\,R^d_p R^v_{\pi 12}\,\chi_{12}^2 \right] \nonumber \\
&& - \,\:H^{F'}(2,\chi_p,\chi_{12},\chi_q,\chi_{12})\,\left[ 1/16\,R^q_{p\pi} 
R^d_p\,\chi_{12}^2 - 1/32\,R^c_q R^d_p\,\chi_{12}^2 \right]
\:-\,1/16\,H^{F'}(2,\chi_p,\chi_{1s},\chi_{2s},\chi_{12})\,R^d_p\,\chi_{ps} \chi_{12} 
\nonumber \\
&& + \,\:5/64\,H^{F'}(5,\chi_p,\chi_p,\chi_{12},\chi_{12})\,(R^d_p)^2 \chi_{12}^2 
\:+\,1/32\,H^{F'}(5,\chi_1,\chi_2,\chi_{12},\chi_{12})\,R^d_1 R^d_2\,\chi_{12}^2
\nonumber \\
&& + \,\:H^{F'}_1(1,\chi_p,\chi_p,\chi_{12},\chi_{12})\,\left[ \chi_{12}^2 + 
1/2\,(R^p_{q\pi})^2 
\,\chi_{12}^2 - 1/2\,R^p_{q\pi} R^c_p\,\chi_{12}^2 + 1/4\,(R^c_p)^2 \chi_{12}^2 \right] 
\nonumber \\
&& + \,\:1/2\,H^{F'}_1(1,\chi_p,\chi_{1s},\chi_{2s},\chi_{12})\,R^p_{q\pi} R^z_{sqp} 
\,\chi_{12}^2
\:-\,1/2\,H^{F'}_1(1,\chi_{ps},\chi_{qs},\chi_p,\chi_{12})\,R^p_{s\pi}\,\chi_{12}^2
\nonumber \\
&& - \,\:1/2\,H^{F'}_1(1,\chi_{ps},\chi_{qs},\chi_\pi,\chi_{12})\,R^\pi_{12} 
R^z_{sp\pi}\,\chi_{12}^2
\:+\,1/4\,H^{F'}_1(1,\chi_{12},\chi_p,\chi_\pi,\chi_{12})\,\left[ R^p_{q\pi} 
R^v_{\pi 12}\,\chi_{12}^2 \right. \nonumber \\
&& - \,\:R^\pi_{pp} R^z_{qp\pi} R^c_p \left. \chi_{12}^2 \right] 
\:-\,H^{F'}_1(1,\chi_{12},\chi_\pi,\chi_\pi,\chi_{12})\,\left[ 1/4\,R^\pi_{12} 
R^v_{\pi 12} \,\chi_{12}^2 + 1/8\,(R^v_{\pi 12})^2 \chi_{12}^2 \right] \nonumber \\
&& + \,\:H^{F'}_1(1,\chi_{12},\chi_1,\chi_2,\chi_{12})\,\left[ 1/4\,R^p_{q\pi} R^c_q 
\,\chi_{12}^2 - 1/2\,R^1_{2\pi} R^2_{1\pi}\,\chi_{12}^2 \right]
\:+\,1/4\,H^{F'}_1(3,\chi_{12},\chi_p,\chi_p,\chi_{12})\,\left[ R^p_{q\pi} R^d_p 
\,\chi_{12}^2 \right. \nonumber \\
&& - \,\:R^c_p R^d_p \left. \chi_{12}^2 \right]
\:+\,1/4\,H^{F'}_1(3,\chi_{12},\chi_p,\chi_q,\chi_{12})\,R^q_{p\pi} R^d_p\,\chi_{12}^2
\:+\,1/4\,H^{F'}_1(3,\chi_{12},\chi_p,\chi_\pi,\chi_{12})\,R^\pi_{12} R^z_{pq\pi} 
R^d_p\,\chi_{12}^2 \nonumber \\
&& - \,\:1/8\,H^{F'}_1(7,\chi_{12},\chi_p,\chi_p,\chi_{12})\,(R^d_p)^2 \chi_{12}^2 
\:-\,3/8\,H^{F'}_{21}(1,\chi_p,\chi_p,\chi_{12},\chi_{12})\,\chi_{12}^2 \nonumber \\ 
&& - \,\:3/16\,H^{F'}_{21}(1,\chi_p,\chi_{1s},\chi_{2s},\chi_{12})\,R^p_{q\pi} 
R^z_{sqp}\,\chi_{12}^2
\:+\,3/16\,H^{F'}_{21}(1,\chi_{ps},\chi_{qs},\chi_p,\chi_{12})\,\left[ R^p_{q\pi} 
\,\chi_{12}^2 + R^p_{s\pi}\,\chi_{12}^2 - R^c_p\,\chi_{12}^2 \right] \nonumber \\
&& + \,\:3/16\,H^{F'}_{21}(1,\chi_{ps},\chi_{qs},\chi_q,\chi_{12})\,R^q_{p\pi} R^z_{spq} 
\,\chi_{12}^2
\:+\,3/16\,H^{F'}_{21}(1,\chi_{ps},\chi_{qs},\chi_\pi,\chi_{12})\,R^\pi_{12} 
R^z_{pq\pi} R^z_{sp\pi}\,\chi_{12}^2 \nonumber \\
&& - \,\:3/16\,H^{F'}_{21}(1,\chi_\pi,\chi_{1s},\chi_{2s},\chi_{12})\,R^\pi_{12} 
R^z_{s1\pi} R^z_{s2\pi}\,\chi_{12}^2
\:+\,H^{F'}_{21}(1,\chi_{12},\chi_p,\chi_p,\chi_{12})\,\left[ 3/16\,\chi_{12}^2 
+ 3/16\,(R^p_{q\pi})^2 \chi_{12}^2 \right. \nonumber \\
&& - \,\:3/16\,R^p_{q\pi} R^c_p\,\chi_{12}^2 
+ 3/64 \left. (R^c_p)^2 \chi_{12}^2 \right]
\:+\,H^{F'}_{21}(1,\chi_{12},\chi_p,\chi_\pi,\chi_{12})\,\left[ - 3/16\,R^p_{q\pi} R^v_{\pi 
12}\,\chi_{12}^2 + 3/32\,R^c_p R^v_{\pi 12}\,\chi_{12}^2 \right] \nonumber \\ 
&& + \,\:3/64\,H^{F'}_{21}(1,\chi_{12},\chi_\pi,\chi_\pi,\chi_{12})\,(R^v_{\pi 12})^2 
\chi_{12}^2
\:+\,H^{F'}_{21}(1,\chi_{12},\chi_1,\chi_2,\chi_{12})\,\left[ - 3/16\,R^p_{q\pi} R^c_q 
\,\chi_{12}^2 + 3/8\,R^1_{2\pi} R^2_{1\pi}\,\chi_{12}^2 \right. \nonumber \\
&& + \,\:3/32\,R^c_1 R^c_2 \left. \chi_{12}^2 \right]
\:+\,3/8\,H^{F'}_{21}(1,\chi_{12},\chi_{12},\chi_{12},\chi_{12})\,\chi_{12}^2
\:+\,3/8\,H^{F'}_{21}(1,\chi_{34},\chi_{p3},\chi_{q4},\chi_{12})\,\chi_{12}^2 \nonumber \\
&& - \,\:3/16\,H^{F'}_{21}(3,\chi_{ps},\chi_p,\chi_{qs},\chi_{12})\,R^d_p\,\chi_{12}^2 
\:-\,H^{F'}_{21}(3,\chi_{12},\chi_p,\chi_p,\chi_{12})\,\left[ 3/16\,R^p_{q\pi} R^d_p 
\,\chi_{12}^2 - 3/32\,R^c_p R^d_p\,\chi_{12}^2 \right] \nonumber \\ 
&& - \,\:H^{F'}_{21}(3,\chi_{12},\chi_p,\chi_q,\chi_{12})\,\left[ 3/16\,R^q_{p\pi} R^d_p 
\,\chi_{12}^2 - 3/32\,R^c_q R^d_p\,\chi_{12}^2 \right] 
\:+\,3/32\,H^{F'}_{21}(3,\chi_{12},\chi_p,\chi_\pi,\chi_{12})\,R^d_p R^v_{\pi 12} 
\,\chi_{12}^2 \nonumber \\
&& + \,\:3/64\,H^{F'}_{21}(7,\chi_{12},\chi_p,\chi_p,\chi_{12})\,(R^d_p)^2 \chi_{12}^2
\:+\,3/32\,H^{F'}_{21}(7,\chi_{12},\chi_1,\chi_2,\chi_{12})\,R^d_1 R^d_2\,\chi_{12}^2.
\label{F0loop_NNLO_nf2_22} 
\end{eqnarray} 

\end{widetext}

\section{Two-Flavor Pseudoscalar Meson Masses}

\begin{figure}[h!]
\begin{center}
\includegraphics[width=\columnwidth]{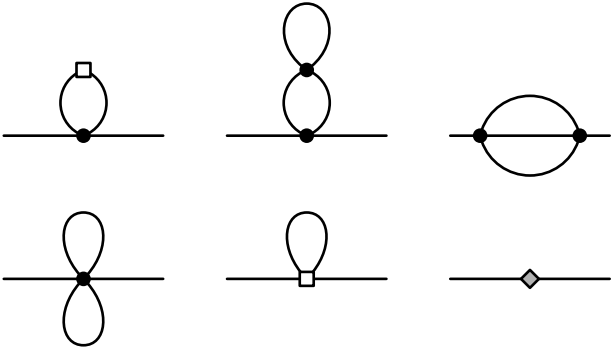}
\caption{Feynman diagrams at ${\mathcal O}(p^6)$ or two-loop for the 
irreducible self-energy contribution $-i\Sigma$ to the pseudoscalar 
meson mass. Filled circles denote vertices of the ${\mathcal L}_2$ 
Lagrangian, whereas open squares and shaded diamonds denote vertices of 
the ${\mathcal L}_4$ and ${\mathcal L}_6$ Lagrangians, respectively.}
\label{feynfig}
\end{center}
\end{figure}

The pseudoscalar meson squared masses $M_{\mathrm{PS}}^2$ are calculated
from the resummation of the irreducible self-energies, according to
\begin{equation}
M_{\mathrm{PS}}^2 = M_0^2 + \Sigma(M_{\mathrm{PS}}^2),
\label{massdef}
\end{equation}
where $M_0$ is the lowest order meson mass, and $\Sigma$ is the sum of 
the irreducible self-energy diagrams. The self-energy contributions 
$-i\Sigma$ at ${\mathcal O}(p^6)$, or NNLO, are shown in 
Fig.~\ref{feynfig}. The pseudoscalar meson mass is a function of the 
same mass inputs and LEC:s as outlined for the decay constants in the 
previous section, although the set of ${\mathcal O}(p^6)$ LEC:s that 
contribute to the meson mass is slightly different from that of the 
decay constant, in particular the number of ${\mathcal O}(p^6)$ LEC:s is 
larger for the meson mass. The results for the pseudoscalar 
meson masses are given in the form
\begin{equation}
M_{\mathrm{PS}}^2 = M_0^2 + \frac{\delta^{(4)\mathrm{vs}}}{F^2} +
\frac{\delta^{(6)\mathrm{vs}}_{\mathrm{ct}} + 
\delta^{(6)\mathrm{vs}}_{\mathrm{loops}}}{F^4} +
\mathcal{O}(p^8),
\label{masseq}
\end{equation}
where the ${\cal O}(p^4)$ and ${\cal O}(p^6)$ contributions have been 
separated. It should be noted that the lowest order mass has not been 
factored out of the results, as this is not meaningful for the results 
with $d_{\mathrm{val}} = 2$. Similarly to the treatment of the decay 
constant, the contribution from the ${\cal O}(p^6)$ counterterms has 
also been separated from the rest of the NNLO result, which consists 
mostly of the contributions from the chiral loops of ${\cal O}(p^6)$, 
although there is also a contribution from the diagrams at ${\mathcal 
O}(p^4)$ through the expansion of Eq.~(\ref{masseq}) to ${\cal O}(p^6)$. 
The superscripts (v) and (s) indicate the values of $d_{\mathrm{val}}$ 
and $d_{\mathrm{sea}}$, respectively. As expected, the infinities in all 
expressions for the meson mass have canceled.

\subsection{Results for $d_{\mathrm{val}} = 1$}

The lowest order mass, $M_0^2$, is for $d_{\mathrm{val}} = 1$ equal to 
$\chi_1$. The NLO, or ${\cal O}(p^4)$ contribution to the meson 
mass for $d_{\mathrm{val}} = 1$ and $d_{\mathrm{sea}} = 2$ is
\begin{eqnarray} 
\delta^{(4)12} & = & - \,\:16\,L^r_{4}\,\chi_1 \chi_{34} 
\:-\,8\,L^r_{5}\,\chi_1^2 \nonumber \\
&& + \,\:32\,L^r_{6}\,\chi_1 \chi_{34}
\:+\,16\,L^r_{8}\,\chi_1^2 \nonumber \\
&& - \,\:1/2\,\bar{A}(\chi_\pi)\,R^\pi_{11}\,\chi_1
\:-\,1/2\,\bar{A}(\chi_1)\,R^c_1\,\chi_1 \nonumber \\
&& - \,\:1/2\,\bar{B}(\chi_1,\chi_1,0)\,R^d_1\,\chi_1, 
\label{M0_NLO_nf2_12} 
\end{eqnarray} 
from which the $d_{\mathrm{sea}} = 1$ case can be obtained 
by carefully taking the limit $\chi_4 \rightarrow \chi_3$. 
In that limit the
$\bar{A}(\chi_\pi)$ term vanishes, and the remaining residues reduce 
such that $R^c_1 \rightarrow 1$, yielding the correct result 
for $d_{\mathrm{val}} = 1$ and $d_{\mathrm{sea}} = 1$. The tree level 
contribution from the ${\cal O}(p^6)$ Lagrangian is
\begin{eqnarray} 
\delta^{(6)12}_{\mathrm{ct}} & = & - \,\:32\,K^r_{17}\,\chi_1^3
\:-\,64\,K^r_{18}\,\chi_1^2 \chi_{34}
\:-\,16\,K^r_{19}\,\chi_1^3 \nonumber \\
&& - \,\:32\,K^r_{20}\,\chi_1^2 \chi_{34}
\:-\,16\,K^r_{21}\,\chi_1 \chi_s^2 \nonumber \\
&& - \,\:64\,K^r_{22}\,\chi_1 \chi_{34}^2
\:-\,16\,K^r_{23}\,\chi_1^3
\:+\,48\,K^r_{25}\,\chi_1^3 \nonumber \\
&& + \,\:K^r_{26}\,\left[ 16\,\chi_1 \chi_s^2 + 64\,\chi_1^2 \chi_{34} \right]
\:+\,192\,K^r_{27}\,\chi_1 \chi_{34}^2 \nonumber \\
&& + \,\:32\,K^r_{39}\,\chi_1^3
\:+\,64\,K^r_{40}\,\chi_1^2 \chi_{34},
\label{M0tree_NNLO_nf2_12} 
\end{eqnarray} 
from which the result for $d_{\mathrm{sea}} = 1$ may again be obtained 
by setting $\chi_4 \rightarrow \chi_3$. The remaining contributions to 
the meson mass at ${\cal O}(p^6)$ we again give separately for 
$d_{\mathrm{sea}} = 1$ and $d_{\mathrm{sea}} = 2$. These contributions 
are

\begin{widetext}

\begin{eqnarray} 
\delta^{(6)11}_{\mathrm{loops}} & = & \pi_{16}\,L^r_{0}\,\left[ 2\,\chi_1 \chi_3^2 
+ 8\,\chi_1^2 \chi_3 - 4\,\chi_1^3 \right]
\:+\,4\,\pi_{16}\,L^r_{1}\,\chi_1^3
\:+\,\pi_{16}\,L^r_{2}\,\left[ 6\,\chi_1 \chi_3^2 + 2\,\chi_1^3 \right]
\:+\,\pi_{16}\,L^r_{3}\,\left[ \chi_1 \chi_3^2 + 6\,\chi_1^2 \chi_3 
- 5\,\chi_1^3 \right] \nonumber \\
&& + \,\:\pi_{16}^2\,\left[ 47/96\,\chi_1 \chi_3^2 + 35/144\,\chi_1^2 \chi_3 
- 17/288\,\chi_1^3 \right] 
\:+\,256\,L^r_{4}L^r_{5}\,\chi_1^2 \chi_3
\:-\,512\,L^r_{4}L^r_{6}\,\chi_1 \chi_3^2
\:-\,256\,L^r_{4}L^r_{8}\,\chi_1^2 \chi_3 \nonumber \\
&& + \,\:128\,L^r_{4}L^r_{11}\,\chi_1^2 \chi_3
\:+\,256\,L^{r2}_{4}\,\chi_1 \chi_3^2
\:-\,256\,L^r_{5}L^r_{6}\,\chi_1^2 \chi_3
\:-\,128\,L^r_{5}L^r_{8}\,\chi_1^3 
\:+\,64\,L^r_{5}L^r_{11}\,\chi_1^3
\:+\,64\,L^{r2}_{5}\,\chi_1^3 \nonumber \\
&& - \,\:256\,L^r_{6}L^r_{11}\,\chi_1^2 \chi_3
\:-\,128\,L^r_{8}L^r_{11}\,\chi_1^3
\:-\,12\,\bar{A}(\chi_1)\,L^r_{0}\,\left[ \chi_1^2 + R^d_1\,\chi_1 \right]
\:+\,8\,\bar{A}(\chi_1)\,L^r_{1}\,\chi_1^2
\:+\,20\,\bar{A}(\chi_1)\,L^r_{2}\,\chi_1^2 \nonumber \\
&& - \,\:12\,\bar{A}(\chi_1)\,L^r_{3}\,\left[ \chi_1^2 + R^d_1\,\chi_1 \right]
\:+\,16\,\bar{A}(\chi_1)\,L^r_{4}\,\chi_1 \chi_3
\:+\,\bar{A}(\chi_1)\,L^r_{5}\,\left[ 16\,\chi_1^2 + 8\,R^d_1\,\chi_1 \right] \nonumber \\ 
&& + \,\:16\,\bar{A}(\chi_1)\,L^r_{6}\,\left[ \chi_1^2 + R^d_1\,\chi_1 \right]
\:+\,32\,\bar{A}(\chi_1)\,L^r_{7}\,R^d_1\,\chi_1
\:-\,32\,\bar{A}(\chi_1)\,L^r_{8}\,\chi_1^2
\:+\,5/8\,\bar{A}(\chi_1)^2 \chi_1 \nonumber \\
&& + \,\:\bar{A}(\chi_1) \bar{B}(\chi_1,\chi_1,0)\,\left[ 3/2\,\chi_1^2 
+ 1/4\,R^d_1\,\chi_1 \right] 
\:+\,1/2\,\bar{A}(\chi_1) \bar{C}(\chi_1,\chi_1,\chi_1,0)\,R^d_1\,\chi_1^2 
\:-\,\bar{A}(\chi_1,\varepsilon)\,\pi_{16}\,\left[ 7/4\,\chi_1^2 \right. \nonumber \\
&& - \,\:1/4 \left. R^d_1\,\chi_1 \right]
\:+\,4/3\,\bar{A}(\chi_{13})\,\pi_{16}\,\chi_1 \chi_3
\:+\,16\,\bar{A}(\chi_{13})\,L^r_{0}\,\chi_1 \chi_{13}
\:+\,40\,\bar{A}(\chi_{13})\,L^r_{3}\,\chi_1 \chi_{13}
\:-\,32\,\bar{A}(\chi_{13})\,L^r_{5}\,\chi_1 \chi_{13} \nonumber \\
&& + \,\:64\,\bar{A}(\chi_{13})\,L^r_{8}\,\chi_1 \chi_{13}
\:-\,\bar{A}(\chi_{13})^2\,\chi_1
\:-\,2\,\bar{A}(\chi_{13}) \bar{B}(\chi_1,\chi_1,0)\,\chi_1 \chi_3
\:-\,2\,\bar{A}(\chi_{13},\varepsilon)\,\pi_{16}\,\chi_1 \chi_3 \nonumber \\
&& + \,\:48\,\bar{A}(\chi_3)\,L^r_{1}\,\chi_1 \chi_3
\:+\,12\,\bar{A}(\chi_3)\,L^r_{2}\,\chi_1 \chi_3
\:-\,48\,\bar{A}(\chi_3)\,L^r_{4}\,\chi_1 \chi_3
\:+\,48\,\bar{A}(\chi_3)\,L^r_{6}\,\chi_1 \chi_3 \nonumber \\
&& + \,\:3/4\,\bar{A}(\chi_3) \bar{B}(\chi_1,\chi_1,0)\,\chi_1 \chi_3
\:-\,3/4\,\bar{A}(\chi_3,\varepsilon)\,\pi_{16}\,\chi_1 \chi_3
\:-\,12\,\bar{B}(\chi_1,\chi_1,0)\,L^r_{0}\,R^d_1\,\chi_1^2 \nonumber \\
&& - \,\:12\,\bar{B}(\chi_1,\chi_1,0)\,L^r_{3}\,R^d_1\,\chi_1^2
\:-\,\bar{B}(\chi_1,\chi_1,0)\,L^r_{4}\,\left[ 24\,\chi_1 \chi_3^2 
- 32\,\chi_1^2 \chi_3 \right] 
\:+\,\bar{B}(\chi_1,\chi_1,0)\,L^r_{5}\,\left[ 4\,\chi_1^3 
+ 24\,R^d_1\,\chi_1^2 \right] \nonumber \\
&& + \,\:\bar{B}(\chi_1,\chi_1,0)\,L^r_{6}\,\left[ 32\,\chi_1 \chi_3^2
- 48\,\chi_1^2 \chi_3 \right]
\:+\,16\,\bar{B}(\chi_1,\chi_1,0)\,L^r_{7}\,(R^d_1)^2 \chi_1
\:-\,\bar{B}(\chi_1,\chi_1,0)\,L^r_{8}\,\left[ 8\,\chi_1^3 
+ 48\,R^d_1\,\chi_1^2 \right. \nonumber \\
&& - \,\:8 \left. (R^d_1)^2 \chi_1 \right]
\:+\,\bar{B}(\chi_1,\chi_1,0)^2\,\left[ 1/2\,R^d_1\,\chi_1^2 
+ 1/8\,(R^d_1)^2 \chi_1 \right] 
\:+\,1/2\,\bar{B}(\chi_1,\chi_1,0) \bar{C}(\chi_1,\chi_1,\chi_1,0)\,(R^d_1)^2 
\chi_1^2 \nonumber \\
&& + \,\:1/4\,\bar{B}(\chi_1,\chi_1,0,\varepsilon)\,\pi_{16}\,R^d_1\,\chi_1^2
\:+\,16\,\bar{C}(\chi_1,\chi_1,\chi_1,0)\,L^r_{4}\,R^d_1\,\chi_1^2 \chi_3
\:+\,8\,\bar{C}(\chi_1,\chi_1,\chi_1,0)\,L^r_{5}\,R^d_1\,\chi_1^3 \nonumber \\
&& - \,\:32\,\bar{C}(\chi_1,\chi_1,\chi_1,0)\,L^r_{6}\,R^d_1\,\chi_1^2 \chi_3
\:-\,16\,\bar{C}(\chi_1,\chi_1,\chi_1,0)\,L^r_{8}\,R^d_1\,\chi_1^3
\:+\,5/6\,H^{F}(1,\chi_1,\chi_1,\chi_1,\chi_1)\,\chi_1^2 \nonumber \\
&& + \,\:H^{F}(1,\chi_1,\chi_{13},\chi_{13},\chi_1)\,\left[ 1/4\,\chi_1 \chi_3 
- \chi_1^2 \right] 
\:+\,3/4\,H^{F}(1,\chi_{13},\chi_{13},\chi_3,\chi_1)\,\chi_1 \chi_3 \nonumber \\
&& + \,\:H^{F}(2,\chi_1,\chi_1,\chi_1,\chi_1)\,R^d_1\,\chi_1^2
\:+\,3/4\,H^{F}(2,\chi_1,\chi_{13},\chi_{13},\chi_1)\,R^d_1\,\chi_1^2
\:+\,1/2\,H^{F}(5,\chi_1,\chi_1,\chi_1,\chi_1)\,(R^d_1)^2 \chi_1^2 \nonumber \\
&& - \,\:4\,H^{F}_1(3,\chi_{13},\chi_1,\chi_{13},\chi_1)\,R^d_1\,\chi_1^2
\:+\,3/4\,H^{F}_{21}(1,\chi_1,\chi_{13},\chi_{13},\chi_1)\,\chi_1^2
\:+\,9/4\,H^{F}_{21}(1,\chi_3,\chi_{13},\chi_{13},\chi_1)\,\chi_1^2 \nonumber \\
&& - \,\:3/4\,H^{F}_{21}(2,\chi_1,\chi_{13},\chi_{13},\chi_1)\,R^d_1\,\chi_1^2,
\label{M0loop_NNLO_nf2_11} 
\end{eqnarray} 

\begin{eqnarray} 
\delta^{(6)12}_{\mathrm{loops}} & = & \pi_{16}\,L^r_{0}\,\left[ 8\,\chi_\pi \chi_1^2 
+ 6\,\chi_\pi^2 \chi_1 - 4\,\chi_1 \chi_3 \chi_4 - 4\,\chi_1^3 \right]
\:+\,4\,\pi_{16}\,L^r_{1}\,\chi_1^3
\:+\,\pi_{16}\,L^r_{2}\,\left[ 6\,\chi_\pi^2 \chi_1 + 2\,\chi_1^3 \right] \nonumber \\
&& + \,\:\pi_{16}\,L^r_{3}\,\left[ 6\,\chi_\pi \chi_1^2 + 4\,\chi_\pi^2 \chi_1 
- 3\,\chi_1 \chi_3 \chi_4 - 5\,\chi_1^3 \right]
\:+\,\pi_{16}^2\,\left[ 35/144\,\chi_\pi \chi_1^2 + 2/3\,\chi_\pi^2 \chi_1 
- 17/96\,\chi_1 \chi_3 \chi_4 \right. \nonumber \\
&& - \,\:17/288 \left.\chi_1^3 \right]
\:+\,256\,L^r_{4}L^r_{5}\,\chi_1^2 \chi_{34}
\:-\,512\,L^r_{4}L^r_{6}\,\chi_1 \chi_{34}^2
\:-\,256\,L^r_{4}L^r_{8}\,\chi_1^2 \chi_{34}
\:+\,128\,L^r_{4}L^r_{11}\,\chi_1^2 \chi_{34} \nonumber \\
&& + \,\:256\,L^{r2}_{4}\,\chi_1 \chi_{34}^2
\:-\,256\,L^r_{5}L^r_{6}\,\chi_1^2 \chi_{34}
\:-\,128\,L^r_{5}L^r_{8}\,\chi_1^3
\:+\,64\,L^r_{5}L^r_{11}\,\chi_1^3
\:+\,64\,L^{r2}_{5}\,\chi_1^3
\:-\,256\,L^r_{6}L^r_{11}\,\chi_1^2 \chi_{34} \nonumber \\
&& - \,\:128\,L^r_{8}L^r_{11}\,\chi_1^3
\:-\,12\,\bar{A}(\chi_\pi)\,L^r_{0}\,R^\pi_{11}\,\chi_\pi \chi_1
\:+\,16\,\bar{A}(\chi_\pi)\,L^r_{1}\,\chi_\pi \chi_1
\:+\,4\,\bar{A}(\chi_\pi)\,L^r_{2}\,\chi_\pi \chi_1
\:-\,12\,\bar{A}(\chi_\pi)\,L^r_{3}\,R^\pi_{11}\,\chi_\pi \chi_1 \nonumber \\
&& - \,\:16\,\bar{A}(\chi_\pi)\,L^r_{4}\,R^c_1\,\chi_\pi \chi_1
\:+\,8\,\bar{A}(\chi_\pi)\,L^r_{5}\,\left[ R^\pi_{11}\,\chi_\pi \chi_1 
+ R^\pi_{11}\,\chi_1^2 \right]
\:+\,16\,\bar{A}(\chi_\pi)\,L^r_{6}\,R^c_1\,\chi_\pi \chi_1 \nonumber \\
&& + \,\:32\,\bar{A}(\chi_\pi)\,L^r_{7}\,R^z_{\pi 341}\,\chi_1 
\:-\,32\,\bar{A}(\chi_\pi)\,L^r_{8}\,R^\pi_{11}\,\chi_1^2 
\:+\,1/8\,\bar{A}(\chi_\pi)^2\,(R^\pi_{11})^2 \chi_1
\:+\,1/4\,\bar{A}(\chi_\pi) \bar{A}(\chi_1)\,R^\pi_{11} R^c_1\,\chi_1 \nonumber \\
&& - \,\:1/3\,\bar{A}(\chi_\pi) \bar{A}(\chi_{34})\,R^\pi_{11}\,\chi_1
\:-\,2/3\,\bar{A}(\chi_\pi) \bar{A}(\chi_{1s})\,R^\pi_{1s} R^z_{1s\pi}\,\chi_1
\:-\,1/4\,\bar{A}(\chi_\pi) \bar{B}(\chi_\pi,\chi_\pi,0)\,R^\pi_{11}\,\chi_1 \chi_{34}
\nonumber \\ 
&& + \,\:1/2\,\bar{A}(\chi_\pi) \bar{B}(\chi_1,\chi_\pi,0)\,R^\pi_{11} 
R^1_{\pi\pi}\,\chi_1^2 
\:+\,\bar{A}(\chi_\pi) \bar{B}(\chi_1,\chi_1,0)\,\left[ 3/4\,R^c_1\,\chi_1^2 
- 1/2\,(R^c_1)^2 \chi_1^2 - 1/4\,R^c_1 R^d_1\,\chi_1 \right] \nonumber \\
&& + \,\:1/2\,\bar{A}(\chi_\pi) \bar{C}(\chi_1,\chi_1,\chi_1,0)\,R^\pi_{11} 
R^d_1\,\chi_1^2
\:-\,1/4\,\bar{A}(\chi_\pi,\varepsilon)\,\pi_{16}\,R^c_1\,\chi_\pi \chi_1 
\:-\,12\,\bar{A}(\chi_1)\,L^r_{0}\,\left[ R^c_1\,\chi_1^2 + R^d_1\,\chi_1 \right]
\nonumber \\
&& + \,\:8\,\bar{A}(\chi_1)\,L^r_{1}\,\chi_1^2
\:+\,20\,\bar{A}(\chi_1)\,L^r_{2}\,\chi_1^2
\:-\,12\,\bar{A}(\chi_1)\,L^r_{3}\,\left[ R^c_1\,\chi_1^2 + R^d_1\,\chi_1 \right]
\:+\,16\,\bar{A}(\chi_1)\,L^r_{4}\,R^c_1\,\chi_1 \chi_{34} \nonumber \\
&& + \,\:\bar{A}(\chi_1)\,L^r_{5}\,\left[ 16\,R^c_1\,\chi_1^2 + 8\,R^d_1\,\chi_1 \right] 
\:+\,\bar{A}(\chi_1)\,L^r_{6}\,\left[ 32\,\chi_1^2 - 16\,R^c_1\,\chi_1 \chi_{34} \right] 
\:+\,32\,\bar{A}(\chi_1)\,L^r_{7}\,R^d_1\,\chi_1 \nonumber \\
&& - \,\:32\,\bar{A}(\chi_1)\,L^r_{8}\,R^c_1\,\chi_1^2
\:+\,\bar{A}(\chi_1)^2\,\left[ 1/2\,\chi_1 + 1/8\,(R^c_1)^2 \chi_1 \right]
\:+\,1/3\,\bar{A}(\chi_1) \bar{A}(\chi_{34})\,R^\pi_{11}\,\chi_1 \nonumber \\
&& + \,\:2/3\,\bar{A}(\chi_1) \bar{A}(\chi_{1s})\,R^\pi_{1s} R^z_{1s\pi}\,\chi_1 
\:+\,1/2\,\bar{A}(\chi_1) \bar{B}(\chi_1,\chi_\pi,0)\,R^\pi_{11} R^c_1\,\chi_1^2
\:+\,\bar{A}(\chi_1) \bar{B}(\chi_1,\chi_1,0)\,\left[ \chi_1^2 + 1/2\,(R^c_1)^2 \chi_1^2 
\right. \nonumber \\
&& + \,\:1/4 \left. R^c_1 R^d_1\,\chi_1 \right]
\:+\,1/2\,\bar{A}(\chi_1) \bar{C}(\chi_1,\chi_1,\chi_1,0)\,R^c_1 R^d_1\,\chi_1^2 
\:-\,\bar{A}(\chi_1,\varepsilon)\,\pi_{16}\,\left[ 1/2\,\chi_1 \chi_{34} 
+ 3/2\,\chi_1^2 - 1/4\,R^c_1\,\chi_1 \chi_{34} \right] \nonumber \\
&& - \,\:1/2\,\bar{A}(\chi_{13}) \bar{A}(\chi_{14})\,\chi_1
\:+\,32\,\bar{A}(\chi_{34})\,L^r_{1}\,\chi_1 \chi_{34} 
\:+\,8\,\bar{A}(\chi_{34})\,L^r_{2}\,\chi_1 \chi_{34}
\:-\,32\,\bar{A}(\chi_{34})\,L^r_{4}\,\chi_1 \chi_{34} \nonumber \\
&& + \,\:32\,\bar{A}(\chi_{34})\,L^r_{6}\,\chi_1 \chi_{34}
\:+\,1/2\,\bar{A}(\chi_{34}) \bar{B}(\chi_\pi,\chi_\pi,0)\,R^\pi_{11}\,\chi_\pi \chi_1
\:-\,\bar{A}(\chi_{34}) \bar{B}(\chi_1,\chi_\pi,0)\,\left[ 2/3\,R^\pi_{11}\,\chi_\pi 
\chi_1 \right. \nonumber \\
&& + \,\:1/3 \left. R^\pi_{11}\,\chi_1^2 \right]
\:+\,1/2\,\bar{A}(\chi_{34}) \bar{B}(\chi_1,\chi_1,0)\,R^1_{\pi\pi}\,\chi_1 \chi_{34} 
\:-\,1/2\,\bar{A}(\chi_{34},\varepsilon)\,\pi_{16}\,\chi_1 \chi_{34}
\:+\,\bar{A}(\chi_{1s})\,\pi_{16}\,\left[ 1/2\,\chi_1 \chi_{34} \right. \nonumber \\
&& + \,\:1/6 \left. \chi_1 \chi_s \right]
\:+\,8\,\bar{A}(\chi_{1s})\,L^r_{0}\,\chi_1 \chi_{1s}
\:+\,20\,\bar{A}(\chi_{1s})\,L^r_{3}\,\chi_1 \chi_{1s}
\:-\,16\,\bar{A}(\chi_{1s})\,L^r_{5}\,\chi_1 \chi_{1s}
\:+\,32\,\bar{A}(\chi_{1s})\,L^r_{8}\,\chi_1 \chi_{1s} \nonumber \\
&& - \,\:1/4\,\bar{A}(\chi_{1s})^2\,\chi_1
\:-\,\bar{A}(\chi_{1s}) \bar{B}(\chi_1,\chi_\pi,0)\,\left[ 2/3\,R^\pi_{1s}\,\chi_1 
\chi_{1s} + 1/3\,R^\pi_{1s}\,\chi_1^2 \right]
\:-\,\bar{A}(\chi_{1s}) \bar{B}(\chi_1,\chi_1,0)\,R^1_{s\pi}\,\chi_1 \chi_s \nonumber \\
&& - \,\:1/2\,\bar{A}(\chi_{1s},\varepsilon)\,\pi_{16}\,\left[ \chi_1 \chi_{34}  
+ \chi_1 \chi_s \right]
\:+\,8\,\bar{B}(\chi_\pi,\chi_\pi,0)\,L^r_{4}\,R^\pi_{11}\,\chi_\pi^2 \chi_1
\:+\,4\,\bar{B}(\chi_\pi,\chi_\pi,0)\,L^r_{5}\,R^\pi_{11}\,\chi_\pi^2 \chi_1 \nonumber \\ 
&& - \,\:16\,\bar{B}(\chi_\pi,\chi_\pi,0)\,L^r_{6}\,R^\pi_{11}\,\chi_\pi^2 \chi_1
\:-\,16\,\bar{B}(\chi_\pi,\chi_\pi,0)\,L^r_{7}\,R^\pi_{11} (R^z_{\pi 4})^2 \chi_1 
\:-\,4\,\bar{B}(\chi_\pi,\chi_\pi,0)\,L^r_{8}\,\left[ R^\pi_{11}\,\chi_1 \chi_3^2 
\right. \nonumber \\
&& + \,\:R^\pi_{11} \left. \chi_1 \chi_4^2 \right]
\:+\,32\,\bar{B}(\chi_1,\chi_\pi,0)\,L^r_{7}\,R^z_{\pi 341} R^d_1\,\chi_1
\:+\,16\,\bar{B}(\chi_1,\chi_\pi,0)\,L^r_{8}\,R^z_{\pi 341} R^d_1\,\chi_1 \nonumber \\
&& + \,\:1/2\,\bar{B}(\chi_1,\chi_\pi,0) \bar{B}(\chi_1,\chi_1,0)\,R^\pi_{11} R^d_1 
\,\chi_1^2
\:-\,12\,\bar{B}(\chi_1,\chi_1,0)\,L^r_{0}\,R^d_1\,\chi_1^2
\:-\,12\,\bar{B}(\chi_1,\chi_1,0)\,L^r_{3}\,R^d_1\,\chi_1^2 \nonumber \\
&& + \,\:\bar{B}(\chi_1,\chi_1,0)\,L^r_{4}\,\left[ 8\,R^c_1\,\chi_1^2 \chi_{34} 
+ 24\,R^d_1\,\chi_1 \chi_{34} \right]
\:+\,\bar{B}(\chi_1,\chi_1,0)\,L^r_{5}\,\left[ 4\,R^c_1\,\chi_1^3 
+ 24\,R^d_1\,\chi_1^2 \right] \nonumber \\
&& - \,\:16\,\bar{B}(\chi_1,\chi_1,0)\,L^r_{6}\,\left[ R^c_1\,\chi_1^2 
\chi_{34} + 2\,R^d_1\,\chi_1 \chi_{34} \right]
\:+\,16\,\bar{B}(\chi_1,\chi_1,0)\,L^r_{7}\,(R^d_1)^2 \chi_1 \nonumber \\
&& + \,\:\bar{B}(\chi_1,\chi_1,0)\,L^r_{8}\,\left[ - 8\,R^c_1\,\chi_1^3 
- 48\,R^d_1\,\chi_1^2 + 8\,(R^d_1)^2 \chi_1 \right]
\:+\,\bar{B}(\chi_1,\chi_1,0)^2\,\left[ 1/2\,R^c_1 R^d_1\,\chi_1^2 
+ 1/8\,(R^d_1)^2 \chi_1 \right] \nonumber \\
&& + \,\:1/2\,\bar{B}(\chi_1,\chi_1,0) \bar{C}(\chi_1,\chi_1,\chi_1,0)\,(R^d_1)^2 
\chi_1^2
\:+\,1/4\,\bar{B}(\chi_1,\chi_1,0,\varepsilon)\,\pi_{16}\,R^d_1\,\chi_1^2
\:+\,16\,\bar{C}(\chi_1,\chi_1,\chi_1,0)\,L^r_{4}\,R^d_1\,\chi_1^2 \chi_{34} \nonumber \\
&& + \,\:8\,\bar{C}(\chi_1,\chi_1,\chi_1,0)\,L^r_{5}\,R^d_1\,\chi_1^3
\:-\,32\,\bar{C}(\chi_1,\chi_1,\chi_1,0)\,L^r_{6}\,R^d_1\,\chi_1^2 \chi_{34}
\:-\,16\,\bar{C}(\chi_1,\chi_1,\chi_1,0)\,L^r_{8}\,R^d_1\,\chi_1^3 \nonumber \\
&& + \,\:1/2\,H^{F}(1,\chi_\pi,\chi_\pi,\chi_1,\chi_1)\,(R^\pi_{11})^2 \chi_1^2 
\:+\,H^{F}(1,\chi_\pi,\chi_1,\chi_1,\chi_1)\,R^\pi_{11} R^c_1\,\chi_1^2 \nonumber \\
&& - \,\:H^{F}(1,\chi_\pi,\chi_{1s},\chi_{1s},\chi_1)\,\left[ 1/2\,R^\pi_{11}\,\chi_1^2 
+ 1/8\,R^v_{\pi 1s}\,\chi_\pi \chi_1 \right]
\:+\,H^{F}(1,\chi_1,\chi_1,\chi_1,\chi_1)\,\left[ 1/3\,\chi_1^2 
+ 1/2\,(R^c_1)^2 \chi_1^2 \right] \nonumber \\
&& - \,\:H^{F}(1,\chi_1,\chi_{1s},\chi_{1s},\chi_1)\,\left[ 3/4\,R^1_{s\pi}\,\chi_1^2 
- 3/8\,R^c_1\,\chi_1^2 + 1/8\,R^d_1\,\chi_1 \right]
\:+\,1/2\,H^{F}(1,\chi_{13},\chi_{14},\chi_{34},\chi_1)\,\chi_1 \chi_{34} \nonumber \\
&& + \,\:H^{F}(2,\chi_1,\chi_\pi,\chi_1,\chi_1)\,R^\pi_{11} R^d_1\,\chi_1^2
\:+\,H^{F}(2,\chi_1,\chi_1,\chi_1,\chi_1)\,R^c_1 R^d_1\,\chi_1^2
\:+\,3/8\,H^{F}(2,\chi_1,\chi_{1s},\chi_{1s},\chi_1)\,R^d_1\,\chi_1^2 \nonumber \\
&& + \,\:1/2\,H^{F}(5,\chi_1,\chi_1,\chi_1,\chi_1)\,(R^d_1)^2 \chi_1^2
\:-\,H^{F}_1(1,\chi_\pi,\chi_{1s},\chi_{1s},\chi_1)\,R^\pi_{1s} R^z_{1s\pi}\,\chi_1^2
\nonumber \\
&& - \,\:2\,H^{F}_1(1,\chi_{1s},\chi_{1s},\chi_1,\chi_1)\,R^\pi_{1s} R^z_{1s\pi}\,\chi_1^2 
\:-\,2\,H^{F}_1(3,\chi_{1s},\chi_1,\chi_{1s},\chi_1)\,R^d_1\,\chi_1^2 \nonumber \\
&& - \,\:3/8\,H^{F}_{21}(1,\chi_\pi,\chi_{1s},\chi_{1s},\chi_1)\,R^v_{\pi 1s}\,\chi_1^2 
\:+\,H^{F}_{21}(1,\chi_1,\chi_{1s},\chi_{1s},\chi_1)\,\left[ 3/4\,R^1_{s\pi}\,\chi_1^2 
- 3/8\,R^c_1\,\chi_1^2 \right] \nonumber \\
&& + \,\:3/2\,H^{F}_{21}(1,\chi_{34},\chi_{13},\chi_{14},\chi_1)\,\chi_1^2
\:-\,3/8\,H^{F}_{21}(2,\chi_1,\chi_{1s},\chi_{1s},\chi_1)\,R^d_1\,\chi_1^2.
\label{M0loop_NNLO_nf2_12} 
\end{eqnarray} 

\end{widetext}

\subsection{Results for $d_{\mathrm{val}} = 2$}

The most general expression for the pseudoscalar meson mass at 
$n_{\mathrm{sea}} = 2$ is the result which has $d_{\mathrm{val}} = 2$ and 
$d_{\mathrm{sea}} = 2$. In the notation of Eq.~(\ref{masseq}), the lowest 
order mass $M_0^2$ is equal to $\chi_{12}$. The contribution at NLO for 
$d_{\mathrm{val}} = 2$ and $d_{\mathrm{sea}} = 2$ is
\begin{eqnarray} 
\delta^{(4)22} & = & - \,\:16\,L^r_{4}\,\chi_{12} \chi_{34}
\:-\,8\,L^r_{5}\,\chi_{12}^2
\:+\,32\,L^r_{6}\,\chi_{12} \chi_{34} \nonumber \\
&& + \,\:16\,L^r_{8}\,\chi_{12}^2
\:-\,1/2\,\bar{A}(\chi_p)\,R^p_{q\pi}\,\chi_{12} \nonumber \\
&& - \,\:1/2\,\bar{A}(\chi_\pi)\,R^\pi_{12}\,\chi_{12},
\label{M0_NLO_nf2_22} 
\end{eqnarray} 
from which the result for $d_{\mathrm{val}} = 2$ and $d_{\mathrm{sea}} = 
1$ is easily obtained by taking the limit $\chi_4 \rightarrow \chi_3$ 
such that $R^p_{q\pi} \rightarrow R^p_q$ and dropping the 
$\bar{A}(\chi_\pi)$ term. The NNLO contribution from 
the ${\cal O}(p^6)$ Lagrangian is given by
\begin{eqnarray} 
\delta^{(6)22}_{\mathrm{ct}} & = & - \,\:32\,K^r_{17}\,\chi_{12}^3
\:-\,64\,K^r_{18}\,\chi_{12}^2 \chi_{34}
\:-\,8\,K^r_{19}\,\chi_p^2 \chi_{12} \nonumber \\
&& - \,\:32\,K^r_{20}\,\chi_{12}^2 \chi_{34}
\:-\,16\,K^r_{21}\,\chi_{12} \chi_s^2 \nonumber \\
&& - \,\:64\,K^r_{22}\,\chi_{12} \chi_{34}^2
\:-\,16\,K^r_{23}\,\chi_1 \chi_{12} \chi_2 \nonumber \\
&& + \,\:24\,K^r_{25}\,\chi_p^2 \chi_{12}
\:+\,K^r_{26}\,\left[ 16\,\chi_{12} \chi_s^2 \right. \nonumber \\
&& + \,\:64 \left. \chi_{12}^2 \chi_{34} \right]
\:+\,192\,K^r_{27}\,\chi_{12} \chi_{34}^2
\:+\,32\,K^r_{39}\,\chi_{12}^3 \nonumber \\
&& + \,\:64\,K^r_{40}\,\chi_{12}^2 \chi_{34}.
\label{M0tree_NNLO_nf2_22} 
\end{eqnarray} 
The result for $d_{\mathrm{val}} = 2$ and $d_{\mathrm{sea}} = 1$ may 
readily be inferred from Eq.~(\ref{M0tree_NNLO_nf2_22}) by taking the 
limit $\chi_4 \rightarrow \chi_3$. The loop contributions at NNLO, 
separately for $d_{\mathrm{sea}} = 1$ and $d_{\mathrm{sea}} = 2$, are
\begin{widetext}

\begin{eqnarray} 
\delta^{(6)21}_{\mathrm{loops}} & = & \pi_{16}\,L^r_{0}\,\left[ 4\,\chi_1 \chi_{12} \chi_2 
+ 2\,\chi_{12} \chi_3^2 + 8\,\chi_{12}^2 \chi_3 - 8\,\chi_{12}^3 \right]
\:+\,4\,\pi_{16}\,L^r_{1}\,\chi_{12}^3
\:+\,\pi_{16}\,L^r_{2}\,\left[ 6\,\chi_{12} \chi_3^2 + 2\,\chi_{12}^3 \right] \nonumber \\ 
&& + \,\:\pi_{16}\,L^r_{3}\,\left[ 3\,\chi_1 \chi_{12} \chi_2 + \chi_{12} \chi_3^2 
+ 6\,\chi_{12}^2 \chi_3 - 8\,\chi_{12}^3 \right]
\:+\,\pi_{16}^2\,\left[ 17/96\,\chi_1 \chi_{12} \chi_2 + 47/96\,\chi_{12} \chi_3^2 
+ 35/144\,\chi_{12}^2 \chi_3 \right. \nonumber \\ 
&& - \,\:17/72 \left. \chi_{12}^3 \right]
\:+\,256\,L^r_{4}L^r_{5}\,\chi_{12}^2 \chi_3
\:-\,512\,L^r_{4}L^r_{6}\,\chi_{12} \chi_3^2
\:-\,256\,L^r_{4}L^r_{8}\,\chi_{12}^2 \chi_3
\:+\,128\,L^r_{4}L^r_{11}\,\chi_{12}^2 \chi_3 \nonumber \\
&& + \,\:256\,L^{r2}_{4}\,\chi_{12} \chi_3^2
\:-\,256\,L^r_{5}L^r_{6}\,\chi_{12}^2 \chi_3
\:-\,128\,L^r_{5}L^r_{8}\,\chi_{12}^3
\:+\,64\,L^r_{5}L^r_{11}\,\chi_{12}^3
\:+\,64\,L^{r2}_{5}\,\chi_{12}^3
\:-\,256\,L^r_{6}L^r_{11}\,\chi_{12}^2 \chi_3 \nonumber \\
&& - \,\:128\,L^r_{8}L^r_{11}\,\chi_{12}^3
\:+\,\bar{A}(\chi_p)\,\pi_{16}\,\left[ - 1/24\,\chi_q \chi_{12} - 1/8\,\chi_{12} 
\chi_3 + 1/4\,R^p_q\,\chi_{12} \chi_3 + 1/12\,R^p_q\,\chi_{12}^2 \right] \nonumber \\
&& - \,\:\bar{A}(\chi_p)\,L^r_{0}\,\left[ 2\,\chi_p \chi_{12} + 8\,R^p_q\,\chi_p 
\chi_{12} + 2\,R^d_p\,\chi_{12} \right]
\:-\,\bar{A}(\chi_p)\,L^r_{3}\,\left[ 5\,\chi_p \chi_{12} + 2\,R^p_q\,\chi_p \chi_{12} 
+ 5\,R^d_p\,\chi_{12} \right] \nonumber \\
&& + \,\:16\,\bar{A}(\chi_p)\,L^r_{4}\,R^p_q\,\chi_{12} \chi_3
\:+\,\bar{A}(\chi_p)\,L^r_{5}\,\left[ 4\,\chi_p \chi_{12} 
\:+\,8\,R^p_q\,\chi_p \chi_{12} \right]
\:-\,16\,\bar{A}(\chi_p)\,L^r_{6}\,R^p_q\,\chi_{12} \chi_3 \nonumber \\
&& + \,\:\bar{A}(\chi_p)\,L^r_{7}\,\left[ 16/3\,R^d_p\,\chi_{p3} 
+ 32/3\,R^d_p\,\chi_{12} \right]
\:-\,\bar{A}(\chi_p)\,L^r_{8}\,\left[ 8\,\chi_p \chi_{12}
+ 16/3\,R^p_q\,\chi_1 \chi_2 + 32/3\,R^p_q\,\chi_{12}^2 + 4/3\,(R^d_p)^2 \right] 
\nonumber \\
&& + \,\:\bar{A}(\chi_p)^2\,\left[ 5/32\,\chi_{12} + 1/8\,(R^p_q)^2 \chi_{12} 
- 1/24\,R^q_p\,\chi_p \right]
\:+\,\bar{A}(\chi_p) \bar{A}(\chi_{p3})\,\left[ 1/24\,\chi_p - 1/8\,\chi_q 
+ 1/24\,R^p_q\,\chi_p \right. \nonumber \\ 
&& + \,\:19/72 \left. R^p_q\,\chi_q \right]
\:+\,\bar{A}(\chi_p) \bar{A}(\chi_{q3})\,\left[ 1/4\,\chi_{12} 
+ 1/6\,R^p_q\,\chi_q - 1/24\,R^q_p\,\chi_p + 1/72\,R^q_p\,\chi_q \right]
\nonumber \\
&& + \,\:\bar{A}(\chi_p) \bar{A}(\chi_{12})\,\left[ 1/18\,\chi_p - 1/36\,\chi_{12} 
\right]
\:-\,1/8\,\bar{A}(\chi_p) \bar{A}(\chi_3)\,\chi_3
\:+\,\bar{A}(\chi_p) \bar{B}(\chi_p,\chi_p,0)\,\left[ 11/24\,R^p_q\,\chi_p \chi_{12} 
\right. \nonumber \\
&& - \,\:1/12 \left. R^p_q\,\chi_p^2 - 1/8\,R^p_q\,\chi_{12}^2 + 1/24\,R^p_q 
R^d_p\,\chi_p \right]
\:-\,\bar{A}(\chi_p) \bar{B}(\chi_q,\chi_q,0)\,\left[ 1/24\,R^p_q R^d_q\,\chi_q
+ 1/16\, R^d_q\,\chi_{12} \right] \nonumber \\
&& + \,\:1/6\,\bar{A}(\chi_p) \bar{B}(\chi_1,\chi_2,0)\,R^q_p\,\chi_p \chi_{12} 
\:+\,1/12\,\bar{A}(\chi_p) \bar{B}(\chi_1,\chi_2,0,k)\,R^q_p\,\chi_p
\:+\,1/4\,\bar{A}(\chi_p,\varepsilon)\,\pi_{16}\,\left[ \chi_p \chi_{12} \right.
\nonumber \\
&& - \,\:R^p_q \left. \chi_{12} \chi_3 \right]
\:-\,\bar{A}(\chi_{p3})\,\pi_{16}\,\left[ 1/3\,\chi_p \chi_{12}
- 2/3\,\chi_{12} \chi_3 - 1/3\,\chi_{12}^2 \right]
\:+\,8\,\bar{A}(\chi_{p3})\,L^r_{0}\,\chi_{p3} \chi_{12} \nonumber \\
&& + \,\:20\,\bar{A}(\chi_{p3})\,L^r_{3}\,\chi_{p3} \chi_{12}
\:-\,16\,\bar{A}(\chi_{p3})\,L^r_{5}\,\chi_{p3} \chi_{12}
\:+\,32\,\bar{A}(\chi_{p3})\,L^r_{8}\,\chi_{p3} \chi_{12} \nonumber \\
&& + \,\:1/4\,\bar{A}(\chi_{p3}) \bar{B}(\chi_q,\chi_q,0)\,R^d_q\,\chi_{12}
\:-\,\bar{A}(\chi_{p3}) \bar{B}(\chi_1,\chi_2,0)\,\left[ 4/9\,\chi_{p3} \chi_{12} 
+ 1/9\,\chi_1 \chi_2 + 1/9\,R^q_p\,\chi_p^2 \right] \nonumber \\
&& + \,\:\bar{A}(\chi_{p3}) \bar{B}(\chi_1,\chi_2,0,k)\,\left[ - 2/9\,\chi_{p3} 
- 2/9\,\chi_{12} + 1/9\,R^q_p\,\chi_q \right]
\:-\,\bar{A}(\chi_{p3},\varepsilon)\,\pi_{16}\,\chi_{12} \chi_3 \nonumber \\
&& + \,\:\bar{A}(\chi_1) \bar{A}(\chi_2)\,\left[ - 1/16\,\chi_{12} - 1/24\,\chi_3 
+ 1/4\,R^1_2 R^2_1\,\chi_{12} \right]
\:+\,8\,\bar{A}(\chi_{12})\,L^r_{1}\,\chi_{12}^2
\:+\,20\,\bar{A}(\chi_{12})\,L^r_{2}\,\chi_{12}^2 \nonumber \\
&& + \,\:32\,\bar{A}(\chi_{12})\,L^r_{6}\,\chi_{12}^2
\:-\,1/4\,\bar{A}(\chi_{12})^2\,\chi_{12}
\:+\,\bar{A}(\chi_{12}) \bar{B}(\chi_1,\chi_2,0)\,\left[ 1/9\,\chi_1 \chi_2 
+ 4/9\,\chi_{12}^2 \right] \nonumber \\
&& + \,\:4/9\,\bar{A}(\chi_{12}) \bar{B}(\chi_1,\chi_2,0,k)\,\chi_{12}
\:-\,2\,\bar{A}(\chi_{12},\varepsilon)\,\pi_{16}\,\chi_{12}^2
\:-\,\bar{A}(\chi_{13}) \bar{A}(\chi_{23})\,\chi_{12}
\:+\,48\,\bar{A}(\chi_3)\,L^r_{1}\,\chi_{12} \chi_3 \nonumber \\
&& + \,\:12\,\bar{A}(\chi_3)\,L^r_{2}\,\chi_{12} \chi_3
\:-\,48\,\bar{A}(\chi_3)\,L^r_{4}\,\chi_{12} \chi_3
\:+\,48\,\bar{A}(\chi_3)\,L^r_{6}\,\chi_{12} \chi_3
\:+\,1/2\,\bar{A}(\chi_3) \bar{B}(\chi_1,\chi_2,0)\,\chi_{12} \chi_3 \nonumber \\ 
&& + \,\:1/4\,\bar{A}(\chi_3) \bar{B}(\chi_1,\chi_2,0,k)\,\chi_3
\:-\,3/4\,\bar{A}(\chi_3,\varepsilon)\,\pi_{16}\,\chi_{12} \chi_3
\:-\,\bar{B}(\chi_p,\chi_p,0)\,\pi_{16}\,\left[ 1/24\,R^d_p\,\chi_q \chi_{12} 
+ 1/8\,R^d_p\,\chi_{12} \chi_3 \right] \nonumber \\
&& - \,\:2\,\bar{B}(\chi_p,\chi_p,0)\,L^r_{0}\,R^d_p\,\chi_p \chi_{12}
\:-\,5\,\bar{B}(\chi_p,\chi_p,0)\,L^r_{3}\,R^d_p\,\chi_p \chi_{12}
\:+\,8\,\bar{B}(\chi_p,\chi_p,0)\,L^r_{4}\,R^p_q\,\chi_p \chi_{12} \chi_3 \nonumber \\ 
&& + \,\:\bar{B}(\chi_p,\chi_p,0)\,L^r_{5}\,\left[ 4\,R^p_q\,\chi_p^3 
+ 2\,R^d_p\,\chi_1 \chi_2 \right] 
\:-\,16\,\bar{B}(\chi_p,\chi_p,0)\,L^r_{6}\,R^p_q\,\chi_p \chi_{12} \chi_3 
\:-\,\bar{B}(\chi_p,\chi_p,0)\,L^r_{8}\,\left[ 8\,R^p_q\,\chi_p^3 \right. \nonumber \\
&& + \,\:4 \left. R^d_p\,\chi_1 \chi_2 \right]
\:+\,\bar{B}(\chi_p,\chi_p,0)^2\,\left[ 9/32\,R^p_q R^d_p\,\chi_p \chi_{12} 
- 1/32\,R^p_q R^d_p\,\chi_q \chi_{12} \right] \nonumber \\
&& + \,\:1/6\,\bar{B}(\chi_p,\chi_p,0) \bar{B}(\chi_1,\chi_2,0)\,R^q_p R^d_p\,\chi_p 
\chi_{12}
\:+\,1/12\,\bar{B}(\chi_p,\chi_p,0) \bar{B}(\chi_1,\chi_2,0,k)\,R^q_p R^d_p\,\chi_p 
\nonumber \\
&& + \,\:1/4\,\bar{B}(\chi_p,\chi_p,0,\varepsilon)\,\pi_{16}\,R^d_p\,\chi_{p3}\chi_{12} 
\:-\,1/16\,\bar{B}(\chi_1,\chi_1,0) \bar{B}(\chi_2,\chi_2,0)\,R^d_1 R^d_2\,\chi_{12} 
\nonumber \\ 
&& + \,\:32/3\,\bar{B}(\chi_1,\chi_2,0)\,L^r_{7}\,R^d_1 R^d_2\,\chi_{12}
\:+\,16/3\,\bar{B}(\chi_1,\chi_2,0)\,L^r_{8}\,R^d_1 R^d_2\,\chi_{12}
\:+\,16/3\,\bar{B}(\chi_1,\chi_2,0,k)\,L^r_{7}\,R^d_1 R^d_2 \nonumber \\
&& + \,\:8/3\,\bar{B}(\chi_1,\chi_2,0,k)\,L^r_{8}\,R^d_1 R^d_2
\:+\,H^{F}(1,\chi_p,\chi_p,\chi_{12},\chi_{12})\,\left[ - 1/4\,\chi_p \chi_{12} 
- 15/32\,\chi_{12}^2 + 3/8\,R^p_q\,\chi_{12}^2 \right. \nonumber \\
&& - \,\:1/8 \left. (R^p_q)^2 \chi_{12}^2 \right]
\:+\,H^{F}(1,\chi_p,\chi_{13},\chi_{23},\chi_{12})\,\left[ 1/4\,\chi_{12}^2 
+ 3/4\,R^q_p\,\chi_p \chi_{12} + 1/4\,R^q_p\,\chi_q \chi_{12} 
- 1/4\,R^q_p\,\chi_{12} \chi_3 \right] \nonumber \\
&& - \,\:H^{F}(1,\chi_1,\chi_{12},\chi_2,\chi_{12})\,\left[ 1/16\,\chi_{12}^2 
- 3/4\,R^1_2 R^2_1\,\chi_{12}^2 \right]
\:+\,1/4\,H^{F}(1,\chi_{12},\chi_{12},\chi_{12},\chi_{12})\,\chi_{12}^2 \nonumber \\
&& + \,\:3/4\,H^{F}(1,\chi_{13},\chi_{23},\chi_3,\chi_{12})\,\chi_{12} \chi_3 
\:-\,H^{F}(2,\chi_p,\chi_p,\chi_{12},\chi_{12})\,\left[ 1/8\,R^p_q R^d_p\,\chi_{12}^2 
- 5/16\,R^d_p\,\chi_{12}^2 \right] \nonumber \\
&& - \,\:H^{F}(2,\chi_p,\chi_{12},\chi_q,\chi_{12})\,\left[ 1/8\,R^q_p 
R^d_p\,\chi_{12}^2 - 1/16\,R^d_p\,\chi_{12}^2 \right]
\:-\,1/4\,H^{F}(2,\chi_p,\chi_{13},\chi_{23},\chi_{12})\,R^d_p\,\chi_{p3} \chi_{12} 
\nonumber \\
&& + \,\:5/32\,H^{F}(5,\chi_p,\chi_p,\chi_{12},\chi_{12})\,(R^d_p)^2 \chi_{12}^2  
\:+\,1/16\,H^{F}(5,\chi_1,\chi_2,\chi_{12},\chi_{12})\,R^d_1 R^d_2\,\chi_{12}^2 \nonumber \\
&& + \,\:H^{F}_1(1,\chi_p,\chi_p,\chi_{12},\chi_{12})\,\left[ 5/2\,\chi_{12}^2 
- R^p_q R^q_p\,\chi_{12}^2 \right]
\:-\,2\,H^{F}_1(1,\chi_p,\chi_{13},\chi_{23},\chi_{12})\,R^q_p\,\chi_{12}^2 \nonumber \\
&& - \,\:2\,H^{F}_1(1,\chi_{p3},\chi_{q3},\chi_p,\chi_{12})\,\chi_{12}^2
\:+\,H^{F}_1(1,\chi_{12},\chi_1,\chi_2,\chi_{12})\,\left[ 1/2\,\chi_{12}^2 - R^1_2 R^2_1 
\,\chi_{12}^2 \right] \nonumber \\
&& - \,\:1/2\,H^{F}_1(3,\chi_{12},\chi_p,\chi_p,\chi_{12})\,R^q_p R^d_p\,\chi_{12}^2
\:+\,1/2\,H^{F}_1(3,\chi_{12},\chi_p,\chi_q,\chi_{12})\,R^q_p R^d_p\,\chi_{12}^2 
\nonumber \\
&& - \,\:1/4\,H^{F}_1(7,\chi_{12},\chi_p,\chi_p,\chi_{12})\,(R^d_p)^2 \chi_{12}^2 
\:-\,3/4\,H^{F}_{21}(1,\chi_p,\chi_p,\chi_{12},\chi_{12})\,\chi_{12}^2 \nonumber \\
&& + \,\:3/4\,H^{F}_{21}(1,\chi_p,\chi_{13},\chi_{23},\chi_{12})\,R^q_p\,\chi_{12}^2
\:+\,3/4\,H^{F}_{21}(1,\chi_{p3},\chi_{q3},\chi_p,\chi_{12})\,R^p_q\,\chi_{12}^2 \nonumber \\
&& - \,\:3/4\,H^{F}_{21}(1,\chi_{p3},\chi_{q3},\chi_q,\chi_{12})\,R^p_q\,\chi_{12}^2 
\:+\,H^{F}_{21}(1,\chi_{12},\chi_p,\chi_p,\chi_{12})\,\left[ 15/32\,\chi_{12}^2 
- 3/8\,R^p_q R^q_p\,\chi_{12}^2 \right] \nonumber \\
&& - \,\:H^{F}_{21}(1,\chi_{12},\chi_1,\chi_2,\chi_{12})\,\left[ 3/16\,\chi_{12}^2 
- 3/4\,R^1_2 R^2_1\,\chi_{12}^2 \right]
\:+\,3/4\,H^{F}_{21}(1,\chi_{12},\chi_{12},\chi_{12},\chi_{12})\,\chi_{12}^2 \nonumber \\ 
&& + \,\:9/4\,H^{F}_{21}(1,\chi_3,\chi_{13},\chi_{23},\chi_{12})\,\chi_{12}^2
\:-\,3/4\,H^{F}_{21}(3,\chi_{p3},\chi_p,\chi_{q3},\chi_{12})\,R^d_p\,\chi_{12}^2 
\nonumber \\
&& - \,\:H^{F}_{21}(3,\chi_{12},\chi_p,\chi_p,\chi_{12})\,\left[ 3/8\,R^p_q 
R^d_p\,\chi_{12}^2 - 3/16\,R^d_p\,\chi_{12}^2 \right] 
\:-\,H^{F}_{21}(3,\chi_{12},\chi_p,\chi_q,\chi_{12})\,\left[ 3/8\,R^q_p R^d_p\,\chi_{12}^2 
\right. \nonumber \\
&& - \,\:3/16 \left.R^d_p\,\chi_{12}^2 \right]
\:+\,3/32\,H^{F}_{21}(7,\chi_{12},\chi_p,\chi_p,\chi_{12})\,(R^d_p)^2 \chi_{12}^2
\:+\,3/16\,H^{F}_{21}(7,\chi_{12},\chi_1,\chi_2,\chi_{12})\,R^d_1 R^d_2\,\chi_{12}^2,
\label{M0loop_NNLO_nf2_21} 
\end{eqnarray} 

\begin{eqnarray} 
\delta^{(6)22}_{\mathrm{loops}} & = & \pi_{16}\,L^r_{0}\,\left[ 8\,\chi_\pi \chi_{12}^2 
+ 6\,\chi_\pi^2 \chi_{12} + 4\,\chi_1 \chi_{12} \chi_2 - 4\,\chi_{12} \chi_3 \chi_4 
- 8\,\chi_{12}^3 \right]
\:+\,4\,\pi_{16}\,L^r_{1}\,\chi_{12}^3
\:+\,\pi_{16}\,L^r_{2}\,\left[ 6\,\chi_\pi^2 \chi_{12} 
+ 2\,\chi_{12}^3 \right] \nonumber \\
&& + \,\:\pi_{16}\,L^r_{3}\,\left[ 6\,\chi_\pi \chi_{12}^2 + 4\,\chi_\pi^2 \chi_{12} 
+ 3\,\chi_1 \chi_{12} \chi_2 - 3\,\chi_{12} \chi_3 \chi_4 - 8\,\chi_{12}^3 \right]
\:+\,\pi_{16}^2\,\left[ 35/144\,\chi_\pi \chi_{12}^2 
+ 2/3\,\chi_\pi^2 \chi_{12} \right. \nonumber \\
&& + \,\:17/96 \left. \chi_1 \chi_{12} \chi_2 -\,17/96\,\chi_{12} \chi_3 \chi_4 
- 17/72\,\chi_{12}^3 \right]
\:+\,256\,L^r_{4}L^r_{5}\,\chi_{12}^2 \chi_{34}
\:-\,512\,L^r_{4}L^r_{6}\,\chi_{12} \chi_{34}^2 \nonumber \\
&& - \,\:256\,L^r_{4}L^r_{8}\,\chi_{12}^2 \chi_{34}
\:+\,128\,L^r_{4}L^r_{11}\,\chi_{12}^2 \chi_{34}
\:+\,256\,L^{r2}_{4}\,\chi_{12} \chi_{34}^2
\:-\,256\,L^r_{5}L^r_{6}\,\chi_{12}^2 \chi_{34}
\:-\,128\,L^r_{5}L^r_{8}\,\chi_{12}^3 \nonumber \\
&& + \,\:64\,L^r_{5}L^r_{11}\,\chi_{12}^3
\:+\,64\,L^{r2}_{5}\,\chi_{12}^3
\:-\,256\,L^r_{6}L^r_{11}\,\chi_{12}^2 \chi_{34}
\:-\,128\,L^r_{8}L^r_{11}\,\chi_{12}^3
\:+\,\bar{A}(\chi_p)\,\pi_{16}\,\left[ 1/4\,R^p_{q\pi}\,\chi_{12} \chi_{34} \right.
\nonumber \\
&& + \,\:1/12 \left. R^p_{q\pi}\,\chi_{12}^2 - 1/24\,R^c_p\,\chi_q \chi_{12}
- 1/8\,R^c_p\,\chi_{12} \chi_{34} \right]
\:-\,\bar{A}(\chi_p)\,L^r_{0}\,\left[ 8\,R^p_{q\pi}\,\chi_p \chi_{12} 
+ 2\,R^c_p\,\chi_p \chi_{12} + 2\,R^d_p\,\chi_{12} \right] \nonumber \\
&& - \,\:\bar{A}(\chi_p)\,L^r_{3}\,\left[ 2\,R^p_{q\pi}\,\chi_p \chi_{12} 
+ 5\,R^c_p\,\chi_p \chi_{12} + 5\,R^d_p\,\chi_{12} \right]
\:+\,16\,\bar{A}(\chi_p)\,L^r_{4}\,R^p_{q\pi}\,\chi_{12} \chi_{34}
\:+\,\bar{A}(\chi_p)\,L^r_{5}\,\left[ 8\,R^p_{q\pi}\,\chi_p \chi_{12} \right. 
\nonumber \\ 
&& + \,\:4 \left. R^c_p\,\chi_p \chi_{12} \right]
\:-\,16\,\bar{A}(\chi_p)\,L^r_{6}\,R^p_{q\pi}\,\chi_{12} \chi_{34} 
\:+\,\bar{A}(\chi_p)\,L^r_{7}\,\left[ 32/3\,R^d_p\,\chi_p + 16/3\,R^d_p\,\chi_q 
- 8/3\,(R^d_p)^2 \right] \nonumber \\
&& - \,\:\bar{A}(\chi_p)\,L^r_{8}\,\left[ 16/3\,R^p_{q\pi}\,\chi_1 \chi_2 
+ 32/3\,R^p_{q\pi}\,\chi_{12}^2 + 8\,R^c_p\,\chi_p \chi_{12} + 4/3\,(R^d_p)^2 \right]
\nonumber \\
&& + \,\:\bar{A}(\chi_p)^2\,\left[ 1/8\,\chi_{12} + 1/8\,(R^p_{q\pi})^2 \chi_{12} 
+ 1/24\,R^p_{q\pi} R^c_p\,\chi_p - 1/24\,(R^c_p)^2 \chi_p 
+ 1/32\,(R^c_p)^2 \chi_{12} \right] \nonumber \\
&& + \,\:\bar{A}(\chi_p) \bar{A}(\chi_{ps})\,\left[ 1/48\,R^p_{q\pi}\,\chi_p 
+ 19/144\,R^p_{q\pi}\,\chi_q - 5/144\,R^p_{s\pi}\,\chi_p - 1/16\,R^p_{s\pi}\,\chi_q 
+ 1/18\,R^c_p\,\chi_p \right] \nonumber \\
&& + \,\:\bar{A}(\chi_p) \bar{A}(\chi_{qs})\,\left[ 1/12\,R^p_{q\pi}\,\chi_q - 
1/48\,R^p_{s\pi} R^z_{qsp}\,\chi_p + 1/144\,R^p_{s\pi} R^z_{qsp}\,\chi_q 
+ 1/8\,R^c_p\,\chi_{12} \right] \nonumber \\
&& + \,\:\bar{A}(\chi_p) \bar{A}(\chi_\pi)\,\left[ 1/24\,R^p_{q\pi} R^\pi_{pp}\,\chi_p 
- 1/40\,R^p_{q\pi} R^\pi_{qq}\,\chi_q + 13/120\,R^p_{q\pi} R^\pi_{12}\,\chi_p 
+ 1/8\,R^p_{q\pi} R^\pi_{12}\,\chi_q - 1/60\,R^p_{\pi\pi} R^\pi_{qq}\,\chi_\pi 
\right. \nonumber \\ 
&& - \,\:1/24 \left. R^\pi_{qq} R^c_p\,\chi_p - 1/24\,R^c_p\,\chi_\pi + 1/96\,R^c_p 
R^v_{\pi 12}\,\chi_p + 1/32\,R^c_p R^v_{\pi 12}\,\chi_q 
- 1/16\,R^c_p R^v_{\pi 12}\,\chi_\pi 
- 1/12\,R^z_{\pi 34p} \right] \nonumber \\
&& + \,\:\bar{A}(\chi_p) \bar{A}(\chi_{12})\,\left[ 1/18\,\chi_p - 1/36\,\chi_{12} 
\right] 
\:-\,\bar{A}(\chi_p) \bar{A}(\chi_{34})\,\left[ 1/12\,R^p_{\pi\pi}\,\chi_{34} 
- 1/9\,R^\pi_{pp}\,\chi_p \right] \nonumber \\
&& + \,\:\bar{A}(\chi_p) \bar{B}(\chi_p,\chi_p,0)\,\left[ 11/24\,R^p_{q\pi} 
R^c_p\,\chi_p \chi_{12} - 1/12\,R^p_{q\pi} R^c_p\,\chi_p^2 - 1/8\,R^p_{q\pi} 
R^c_p\,\chi_{12}^2 + 1/24\,R^p_{q\pi} R^d_p\,\chi_p \right] \nonumber \\
&& + \,\:\bar{A}(\chi_p) \bar{B}(\chi_p,\chi_\pi,0)\,\left[ 1/6\,R^\pi_{12} 
R^c_p\,\chi_p^2 + 1/12\,R^\pi_{12} R^c_p\,\chi_1 \chi_2 \right]
\:-\,\bar{A}(\chi_p) \bar{B}(\chi_q,\chi_q,0)\,\left[ 1/24\,R^p_{q\pi} R^d_q\,\chi_q 
\right. \nonumber \\
&& + \,\:1/16 \left. R^c_p R^d_q\,\chi_{12} \right]
\:+\,1/6\,\bar{A}(\chi_p) \bar{B}(\chi_1,\chi_2,0)\,R^q_{p\pi} R^c_p\,\chi_p \chi_{12} 
\:+\,1/12\,\bar{A}(\chi_p) \bar{B}(\chi_1,\chi_2,0,k)\,R^q_{p\pi} R^c_p\,\chi_p 
\nonumber \\
&& + \,\:1/4\,\bar{A}(\chi_p,\varepsilon)\,\pi_{16}\,\left[ \chi_p \chi_{12} 
- R^p_{q\pi}\,\chi_{12} \chi_{34} - R^\pi_{pp}\,\chi_{12} \chi_{34} \right]
\:+\,\bar{A}(\chi_{ps})\,\pi_{16}\,\left[ 1/6\,\chi_q \chi_{12} + 1/4\,\chi_{12} \chi_{34} 
+ 1/12\,\chi_{12} \chi_s \right. \nonumber \\ 
&& - \,\:1/6 \left. \chi_{12}^2 \right]
\:+\,4\,\bar{A}(\chi_{ps})\,L^r_{0}\,\chi_{ps} \chi_{12}
\:+\,10\,\bar{A}(\chi_{ps})\,L^r_{3}\,\chi_{ps} \chi_{12}
\:-\,8\,\bar{A}(\chi_{ps})\,L^r_{5}\,\chi_{ps} \chi_{12}
\:+\,16\,\bar{A}(\chi_{ps})\,L^r_{8}\,\chi_{ps} \chi_{12} \nonumber \\
&& + \,\:\bar{A}(\chi_{ps}) \bar{A}(\chi_\pi)\,\left[ 1/18\,R^\pi_{pp}\,\chi_p 
- 5/144\,R^\pi_{ps}\,\chi_p - 1/16\,R^\pi_{ps}\,\chi_q + 1/8\,R^\pi_{qq}\,\chi_{12} 
- 5/48\,R^\pi_{qs}\,\chi_p \right. \nonumber \\
&& - \,\:19/144 \left. R^\pi_{qs} R^z_{ps\pi}\,\chi_q + 1/48\,R^\pi_{12}\,\chi_p \right]
\:-\,\bar{A}(\chi_{ps}) \bar{B}(\chi_p,\chi_\pi,0)\,\left[ 2/9\,R^\pi_{qs}\,\chi_p 
\chi_{ps} + 1/9\,R^\pi_{qs}\,\chi_p \chi_{12} \right. \nonumber \\
&& + \,\:1/9 \left. R^\pi_{qs}\,\chi_q \chi_{ps} 
\:+\,1/18\,R^\pi_{qs} R^z_{ps\pi}\,\chi_1 \chi_2 + 1/18\,R^\pi_{12}\,\chi_p^2 \right]
\:+\,1/8\,\bar{A}(\chi_{ps}) \bar{B}(\chi_q,\chi_q,0)\,R^d_q\,\chi_{12} \nonumber \\
&& - \,\:\bar{A}(\chi_{ps}) \bar{B}(\chi_1,\chi_2,0)\,\left[ 1/18\,R^q_{p\pi}\,\chi_p^2 
+ 2/9\,R^q_{s\pi}\,\chi_{ps} \chi_{12} + 1/18\,R^q_{s\pi}\,\chi_1 \chi_2 \right]
\nonumber \\
&& + \,\:\bar{A}(\chi_{ps}) \bar{B}(\chi_1,\chi_2,0,k)\,\left[ 1/18\,R^q_{p\pi}\,\chi_q 
- 1/9\,R^q_{s\pi}\,\chi_{ps} - 1/9\,R^q_{s\pi}\,\chi_{12} \right]
\:-\,1/4\,\bar{A}(\chi_{ps},\varepsilon)\,\pi_{16}\,\left[ \chi_{12} \chi_{34} 
+ \chi_{12} \chi_s \right] \nonumber \\
&& - \,\:1/4\,\bar{A}(\chi_{p3}) \bar{A}(\chi_{q4})\,\chi_{12}
\:-\,1/12\,\bar{A}(\chi_\pi)\,\pi_{16}\,\left[ R^v_{\pi 12}\,\chi_\pi \chi_{12}
\:+\,R^v_{\pi 12}\,\chi_{12}^2 \right] \nonumber \\
&& - \,\:\bar{A}(\chi_\pi)\,L^r_{0}\,\left[ 12\,R^\pi_{12}\,\chi_\pi \chi_{12} 
+ 2\,R^v_{\pi 12}\,\chi_\pi \chi_{12} \right]
\:+\,16\,\bar{A}(\chi_\pi)\,L^r_{1}\,\chi_\pi \chi_{12}
\:+\,4\,\bar{A}(\chi_\pi)\,L^r_{2}\,\chi_\pi \chi_{12} \nonumber \\
&& - \,\:\bar{A}(\chi_\pi)\,L^r_{3}\,\left[ 12\,R^\pi_{12}\,\chi_\pi \chi_{12} 
+ 5\,R^v_{\pi 12}\,\chi_\pi \chi_{12} \right]
\:-\,16\,\bar{A}(\chi_\pi)\,L^r_{4}\,\left[ \chi_\pi \chi_{12} 
- R^\pi_{12}\,\chi_\pi \chi_{12} \right] \nonumber \\
&& + \,\:\bar{A}(\chi_\pi)\,L^r_{5}\,\left[ 4\,R^\pi_{pp}\,\chi_\pi \chi_{12} 
+ 8\,R^\pi_{12}\,\chi_{12}^2 \right]
\:+\,16\,\bar{A}(\chi_\pi)\,L^r_{6}\,\left[ \chi_\pi \chi_{12} 
- R^\pi_{12}\,\chi_\pi \chi_{12} \right] \nonumber \\
&& + \,\:\bar{A}(\chi_\pi)\,L^r_{7}\,\left[ 64/3\,R^\pi_{12}\,\chi_\pi \chi_{12} 
- 64/3\,R^\pi_{12}\,\chi_{12}^2 + 16/3\,R^z_{\pi 34p}\,\chi_q 
- 8/3\,(R^z_{\pi 4})^2 R^v_{\pi 12} \right] \nonumber \\ 
&& - \,\:\bar{A}(\chi_\pi)\,L^r_{8}\,\left[ 32/3\,R^\pi_{12}\,\chi_1 \chi_2 
+ 64/3\,R^\pi_{12}\,\chi_{12}^2 + 16/3\,R^v_{\pi 12}\,\chi_\pi^2
+ 4\,R^v_{\pi 12}\,\chi_1 \chi_2 - 4/3\,R^v_{\pi 12}\,\chi_3 \chi_4 \right]
\nonumber \\
&& + \,\:\bar{A}(\chi_\pi)^2\,\left[ 1/8\,(R^\pi_{12})^2 \chi_{12} - 1/24\,R^\pi_{12} 
R^v_{\pi 12}\,\chi_\pi + 1/24\,R^\pi_{12} R^v_{\pi 12}\,\chi_{12}
- 1/24\,R^v_{\pi 12}\,\chi_\pi \right. \nonumber \\
&& - \,\:1/24 \left. (R^v_{\pi 12})^2 \chi_\pi 
+ 1/32\,(R^v_{\pi 12})^2 \chi_{12} \right] 
\:-\,\bar{A}(\chi_\pi) \bar{A}(\chi_{34})\,\left[ 1/3\,R^\pi_{12}\,\chi_{12} 
+ 1/36\,R^v_{\pi 12}\,\chi_\pi \right] \nonumber \\
&& + \,\:\bar{A}(\chi_\pi) \bar{B}(\chi_p,\chi_p,0)\,\left[ 1/4\,R^p_{q\pi} 
R^\pi_{pp}\,\chi_p^2 - 1/12\,R^\pi_{pp} R^d_p\,\chi_p - 1/24\,R^\pi_{qq} R^d_p\,\chi_p 
+ 1/96\,R^d_p R^v_{\pi 12}\,\chi_p \right. \nonumber \\
&& + \,\:1/32 \left. R^d_p R^v_{\pi 12}\,\chi_q 
- 1/16\,R^d_p R^v_{\pi 12}\,\chi_\pi \right]
\:+\,\bar{A}(\chi_\pi) \bar{B}(\chi_p,\chi_\pi,0)\,\left[ 1/6\,R^p_{\pi\pi} R^\pi_{12} 
\,\chi_p \chi_{12} + 1/12\,R^p_{\pi\pi} R^\pi_{12}\,\chi_p^2 \right] \nonumber \\
&& - \,\:1/4\,\bar{A}(\chi_\pi) \bar{B}(\chi_\pi,\chi_\pi,0)\,R^\pi_{12}\,\chi_{12} 
\chi_{34} 
\:+\,1/6\,\bar{A}(\chi_\pi) \bar{B}(\chi_1,\chi_2,0)\,R^1_{\pi\pi} R^2_{\pi\pi} 
\,\chi_{12} \chi_{34} \nonumber \\
&& + \,\:1/12\,\bar{A}(\chi_\pi) \bar{B}(\chi_1,\chi_2,0,k)\,R^1_{\pi\pi} 
R^2_{\pi\pi}\,\chi_{34}
\:+\,\bar{A}(\chi_\pi,\varepsilon)\,\pi_{16}\,\left[ - 1/4\,\chi_\pi \chi_{12} 
+ 1/4\,R^\pi_{12}\,\chi_\pi \chi_{12} + 1/4\,R^v_{\pi 12}\,\chi_\pi \chi_{12} \right] 
\nonumber \\
&& + \,\:\bar{A}(\chi_1) \bar{A}(\chi_2)\,\left[ - 1/24\,R^p_{q\pi} R^c_q\,\chi_q 
+ 1/4\,R^1_{2\pi} R^2_{1\pi}\,\chi_{12} - 1/16\,R^c_1 R^c_2\,\chi_{12} \right]
\:+\,8\,\bar{A}(\chi_{12})\,L^r_{1}\,\chi_{12}^2 \nonumber \\
&& + \,\:20\,\bar{A}(\chi_{12})\,L^r_{2}\,\chi_{12}^2
\:+\,32\,\bar{A}(\chi_{12})\,L^r_{6}\,\chi_{12}^2
\:-\,1/4\,\bar{A}(\chi_{12})^2\,\chi_{12}
\:+\,\bar{A}(\chi_{12}) \bar{B}(\chi_1,\chi_2,0)\,\left[ 1/9\,\chi_1 \chi_2 
+ 4/9\,\chi_{12}^2 \right] \nonumber \\
&& + \,\:4/9\,\bar{A}(\chi_{12}) \bar{B}(\chi_1,\chi_2,0,k)\,\chi_{12}
\:-\,2\,\bar{A}(\chi_{12},\varepsilon)\,\pi_{16}\,\chi_{12}^2
\:+\,32\,\bar{A}(\chi_{34})\,L^r_{1}\,\chi_{12} \chi_{34}
\:+\,8\,\bar{A}(\chi_{34})\,L^r_{2}\,\chi_{12} \chi_{34} \nonumber \\
&& - \,\:32\,\bar{A}(\chi_{34})\,L^r_{4}\,\chi_{12} \chi_{34}
\:+\,32\,\bar{A}(\chi_{34})\,L^r_{6}\,\chi_{12} \chi_{34}
\:-\,\bar{A}(\chi_{34}) \bar{B}(\chi_p,\chi_\pi,0)\,\left[ 2/9\,R^\pi_{12}\,\chi_p \chi_\pi 
+ 1/9\,R^\pi_{12}\,\chi_q \chi_\pi \right. \nonumber \\ 
&& + \,\:1/6 \left. R^\pi_{12}\,\chi_1 \chi_2 \right] 
\:+\,1/2\,\bar{A}(\chi_{34}) \bar{B}(\chi_\pi,\chi_\pi,0)\,R^\pi_{12}\,\chi_\pi \chi_{12} 
\:+\,\bar{A}(\chi_{34}) \bar{B}(\chi_1,\chi_2,0)\,\left[ 1/3\,\chi_{12} \chi_{34} 
+ 1/9\,R^\pi_{12}\,\chi_1 \chi_2 \right. \nonumber \\
&& + \,\:1/3 \left. R^\pi_{12}\,\chi_{12} \chi_{34} - 2/9\,R^\pi_{12}\,\chi_{12}^2 
\right] 
\:+\,\bar{A}(\chi_{34}) \bar{B}(\chi_1,\chi_2,0,k)\,\left[ 1/6\,\chi_{34} 
+ 1/9\,R^\pi_{12}\,\chi_{12} + 1/6\,R^\pi_{12}\,\chi_{34} \right] \nonumber \\
&& - \,\:1/2\,\bar{A}(\chi_{34},\varepsilon)\,\pi_{16}\,\chi_{12} \chi_{34} 
\:-\,1/4\,\bar{A}(\chi_{1s}) \bar{A}(\chi_{2s})\,\chi_{12}
\:-\,\bar{B}(\chi_p,\chi_p,0)\,\pi_{16}\,\left[ 1/24\,R^d_p\,\chi_q \chi_{12} 
+ 1/8\,R^d_p\,\chi_{12} \chi_{34} \right] \nonumber \\
&& - \,\:2\,\bar{B}(\chi_p,\chi_p,0)\,L^r_{0}\,R^d_p\,\chi_p \chi_{12}
\:-\,5\,\bar{B}(\chi_p,\chi_p,0)\,L^r_{3}\,R^d_p\,\chi_p \chi_{12}
\:+\,8\,\bar{B}(\chi_p,\chi_p,0)\,L^r_{4}\,R^p_{q\pi}\,\chi_p \chi_{12} \chi_{34} 
\nonumber \\
&& + \,\:\bar{B}(\chi_p,\chi_p,0)\,L^r_{5}\,\left[ 4\,R^p_{q\pi}\,\chi_p^3 
+ 2\,R^d_p\,\chi_1 \chi_2 \right]
\:-\,16\,\bar{B}(\chi_p,\chi_p,0)\,L^r_{6}\,R^p_{q\pi}\,\chi_p \chi_{12} \chi_{34}
\nonumber \\
&& - \,\:\bar{B}(\chi_p,\chi_p,0)\,L^r_{8}\,\left[ 8\,R^p_{q\pi}\,\chi_p^3
+ 4\,R^d_p\,\chi_1 \chi_2 \right]
\:+\,\bar{B}(\chi_p,\chi_p,0)^2\,\left[ 9/32\,R^p_{q\pi} R^d_p\,\chi_p \chi_{12} 
- 1/32\,R^p_{q\pi} R^d_p\,\chi_q \chi_{12} \right] \nonumber \\
&& + \,\:\bar{B}(\chi_p,\chi_p,0) \bar{B}(\chi_p,\chi_\pi,0)\,\left[ 1/6\,R^\pi_{12} 
R^d_p\,\chi_p^2 + 1/12\,R^\pi_{12} R^d_p\,\chi_1 \chi_2 \right]
\:+\,1/6\,\bar{B}(\chi_p,\chi_p,0) \bar{B}(\chi_1,\chi_2,0)\,R^q_{p\pi} R^d_p 
\,\chi_p \chi_{12} \nonumber \\
&& + \,\:1/12\,\bar{B}(\chi_p,\chi_p,0) \bar{B}(\chi_1,\chi_2,0,k)\,R^q_{p\pi} 
R^d_p\,\chi_p 
\:+\,1/8\,\bar{B}(\chi_p,\chi_p,0,\varepsilon)\,\pi_{16}\,\left[ R^d_p\,\chi_p \chi_{12} 
+ R^d_p\,\chi_{12} \chi_{34} \right] \nonumber \\
&& + \,\:\bar{B}(\chi_p,\chi_\pi,0)\,L^r_{7}\,\left[ 16/3\,R^d_p R^z_{\pi 34q}\,\chi_p 
+ 32/3\,R^d_p R^z_{\pi 34q}\,\chi_{12} \right]
\:+\,\bar{B}(\chi_p,\chi_\pi,0)\,L^r_{8}\,\left[ 8/3\,R^d_p R^z_{\pi 34q}\,\chi_p \right.
\nonumber \\
&& + \,\:16/3 \left. R^d_p R^z_{\pi 34q}\,\chi_{12} \right]
\:+\,8\,\bar{B}(\chi_\pi,\chi_\pi,0)\,L^r_{4}\,R^\pi_{12}\,\chi_\pi^2 \chi_{12} 
\:+\,4\,\bar{B}(\chi_\pi,\chi_\pi,0)\,L^r_{5}\,R^\pi_{12}\,\chi_\pi^2 \chi_{12} 
\nonumber \\
&& - \,\:16\,\bar{B}(\chi_\pi,\chi_\pi,0)\,L^r_{6}\,R^\pi_{12}\,\chi_\pi^2 \chi_{12}
\:-\,16\,\bar{B}(\chi_\pi,\chi_\pi,0)\,L^r_{7}\,R^\pi_{12}\,(R^z_{\pi 4})^2 \chi_{12} 
\:-\,4\,\bar{B}(\chi_\pi,\chi_\pi,0)\,L^r_{8}\,\left[ R^\pi_{12}\,\chi_{12} \chi_3^2 
\right. \nonumber \\
&& + \,\:R^\pi_{12} \left. \chi_{12} \chi_4^2 \right]
\:-\,1/16\,\bar{B}(\chi_1,\chi_1,0) \bar{B}(\chi_2,\chi_2,0)\,R^d_1 R^d_2\,\chi_{12}
\:+\,32/3\,\bar{B}(\chi_1,\chi_2,0)\,L^r_{7}\,R^d_1 R^d_2\,\chi_{12} \nonumber \\ 
&& + \,\:16/3\,\bar{B}(\chi_1,\chi_2,0)\,L^r_{8}\,R^d_1 R^d_2\,\chi_{12}
\:+\,16/3\,\bar{B}(\chi_1,\chi_2,0,k)\,L^r_{7}\,R^d_1 R^d_2
\:+\,8/3\,\bar{B}(\chi_1,\chi_2,0,k)\,L^r_{8}\,R^d_1 R^d_2 \nonumber \\
&& + \,\:H^{F}(1,\chi_p,\chi_p,\chi_{12},\chi_{12})\,\left[ - 1/4\,\chi_p \chi_{12} 
- 3/8\,\chi_{12}^2 - 1/8\,(R^p_{q\pi})^2 \chi_{12}^2 + 3/8\,R^p_{q\pi} R^c_p\,\chi_{12}^2 
- 3/32\,(R^c_p)^2 \chi_{12}^2 \right] \nonumber \\
&& + \,\:H^{F}(1,\chi_p,\chi_{1s},\chi_{2s},\chi_{12})\,\left[ - 
5/8\,R^p_{q\pi}\,\chi_q \chi_{12} + 1/4\,R^p_{q\pi}\,\chi_{ps} \chi_{12} 
+ 3/16\,R^p_{s\pi}\,\chi_q \chi_{12} 
+ 7/16\,R^p_{s\pi}\,\chi_{12} \chi_s \right. \nonumber \\
&& - \,\:1/8 \left. R^c_p\,\chi_{ps} \chi_{12} \right]
\:+\,1/4\,H^{F}(1,\chi_{p3},\chi_{q4},\chi_{34},\chi_{12})\,\chi_{12} \chi_{34}
\:+\,H^{F}(1,\chi_\pi,\chi_p,\chi_{12},\chi_{12})\,\left[ 1/2\,R^p_{q\pi} R^\pi_{12} 
\,\chi_{12}^2 \right. \nonumber \\
&& - \,\:1/8 \left. R^p_{q\pi} R^v_{\pi 12}\,\chi_{12}^2 + 1/4\,R^\pi_{pp} 
R^c_p\,\chi_{12}^2 + 1/16\,R^c_p R^v_{\pi 12}\,\chi_{12}^2 \right]
\:+\,H^{F}(1,\chi_\pi,\chi_\pi,\chi_{12},\chi_{12})\,\left[ 1/2\,(R^\pi_{12})^2 \chi_{12}^2 
\right. \nonumber \\ 
&& + \,\:1/2 \left. R^\pi_{12} R^v_{\pi 12}\,\chi_{12}^2 
+ 5/32\,(R^v_{\pi 12})^2\,\chi_{12}^2 \right]
\:+\,H^{F}(1,\chi_\pi,\chi_{1s},\chi_{2s},\chi_{12})\,\left[ - 3/16\,R^\pi_{ps}\,\chi_p 
\chi_{12} - 5/16\,R^\pi_{ps}\,\chi_q \chi_{12} \right. \nonumber \\
&& + \,\:1/2 \left. R^\pi_{12}\,\chi_{12}^2 - 1/16\,R^v_{\pi ps}\,\chi_\pi \chi_{12} 
- 1/16\,R^v_{\pi 12}\,\chi_{12} \chi_s \right]
\:+\,H^{F}(1,\chi_1,\chi_{12},\chi_2,\chi_{12})\,\left[ - 1/8\,R^p_{q\pi} 
R^c_q\,\chi_{12}^2 \right. \nonumber \\ 
&& + \,\:3/4 \left. R^1_{2\pi} R^2_{1\pi}\,\chi_{12}^2 
+ 1/16\,R^c_1 R^c_2\,\chi_{12}^2 \right]
\:+\,1/4\,H^{F}(1,\chi_{12},\chi_{12},\chi_{12},\chi_{12})\,\chi_{12}^2 \nonumber \\
&& - \,\:H^{F}(2,\chi_p,\chi_p,\chi_{12},\chi_{12})\,\left[ 1/8\,R^p_{q\pi} 
R^d_p\,\chi_{12}^2 - 5/16\,R^c_p R^d_p\,\chi_{12}^2 \right]
\:+\,H^{F}(2,\chi_p,\chi_\pi,\chi_{12},\chi_{12})\,\left[ 1/4\,R^\pi_{pp} 
R^d_p\,\chi_{12}^2 \right. \nonumber \\
&& + \,\:1/16 \left. R^d_p R^v_{\pi 12}\,\chi_{12}^2 \right]
\:-\,H^{F}(2,\chi_p,\chi_{12},\chi_q,\chi_{12})\,\left[ 1/8\,R^q_{p\pi} R^d_p\,\chi_{12}^2 
- 1/16\,R^c_q R^d_p\,\chi_{12}^2 \right] \nonumber \\
&& - \,\:1/8\,H^{F}(2,\chi_p,\chi_{1s},\chi_{2s},\chi_{12})\,R^d_p\,\chi_{ps} \chi_{12} 
\:+\,5/32\,H^{F}(5,\chi_p,\chi_p,\chi_{12},\chi_{12})\,(R^d_p)^2 \chi_{12}^2 \nonumber \\
&& + \,\:1/16\,H^{F}(5,\chi_1,\chi_2,\chi_{12},\chi_{12})\,R^d_1 R^d_2\,\chi_{12}^2
\:+\,H^{F}_1(1,\chi_p,\chi_p,\chi_{12},\chi_{12})\,\left[ 2\,\chi_{12}^2 
+ (R^p_{q\pi})^2 \chi_{12}^2 - R^p_{q\pi} R^c_p\,\chi_{12}^2 \right. \nonumber \\
&& + \,\:1/2 \left. (R^c_p)^2 \chi_{12}^2 \right] 
\:+\,H^{F}_1(1,\chi_p,\chi_{1s},\chi_{2s},\chi_{12})\,R^p_{q\pi} R^z_{sqp}\,\chi_{12}^2 
\:-\,H^{F}_1(1,\chi_{ps},\chi_{qs},\chi_p,\chi_{12})\,R^p_{s\pi}\,\chi_{12}^2 \nonumber \\ 
&& - \,\:H^{F}_1(1,\chi_{ps},\chi_{qs},\chi_\pi,\chi_{12})\,R^\pi_{12} 
R^z_{sp\pi}\,\chi_{12}^2
\:+\,1/2\,H^{F}_1(1,\chi_{12},\chi_p,\chi_\pi,\chi_{12})\,\left[ R^p_{q\pi} R^v_{\pi 12} 
\,\chi_{12}^2 - R^\pi_{pp} R^z_{qp\pi} R^c_p\,\chi_{12}^2 \right] \nonumber \\
&& - \,\:H^{F}_1(1,\chi_{12},\chi_\pi,\chi_\pi,\chi_{12})\,\left[ 1/2\,R^\pi_{12} 
R^v_{\pi 12}\,\chi_{12}^2 + 1/4\,(R^v_{\pi 12})^2 \chi_{12}^2 \right]
\:+\,H^{F}_1(1,\chi_{12},\chi_1,\chi_2,\chi_{12})\,\left[ 1/2\,R^p_{q\pi} 
R^c_q\,\chi_{12}^2 \right. \nonumber \\
&& - \,\:R^1_{2\pi} R^2_{1\pi} \left. \chi_{12}^2 \right]
\:+\,1/2\,H^{F}_1(3,\chi_{12},\chi_p,\chi_p,\chi_{12})\,\left[ R^p_{q\pi} R^d_p 
\,\chi_{12}^2 - R^c_p R^d_p\,\chi_{12}^2 \right] \nonumber \\
&& + \,\:1/2\,H^{F}_1(3,\chi_{12},\chi_p,\chi_q,\chi_{12})\,R^q_{p\pi} 
R^d_p\,\chi_{12}^2 
\:+\,1/2\,H^{F}_1(3,\chi_{12},\chi_p,\chi_\pi,\chi_{12})\,R^\pi_{12} R^z_{pq\pi} 
R^d_p\,\chi_{12}^2 \nonumber \\
&& - \,\:1/4\,H^{F}_1(7,\chi_{12},\chi_p,\chi_p,\chi_{12})\,(R^d_p)^2 \chi_{12}^2 
\:-\,3/4\,H^{F}_{21}(1,\chi_p,\chi_p,\chi_{12},\chi_{12})\,\chi_{12}^2 \nonumber \\
&& - \,\:3/8\,H^{F}_{21}(1,\chi_p,\chi_{1s},\chi_{2s},\chi_{12})\,R^p_{q\pi} 
R^z_{sqp}\,\chi_{12}^2
\:+\,3/8\,H^{F}_{21}(1,\chi_{ps},\chi_{qs},\chi_p,\chi_{12})\,\left[ R^p_{q\pi} 
\,\chi_{12}^2 + R^p_{s\pi}\,\chi_{12}^2 \right. \nonumber \\ 
&& - \,\:R^c_p \left.\chi_{12}^2 \right] 
\:+\,3/8\,H^{F}_{21}(1,\chi_{ps},\chi_{qs},\chi_q,\chi_{12})\,R^q_{p\pi} R^z_{spq} 
\,\chi_{12}^2
\:+\,3/8\,H^{F}_{21}(1,\chi_{ps},\chi_{qs},\chi_\pi,\chi_{12})\,R^\pi_{12} 
R^z_{pq\pi} R^z_{sp\pi}\,\chi_{12}^2 \nonumber \\
&& - \,\:3/8\,H^{F}_{21}(1,\chi_\pi,\chi_{1s},\chi_{2s},\chi_{12})\,R^\pi_{12} 
R^z_{s1\pi} R^z_{s2\pi}\,\chi_{12}^2
\:+\,H^{F}_{21}(1,\chi_{12},\chi_p,\chi_p,\chi_{12})\,\left[ 3/8\,\chi_{12}^2 
+ 3/8\,(R^p_{q\pi})^2 \chi_{12}^2 \right. \nonumber \\
&& - \,\:3/8 \left. R^p_{q\pi} R^c_p\,\chi_{12}^2 + 3/32\,(R^c_p)^2 \chi_{12}^2 \right] 
\:-\,H^{F}_{21}(1,\chi_{12},\chi_p,\chi_\pi,\chi_{12})\,\left[ 3/8\,R^p_{q\pi} R^v_{\pi 12} 
\,\chi_{12}^2 - 3/16\,R^c_p R^v_{\pi 12}\,\chi_{12}^2 \right] \nonumber \\
&& + \,\:3/32\,H^{F}_{21}(1,\chi_{12},\chi_\pi,\chi_\pi,\chi_{12})\,(R^v_{\pi 
12})^2\,\chi_{12}^2
\:+\,H^{F}_{21}(1,\chi_{12},\chi_1,\chi_2,\chi_{12})\,\left[ - 3/8\,R^p_{q\pi} R^c_q 
\,\chi_{12}^2 + 3/4\,R^1_{2\pi} R^2_{1\pi}\,\chi_{12}^2 \right. \nonumber \\
&& + \,\:3/16 \left. R^c_1 R^c_2\,\chi_{12}^2 \right]
\:+\,3/4\,H^{F}_{21}(1,\chi_{12},\chi_{12},\chi_{12},\chi_{12})\,\chi_{12}^2
\:+\,3/4\,H^{F}_{21}(1,\chi_{34},\chi_{p3},\chi_{q4},\chi_{12})\,\chi_{12}^2 \nonumber \\
&& - \,\:3/8\,H^{F}_{21}(3,\chi_{ps},\chi_p,\chi_{qs},\chi_{12})\,R^d_p\,\chi_{12}^2
\:-\,3/8\,H^{F}_{21}(3,\chi_{12},\chi_p,\chi_p,\chi_{12})\,\left[ R^p_{q\pi} R^d_p 
\,\chi_{12}^2 - 1/2\,R^c_p R^d_p\,\chi_{12}^2 \right] \nonumber \\
&& - \,\:H^{F}_{21}(3,\chi_{12},\chi_p,\chi_q,\chi_{12})\,\left[ 3/8\,R^q_{p\pi} R^d_p 
\,\chi_{12}^2 - 3/16\,R^c_q R^d_p\,\chi_{12}^2 \right]
\:+\,3/16\,H^{F}_{21}(3,\chi_{12},\chi_p,\chi_\pi,\chi_{12})\,R^d_p R^v_{\pi 12} 
\,\chi_{12}^2 \nonumber \\
&& + \,\:3/32\,H^{F}_{21}(7,\chi_{12},\chi_p,\chi_p,\chi_{12})\,(R^d_p)^2 \chi_{12}^2 
\:+\,3/16\,H^{F}_{21}(7,\chi_{12},\chi_1,\chi_2,\chi_{12})\,R^d_1 R^d_2\,\chi_{12}^2.
\label{M0loop_NNLO_nf2_22} 
\end{eqnarray} 

\end{widetext}

\section{Numerical Analysis and Conclusions}

The NNLO results given in the preceding section depend on a number of 
${\cal O}(p^4)$ and ${\cal O}(p^6)$ LEC:s. Eventually, their proper values, 
from which the physical LEC:s $l^r_i$ and $c^r_i$ can be obtained, 
should be determined by a fit of the NNLO expressions to lattice QCD data. 
For the present purpose, in order to have some reasonable estimates of 
the LEC:s, we simply use the values obtained from experiment in 
Ref.~\cite{ABT2}, which used three-flavor continuum $\chi$PT to NNLO.

The combination of LEC:s used in this paper corresponds to 'fit~10' of 
Ref.~\cite{ABT2}, which had $F_0 = 87.7$~MeV and $\mu = 770$~MeV. The 
parameters which were not determined by that fit, namely $L^r_0,L^r_4$ 
and $L^r_6$, have been set to zero at the abovementioned scale $\mu$, as 
have all the $K^r_i$. Also, $L^r_{11}$ has been treated 
according to the condition $L^r_{11} = - l^r_4 / 4 = -(2 L_4^r+L_5^r)$. 
It should be noted that $L^r_0$ cannot be determined from experiment, 
but is obtainable from partially quenched simulations. Some recent 
results on $L^r_4$ and $L^r_6$ may be found in Ref.~\cite{piK}. If we 
temporarily reintroduce the superscripts $(2pq)$, the choice of LEC:s 
for the numerical analysis is

\begin{eqnarray}
F &=&  F_0\:\Big|_{\mbox{\small fit 10}}\,,
\nonumber\\[1mm]
L_{i=1\ldots 10}^{r(2pq)} &=& 
L_{i=1\ldots 10}^r\:\Big|_{\mbox{\small fit 10}}\,,
\nonumber\\
L^{r(2pq)}_{11}& =& -\left(2 L_4^r+L_5^r\right)
\Big|_{\mbox{\small fit 10}}\,,
\nonumber\\
L^{r(2pq)}_{0}& =&L^{r(2pq)}_{12} =K^{r(2pq)}_{i} =0\,.
\end{eqnarray}

\newpage

In order to set the scale for the lowest order masses, for a quark mass 
of 100~MeV we have roughly $\chi \sim 0.01$~GeV$^2$ and for 500~MeV we 
have $\chi \sim 0.25$~GeV$^2$. A crude estimate of the range of validity 
of PQ$\chi$PT is for all $\chi_i,\chi_{ij} \lesssim 0.3$~GeV$^2$.

\begin{figure}[ht!]
\begin{center}
\includegraphics[width=\columnwidth]{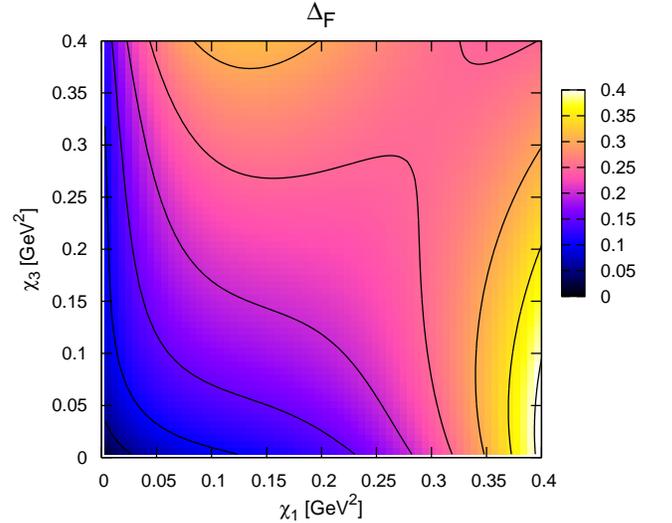}
\caption{The NNLO modification $\Delta_F$ of the decay constant for 
$d_{\mathrm{val}} = 1$ and $d_{\mathrm{sea}} = 1$ as a function 
of the valence quark mass parameter $\chi_1$ and the sea-quark mass 
parameter $\chi_3$. The function plotted represents the sum of the 
${\cal O}(p^4)$ and ${\cal O}(p^6)$ contributions.}
\label{F02dim}
\end{center}
\end{figure}

\newpage

\begin{widetext}
\begin{figure*}[ht!]
\begin{center}
\includegraphics[width=.87\textwidth]{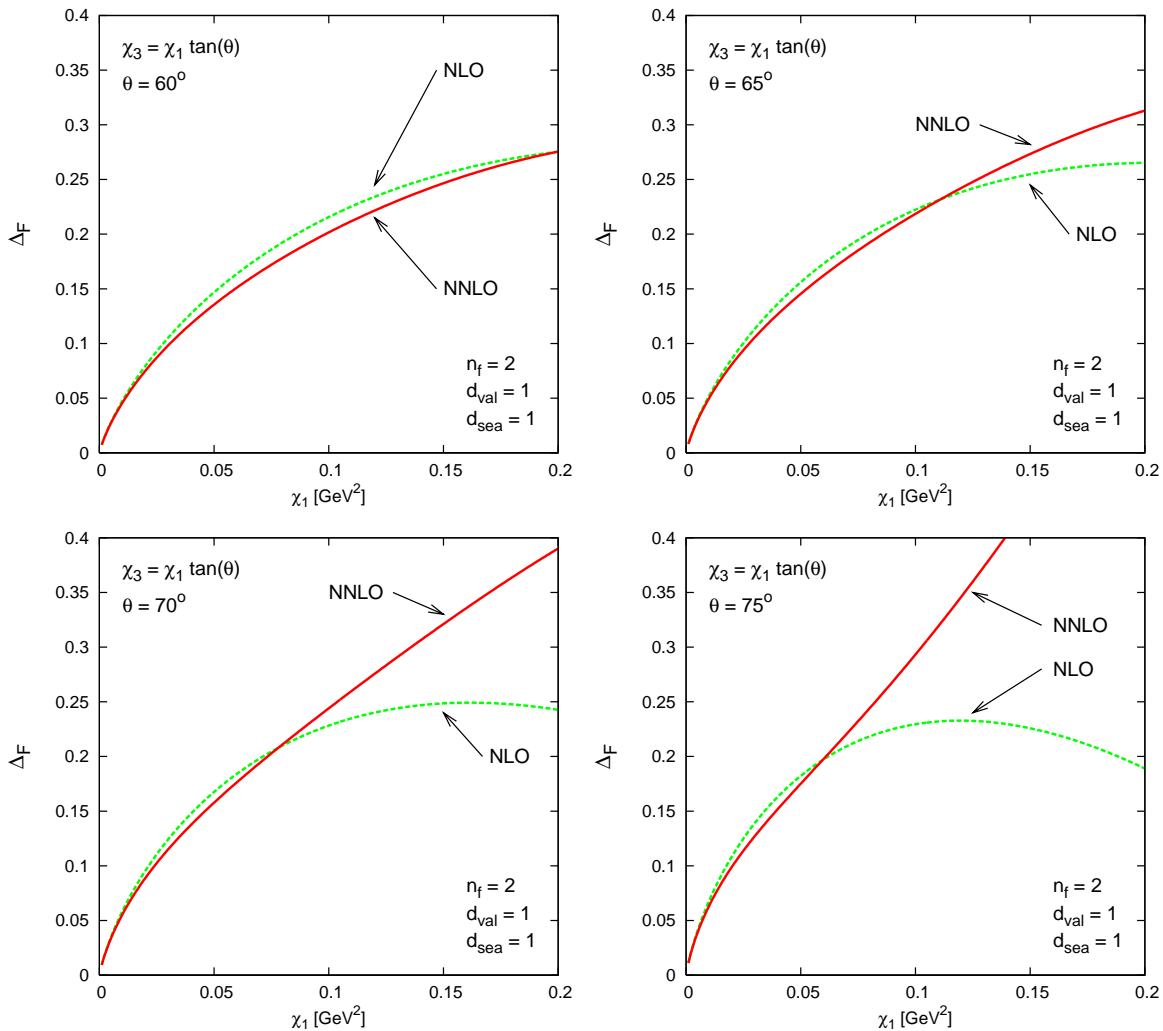}
\caption{Comparison of the NLO and NNLO corrections $\Delta_F$ to the 
decay constant for $d_{\mathrm{val}} = 1$ and $d_{\mathrm{sea}} = 1$. 
The curves labeled 'NLO' represent the ${\cal O}(p^4)$ contribution to 
the decay constant, and those labeled 'NNLO' represent the sum of the 
${\cal O}(p^4)$ and ${\cal O}(p^6)$ contributions. The dependence on the 
sea-quark mass $\chi_3$ has been parameterized as $\chi_3 = 
\chi_1\tan(\theta)$. Present Lattice QCD simulations are performed in 
the region where $\theta > 45^{\circ}$.}
\label{F0ray}
\end{center}
\end{figure*}
\end{widetext}

\subsection{Numerical Results for Decay Constants}

The modification of the decay constant, when corrections up to NNLO 
are taken into account, is parameterized in terms of the relative change 
$\Delta_F$, which is defined as
\begin{eqnarray}
F_{\mathrm{phys}} &=& F\,\left(1 + \Delta_F \right),
\label{deltF}
\end{eqnarray}
so that $\Delta_F$ is expected to disappear in the chiral limit, when 
both sea and valence quark masses are set to zero. That this condition 
is satisfied can be seen directly from Fig.~\ref{F02dim}. However, if 
the sea-quark mass is set to zero for a constant valence quark mass, the 
decay constant approaches a constant value. This behavior is expected 
and may be attributed to the unphysical quenched chiral logarithms and 
similar effects.

The numerical results for the pseudoscalar meson decay constants are 
shown in Figs.~\ref{F02dim} and~\ref{F0ray}. For simplicity, we have 
considered the case with $d_{\mathrm{val}} = 1$ and $d_{\mathrm{sea}} = 
1$, i.e. the valence quark masses and sea-quark masses are degenerate, 
but different from each other. Thus the results presented correspond to 
Eq.~(\ref{F0_NLO_nf2_12}) for NLO, and Eqs.~(\ref{F0tree_NNLO_nf2_12}) 
and~(\ref{F0loop_NNLO_nf2_11}) for NNLO. 

\begin{figure*}[ht!]
\begin{center}
\includegraphics[width=.87\textwidth]{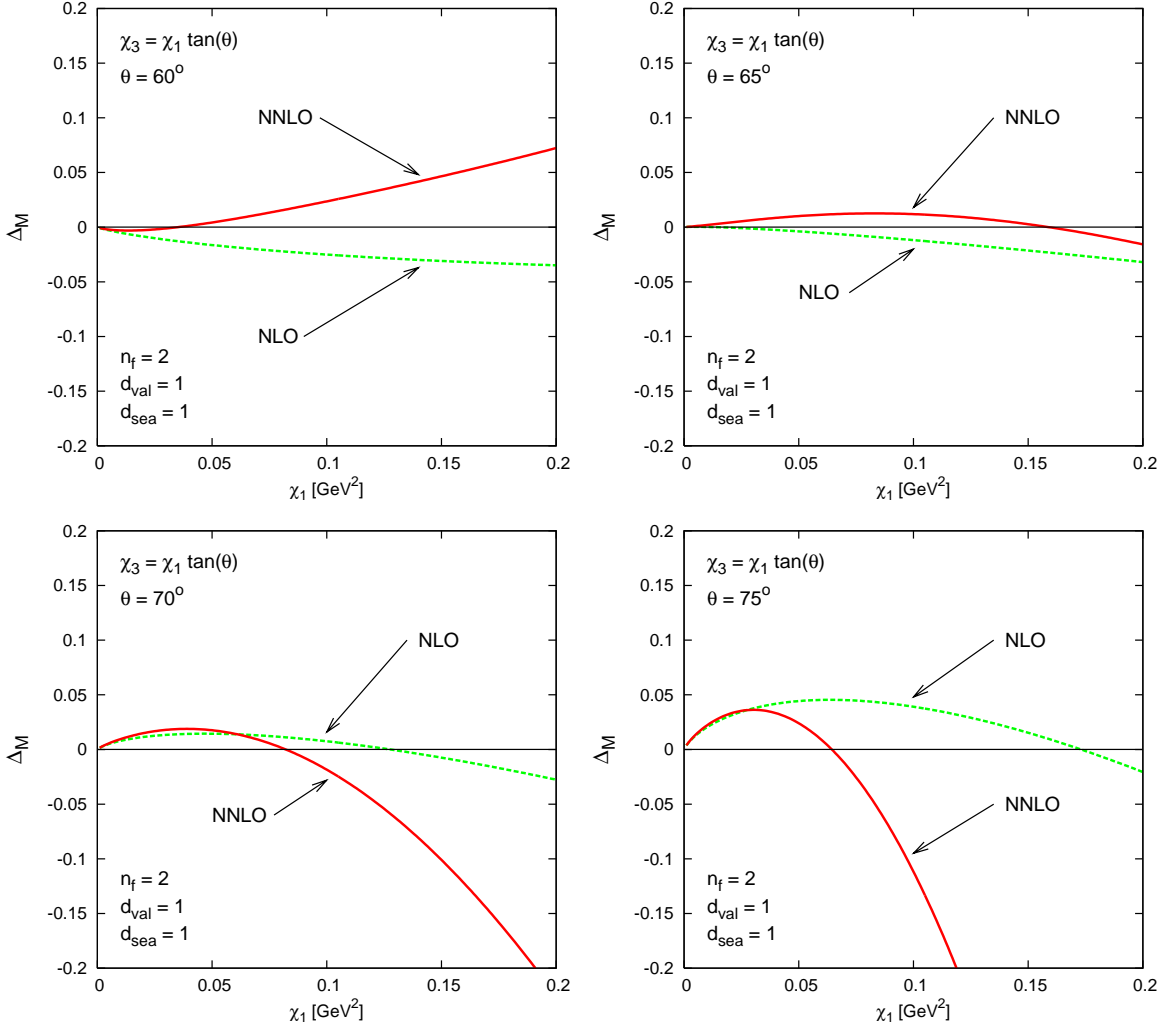}
\caption{Comparison of the NLO and NNLO corrections $\Delta_M$ to the 
meson mass for $d_{\mathrm{val}} = 1$ and $d_{\mathrm{sea}} = 1$. 
The curves labeled 'NLO' represent the ${\cal O}(p^4)$ contribution to 
the meson mass, and those labeled 'NNLO' represent the sum of the 
${\cal O}(p^4)$ and ${\cal O}(p^6)$ contributions. The dependence on the 
sea-quark mass $\chi_3$ has been parameterized as $\chi_3 = 
\chi_1\tan(\theta)$. Present Lattice QCD simulations are performed in 
the region where $\theta > 45^{\circ}$.}
\label{M0ray}
\end{center}
\end{figure*}

The convergence of the chiral expansion for the decay constant is 
investigated in Fig.~\ref{F0ray} for different combinations of sea and 
valence quark masses. We have chosen the sea-quark mass larger than the 
valence quark mass in our examples since this is the most likely regime 
for lattice QCD simulations. Overall, Fig.~\ref{F02dim} shows that the 
NNLO corrections to the decay constant are significant at least for our 
choice of LEC:s. Nevertheless, the comparison between the NLO and NNLO 
curves in Fig.~\ref{F0ray} also shows that there exists a rather large 
range of input quark masses for which the convergence of the chiral 
expansion is reasonable. The cases with less degenaracies in the input 
quark masses could also be considered, but such plots are difficult to 
present in an instructive way since these functions depend on up to four 
different quark masses. One possible way to proceed is to look at 
various special cases where the ratio between the sea or valence quark 
masses is fixed. To illustrate the effects of nondegenerate sea and 
valence quarks, such plots are given in the next subsection for the 
meson mass.

\subsection{Numerical Results for Meson Masses}

\begin{figure}[ht!]
\begin{center}
\includegraphics[width=\columnwidth]{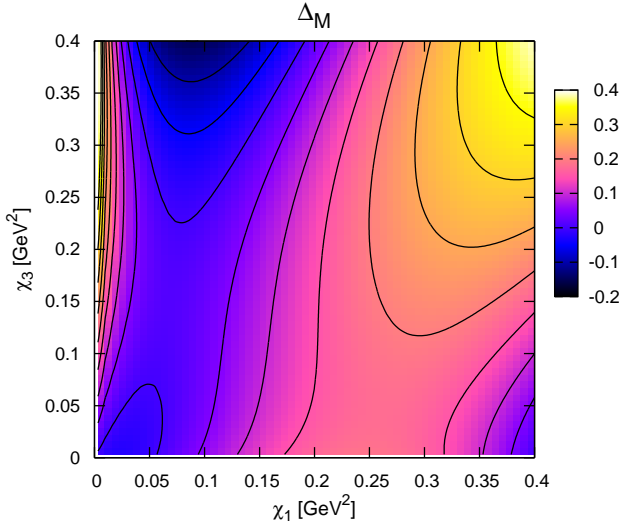}
\caption{The NNLO modification $\Delta_M$ of the meson mass for 
$d_{\mathrm{val}} = 1$ and $d_{\mathrm{sea}} = 1$ as a function 
of the valence quark mass parameter $\chi_1$ and the sea-quark mass 
parameter $\chi_3$. The function plotted represents the sum of the 
${\cal O}(p^4)$ and ${\cal O}(p^6)$ contributions. Note the divergence 
of the result near $\chi_1 = 0$, which vanishes for $\chi_3 = 0$.}
\label{M02dim}
\end{center}
\end{figure}

\begin{figure}[ht!]
\begin{center}
\includegraphics[width=\columnwidth]{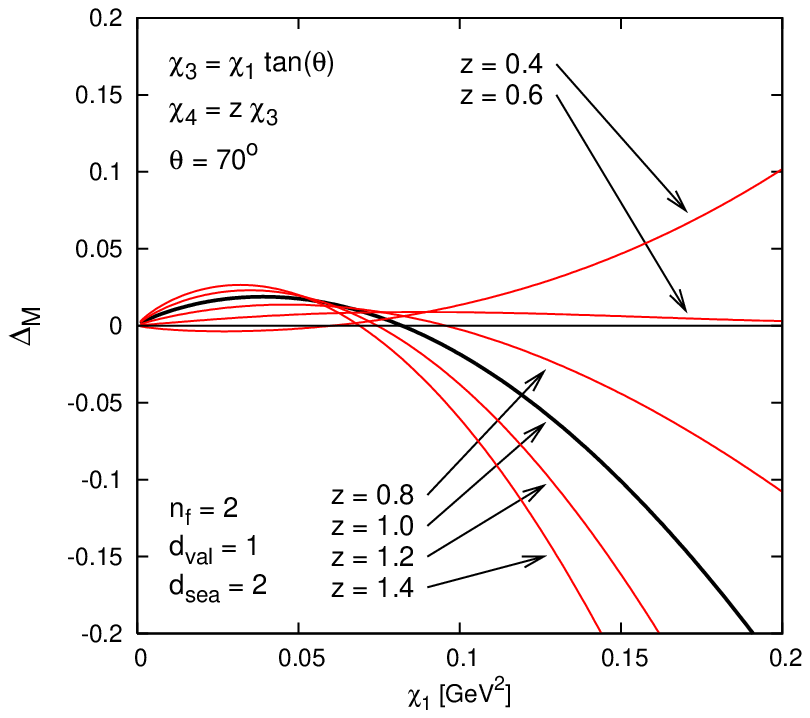}
\caption{The modification $\Delta_M$ of the NNLO meson mass for 
$d_{\mathrm{val}} = 1$ and $d_{\mathrm{sea}} = 2$. The sea quark mass 
dependence has been parameterized as $\chi_4 = z\chi_3$ and the 
relation between the sea and valence quark masses as $\chi_3 = 
\chi_1 \tan(\theta)$.}
\label{M01+2}
\end{center}
\end{figure}

\begin{figure}[ht!]
\begin{center}
\includegraphics[width=\columnwidth]{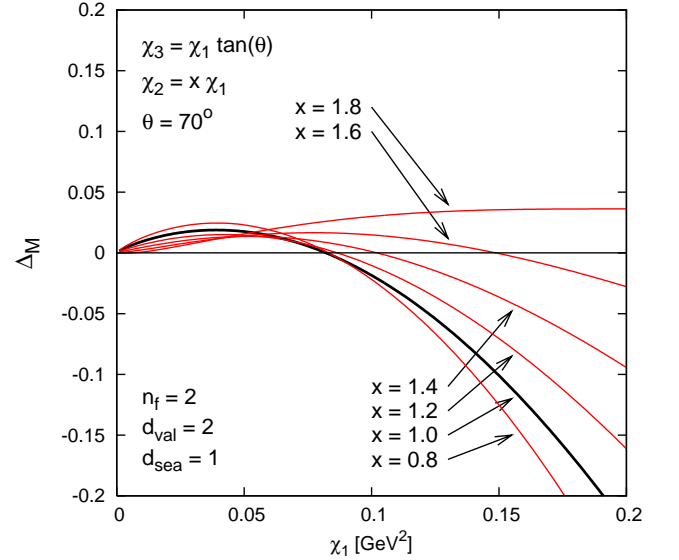}
\caption{The modification $\Delta_M$ of the NNLO meson mass for 
$d_{\mathrm{val}} = 2$ and $d_{\mathrm{sea}} = 1$. The valence quark 
mass dependence has been parameterized as $\chi_2 = x\chi_1$ and the 
relation between the sea and valence quark masses as $\chi_3 = 
\chi_1 \tan(\theta)$.}
\label{M02+1}
\end{center}
\end{figure}

The numerical results for the pseudoscalar meson masses are shown 
in Figs.~\ref{M0ray} through~\ref{M02+1}. Part of the presentation is 
similar to that of the decay constant results in the preceding section, 
i.e. results are given for $d_{\mathrm{val}} = 1$ and $d_{\mathrm{sea}} 
= 1$. These results correspond to Eq.~(\ref{M0_NLO_nf2_12}) for NLO, and 
Eqs.~(\ref{M0tree_NNLO_nf2_12}) and~(\ref{M0loop_NNLO_nf2_11}) for NNLO. 
In addition some results are also presented for $d_{\mathrm{val}} = 
1$, $d_{\mathrm{sea}} = 2$ in Fig.~\ref{M01+2} and for $d_{\mathrm{val}} 
= 2$, $d_{\mathrm{sea}} = 1$ in Fig.~\ref{M02+1}. The modification of 
the meson mass when NNLO corrections are taken into account is 
parameterized in terms of the quantity $\Delta_M$, which 
is defined as
\begin{eqnarray}
M^2_{\mathrm{phys}} &=& M^2_0\,\left(1 + \Delta_M \right),
\label{deltM}
\end{eqnarray}
so that $\Delta_M$ represents a relative deviation from the lowest order 
result $M^2_0$. With this definition, $\Delta_M$ is expected to disappear in 
the chiral limit, when both sea and valence quark masses approach zero at a
constant ratio. 

In general, the appearance of Fig.~\ref{M02dim} indicates 
that the NNLO contribution to the meson mass is more pronounced than for the 
decay constant, which is due to the fact that the NLO contribution to the 
meson mass vanishes for certain values of the input quark masses. Another 
noteworthy feature of Fig.~\ref{M02dim} is the logarithmic divergence for 
small values of the valence quark mass $\chi_1$. This feature, which is 
well-known at NLO and persists at NNLO, is due to the unphysical quenched 
chiral logarithms. This divergence disappears if the sea and valence quark 
masses are sent to zero in constant ratio, which is also confirmed by 
Figs.~\ref{M0ray} through~\ref{M02+1}.

There is no {\it a priori} reason to believe that the convergence of the 
meson mass should be intrinsically worse than that of the decay 
constant. Nevertheless, the combined NLO + NNLO result in 
Fig.~\ref{M02dim} vanishes in several places, with the NLO and NNLO
contributions canceling each other. The decay constants showed a more 
uniform convergence for the masses and input LEC:s used here.
In this respect the curves shown in Fig.~\ref{M0ray} are 
less instructive as to the convergence of the chiral expansion than 
those shown in Fig.~\ref{F0ray}. Admittedly, however, the inclusion of 
NNLO effects into the pseudoscalar meson mass in PQ$\chi$PT is a major 
effect which changes features of the NLO calculation. 

It is also noteworthy that the convergence of the chiral expansion 
for the quantities considered apparently depends rather strongly on 
the values of the LEC:s. For example, the obvious trial choice of 
setting all the $L_i^r$ to zero leads, incidentally, to a much larger 
NNLO contribution which converges less well also for the decay 
constants. Setting several of the $K_i^r$ to nonzero values can also 
change this picture quite dramatically. Thus a definite statement 
about convergence can only be made when some estimates are available 
for the $K_i^r$.

\subsection{Determination of LEC:s from Lattice QCD Simulations}

In the previous sections, a sample of numerical results have been 
presented in order to illustrate the effects of the NNLO 
contributions when the LEC:s are known. Since the intended area of 
application of the NNLO formulas given in this paper is the 
determination of LEC:s from PQQCD Lattice simulations, some attention 
should also be devoted to finding efficient methods for this task. 
Again, it is understood that the $L^{r(2pq)}_i$ and $K^{r(2pq)}_i$ 
are simply denoted by $L_i^r$ and $K_i^r$. 

At NLO in PQ$\chi$PT, the expressions for the pseudoscalar meson 
masses and decay constants can be cleanly separated into a tree-level 
contribution which is a function of the $L_i^r$ and a remaining part, 
independent of the $L_i^r$, which involves the chiral logarithms. 
Thus at NLO it is possible to develop various optimal strategies for 
the purpose of obtaining estimates of the $L_i^r$ by fits to Lattice 
QCD data~\cite{Sharpe1}. At NLO, the dependence on the $L_i^r$
is proportional to $2L_4^r\,\chi_{34} \,+\, L_5^r\,\chi_{12}$ for the 
decay constant and $(4L_6^r \,-\, 2L_4^r)\,\chi_{34} + 
(2L_8^r \,-\, L_5^r)\,\chi_{12}$ for the masses, from which an 
overall factor $\chi_{12}$ has been removed. The LEC:s can thus be 
determined from the 1+1 case by using a constant sea quark mass and
varying the valence quark mass. The more complicated cases with 
nondegenerate quarks, i.e. $\chi_1\ne\chi_2$ or $\chi_3\ne\chi_4$ are 
not needed at NLO.


A similar analysis which involves the determination of LEC:s from a 
fit to Lattice QCD simulations using NNLO PQ$\chi$PT is more 
challenging, not only because of the complexity of the NNLO 
expressions, but mainly since the chiral logarithms now involve the 
$L_i^r$. Furthermore the coefficient functions of the $L_i^r$ at NNLO 
have a more complicated dependence on both sea and valence quark 
masses. On the other hand, the situation for the $K_i^r$ at NNLO is 
similar to that of the $L_i^r$ at NLO, although the much larger 
number of $K_i^r$ parameters, along with the appearance of 
contributions bilinear in the $L_i^r$ certainly complicates the 
analysis.

In general, at ${\cal O}(p^6)$ the expressions for the pseudoscalar 
meson masses and decay constants depend on 12~$K_i^r$ parameters, 5 
for the decay constants and an additional 7 for the masses. The 
dependence of the decay constant on the $K_i^r$ is proportional to
\begin{eqnarray}
\label{eq:Kidecay}
&& 16\,\chi_{12}^2\,K_{19}^r 
  \:-\,8\,\chi_1\chi_2\,\left( K_{19}^r - K_{23}^r \right)
  \:+\,16\,\chi_{12}\chi_{34}\,K_{20}^r \nonumber \\
&& - \,\:16\,\chi_3\chi_4\,K_{21}^r
  \:+\,32\,\chi_{34}^2\,\left(K_{21}^r + K_{22}^r\right),
\end{eqnarray}
from which all 5 appearing constants, $K_{19 \ldots 23}^r$ can be 
determined but at least one case with nondegenerate sea quark masses 
needs to be calculated. From the 1+1 case, three combinations can be 
determined. Also, to separate $K_{19}^r$ from $K_{23}^r$ a case with 
nondegenerate valence quark masses, $d_{\mathrm{val}}=2$, is needed, 
otherwise only the combination $K_{19}^r \,+\, K_{23}^r$ can be 
obtained. Likewise nondegenerate sea quark masses, 
$d_{\mathrm{sea}}=2$, are needed to disentangle $K_{21}^r$ from 
$K_{22}^r$, simulations with degenerate sea quark masses allow only 
the determination of $K_{21}^r \,+\, 2K_{22}^r$. Similarly, the 
dependence of the meson masses on the $K_i^r$ is
\begin{eqnarray}
\label{eq:Kimass}
&& - \,\:\chi_{12}^2\,\left( 32\,K_{17}^r + 32\,K_{19}^r 
- 96\,K_{25}^r - 32\,K_{39}^r \right) \nonumber \\
&& + \,\:\chi_1\chi_2\,\left( 16\,K_{19}^r - 16\,K_{23}^r 
- 48\,K_{25}^r \right) \nonumber \\
&& - \,\:\chi_{12}\chi_{34}\,\left( 64\,K_{18}^r + 32\,K_{20}^r 
- 64\,K_{26}^r - 64\,K_{40}^r \right) \nonumber \\
&& + \,\:\chi_3\chi_4\,\left( 32\,K_{21}^r - 32\,K_{26}^r \right)
\nonumber \\ 
&& - \,\:\chi_{34}^2\,\left( 64\,K_{21}^r + 64\,K_{22}^r - 
64\,K_{26}^r - 192\,K_{27}^r \right), \hspace{.7cm}
\end{eqnarray}
from which an overall factor of $\chi_{12}$ has again been removed.
From the 1+1 case we can obtain 3 combinations, one more from
a case with $d_{\mathrm{val}}=2$ and one more from a case with 
$d_{\mathrm{sea}}=2$. The particular combinations can be easily read off
from Eq.~(\ref{eq:Kimass}).
Thus, in order to obtain independent information on as many of the 
$K_i^r$:s as possible, simulations with nondegenerate quark masses 
are clearly preferrable. On the other hand, a possible strategy for 
the determination of the $L_i^r$ from NNLO fits is to use only one 
distinct sea and valence quark mass, since then the number of 
distinct combinations of $K_i^r$ parameters is minimal, while the 
possibility of distinguishing all of the $L_{4 \ldots 6}^r$ and 
$L_8^r$ is retained. 

At NNLO, additional $L_i^r$ are present in the 
expressions for the decay constant and the meson mass, namely $L_7^r$
and the $L_{0 \ldots 3}^r$. Of these, $L_7^r$ can be determined from 
the diagonal mesons~\cite{Sharpe1} by consideration of the 
coefficient of the double pole. This calculation is not yet performed 
up to NNLO. The other four $L_i^r$ are those relevant at NLO for 
meson-meson scattering. Their values have to be used as input but 
could in principle be separated from the $K_i^r$ since they multiply 
nonanalytic dependences on the quark masses. A detailed study of this 
is beyond the scope of the present work.

The present analysis shows that while the number of free 
parameters in a fit of the NNLO expressions to Lattice QCD data is 
large, the possibility of producing a large amount of simulation data 
with different combinations of sea and valence quark masses should 
compensate for this. Furthermore, it should be kept in mind that many 
of the in principle free parameters $L_i^r$ are already quite 
accurately known from previous work.

\subsection{Conclusions}

In summary, we have calculated the pseudoscalar meson masses and decay 
constants to NNLO in Partially Quenched Chiral Perturbation Theory for 
two flavors of sea-quarks. Explicit analytical formulas of reasonable 
length and complexity have been given, along with a numerical analysis 
which demonstrates the numerical implementation of the results. Although 
the NNLO results appear to be of a reasonable magnitude, definite 
statements concerning the convergence of the chiral expansion have to be 
postponed, as the magnitude of the results depends rather sensitively on 
a number of largely unknown LEC:s.

An implementation of the present results will be made available by the 
authors in the near future~\cite{website}. The availability of these 
results makes it possible to perform a NNLO chiral extrapolation from 
partially quenched lattice data. However, usage of the present results 
requires Lattice QCD data on the pseudoscalar meson masses and decay 
constants which has been extrapolated to the continuum limit and to 
infinite volume at different values of the quark masses. Although such 
data are not yet available, this situation will undoubtedly change in 
the near future. The ultimate goal of such fits to lattice data is thus 
to provide reliable estimates of the (unphysical) low-energy constants 
of PQ$\chi$PT, from which the values of the physical ones can then be 
extracted.

\section*{Acknowledgments}

The program FORM 3.0 has been used extensively in these calculations
\cite{FORM}. This work is supported by the European Union TMR network,
Contract No. HPRN-CT-2002-00311  (EURIDICE). TL wishes to thank Eduardo 
de Rafael, Marc Knecht, Laurent Lellouch and J\"urg Gasser for 
instructive discussions, and F.~Farchioni for correspondence. TL also 
thanks the Mikael Bj\"ornberg memorial foundation for a travel grant.

\end{document}